\documentclass{emulateapj}

\usepackage{graphics}
\usepackage{epsfig}
\usepackage{natbib}
\usepackage[hidelinks]{hyperref}
\usepackage{wrapfig}
\usepackage{csvsimple}
\usepackage{adjustbox}
\usepackage{amsmath}
\usepackage{graphicx}
\usepackage[toc,page]{appendix}
\usepackage{booktabs}
\usepackage{blkarray}

\newcommand{\spi}{$M_{plan}$/$a_{plan}$}

\shorttitle{FUV activity on F, G, K, and M Exoplanet Host Stars}
\shortauthors{France et al.}
\begin{document}

\title{Far-Ultraviolet Activity Levels of F, G, K, and M dwarf Exoplanet Host Stars\altaffilmark{*}}


\author{
Kevin France\altaffilmark{1},
Nicole Arulanantham\altaffilmark{1},
 Luca Fossati\altaffilmark{2}, 
Antonino F. Lanza\altaffilmark{3}, 
R. O. Parke Loyd\altaffilmark{4}, 
Seth Redfield\altaffilmark{5},
P. Christian Schneider\altaffilmark{6}
} 

\altaffiltext{*}{Based on observations made with the NASA/ESA $Hubble$~$Space$~$Telescope$, obtained from the data archive at the Space Telescope Science Institute. STScI is operated by the Association of Universities for Research in Astronomy, Inc. under NASA contract NAS 5-26555.}

\altaffiltext{1}{Laboratory for Atmospheric and Space Physics, University of Colorado, 600 UCB, Boulder, CO 80309; USA,  kevin.france@colorado.edu}
\altaffiltext{2}{Space Research Institute, Austrian Academy of Sciences, Schmiedlstrasse 6, A-8042 Graz, Austria}
\altaffiltext{3}{INAF - Osservatorio Astrofisico di Catania, Via S. Sofia, 78, 95123 Catania, Italy}
 \altaffiltext{4}{School of Earth and Space Exploration, Interplanetary Initiative, Arizona State University, Tempe, AZ, USA}	
\altaffiltext{5}{Astronomy Department and Van Vleck Observatory, Wesleyan University, Middletown, CT 06459-0123, USA}
\altaffiltext{6}{Hamburger Sternwarte, Gojenbergsweg 112, 21029 Hamburg, Germany}



\begin{abstract}

We present a survey of far-ultraviolet (FUV; 1150~--~1450~\AA) emission line spectra from 71 planet-hosting and 33 non-planet-hosting F, G, K, and M dwarfs with the goals of characterizing their range of FUV activity levels, calibrating the FUV activity level to the 90~--~360~\AA\ extreme-ultraviolet (EUV) stellar flux, and investigating the potential for FUV  emission lines to probe star-planet interactions (SPIs).   We build this emission line sample from a combination of new and archival observations with the {\it Hubble Space Telescope}-COS and -STIS instruments, targeting the chromospheric and transition region emission lines of \ion{Si}{3}, \ion{N}{5}, \ion{C}{2}, and \ion{Si}{4}.   

We find that the exoplanet host stars, on average, display factors of 5~--~10 lower UV activity levels compared with the non-planet hosting sample; this is explained by a combination of observational and astrophysical biases in the selection of stars for radial-velocity planet searches.   We demonstrate that UV activity-rotation relation in the full F~--~M star sample is characterized by a power-law decline (with index $\alpha$~$\approx$~$-$1.1),  starting at rotation periods $\gtrsim$~3.5 days.    Using \ion{N}{5} or \ion{Si}{4} spectra and a knowledge of the star's bolometric flux, we present a new analytic relationship to estimate the intrinsic stellar EUV irradiance in the 90~--~360~\AA\ band with an  accuracy of roughly a factor of $\approx$~2.    Finally,  we study the correlation between SPI strength and UV activity in the context of a principal component analysis that controls for the sample biases.  We find that SPIs are not a statistically significant contributor to the observed UV activity levels.

\end{abstract}



\section{Introduction}

The success of planet searches employing radial velocity techniques and transit photometry has demonstrated that $\sim$~300~--~400 stars in the solar neighborhood ($d$~$<$~50 pc) host confirmed planetary systems. $TESS$ will expand this list dramatically in the next several years.  With so many planets now discovered, the next step towards the study of ``comparative planetology'' is the characterization of the physical processes that shape these worlds.  Of particular interest are the environmental parameters that control the physical and chemical state of potentially inhabited rocky planets around cool stars (M~--~F dwarfs; $T_{eff}$~$\approx$~2500~--~6000 K). 
These include the high-energy photon and particle environment~\citep{segura10,tilley18}, as well as the potential for stellar and planetary magnetospheres to interact~\citep{garraffo16}.  These ``exoplanet space weather''  effects may ultimately control the habitability of these systems (e.g., Airapetian et al. 2017)\nocite{airapetian17}.  NASA and ESA are currently studying design reference missions for the detection and/or spectroscopic characterization of potentially habitable rocky planets (e.g., Rauer et al. 2014; Mennesson et al. 2016; France et al. 2016a; Roberge et al. 2017).  However, rocky planets around M dwarfs will likely be the only potentially habitable planets whose atmospheres can be probed for signs of life (with JWST and ELTs) prior to a Large UVOIR mission in the 2030s~--~2040s (Deming et al. 2009; Belu et al. 2011; Snellen et al. 2015).\nocite{deming09,belu11,snellen15}   We need to characterize the radiation and magnetic environments of our stellar neighbors so that spectroscopic observations of their planets can be confidently interpreted.  \nocite{rauer14,habex16,france16b,roberge17} 

\subsection{The Importance of the Host Star} 

It is now clear that the planetary effective surface temperature alone is insufficient to characterize the habitable zone (HZ) and accurately interpret atmospheric gases with a potentially biological origin.  The UV stellar spectrum is required to understand HZ atmospheres, as it both drives and regulates atmospheric heating and chemistry on Earth-like planets and is critical to the long-term stability of terrestrial atmospheres.  Our quest to discover and characterize biological signatures on rocky planets must consider the star-planet system as a whole, including the interaction between the stellar photons, particles, and the exoplanetary atmosphere.   The dependence of abiotic formation of ``biomarker''  molecules (e.g., O$_{2}$, O$_{3}$, CH$_{4}$, and CO$_{2}$;  e.g., Kaltenegger et al.  2007; Seager et al.  2009) on the stellar far- and near-UV irradiance (approximately 912~--1700~\AA\ and 1700~--~3200~\AA\, respectively), particularly around M dwarfs, has been well-documented (e.g., Hu et al. 2012; Tian et al. 2014; Harman et al. 2016; Shields et al. 2016).\nocite{shields16}  

In addition,  the long-term stability of the atmospheres of rocky planets is driven by the ionizing radiation and particle output of their host stars.  Atmospheric escape is a key factor shaping the evolution and distribution of low-mass planets (e.g., Owen \& Wu 2013; Lopez \& Fortney 2013) and their habitability (Lammer et al. 2009; Cockell et al. 2016).   Extreme-UV (EUV; 100~$\lesssim$~$\lambda$ $\lesssim$~911~\AA) photons from the central star drive thermospheric heating, and this may lead to significant atmospheric escape~\citep{tian08,murray09,vidal03,bourrier13,ehrenreich15,spake18}.  Ionization by EUV photons and the subsequent loss of atmospheric ions to stellar wind pick-up can also drive extensive atmospheric mass-loss on geologic time scales (e.g., Rahmati et al. 2014 and references therein).\nocite{rahmati14}   Stellar FUV observations serve as means for predicting the ionizing (extreme-UV) flux from cool stars,  either through the use of solar scaling relations~\citep{linsky14,youngblood16} or more detailed differential emission measure (DEM) techniques (e.g.,  Louden et al. 2017).\nocite{louden17,murray09,owen13,lopez13}

\subsection{Exoplanetary Magnetic Fields and Star-Planet Interactions} 

​A planet and its host star may interact in many ways, with most studies focusing on their photon+particle, gravitational, and magnetic field interactions (e.g., Cuntz et al. 2000). 
A central question for a planet's ability to retain an atmosphere  is  ``what is the role of magnetic fields?''\citep{adams11,nascimento16}.
Searches for exoplanetary magnetic fields have not yielded any firm detections to date~\citep{greissmeier15}.  Magnetic fields play a crucial role in protecting surface life from damaging high-energy particles from stellar winds and coronal mass ejections (CMEs; Lammer et al. 2012) as well as promoting the long-term stability of planetary atmospheres (Tian 2015).  In the solar system, Earth is the only ``habitable zone'' rocky planet (roughly comprising Venus, Earth, and Mars) that was able to retain its water and the only planet out of the three that has a substantial magnetic field today.\nocite{lammer12,tian15}  


 Magnetic star-planet interactions (SPIs) have gained interest in the community because they might provide a way to detect and measure planetary magnetic fields (e.g., Vidotto et al. 2010; Lanza 2015; Cauley et al. 2015; Rogers 2017).\nocite{fossati18,pillitteri14,cuntz00,vidotto10,lanza15,cauley15,rogers17}  The presence of a planetary magnetic field may induce interactions that can generate planetary radio emission (Zarka 2007; Ignace et al. 2010; Vidotto et al. 2012), early-ingress NUV light curves~\citep{fossati10,vidotto10,cauley15}, enhanced flare activity~\citep{pillitteri15}, and FUV aurorae (Yelle 2004; Menager et al. 2013). Radio emission from these systems remains inconclusive~\citep{bastian18}, and  NUV light curve interpretations are debated~\citep{turner16b,turner16a}.
 Close-in giant planets are predicted to have substantial magnetic field strengths  (Christensen et al. 2009), however, auroral emission from exoplanets has not been conclusively detected so far (e.g., Bastian et al. 2000; Lazio et al. 2004; France et al. 2010; Hallinan et al. 2013; Lecavelier des Etangs et al. 2013; Kruzcek et al. 2017).  Enhanced flare activity in favorable star-planet systems~\citep{lanza18} appears promising and phase-resolved observations may provide more direct clues on the properties of exoplanetary magnetism.  
 \nocite{christiansen09,zarka07,ignace10,vidotto12,yelle04,
menager13,bastian00,lazio04,france10,hallinan13,lecavelier13,kruczek17}

Exoplanetary magnetic fields may be indirectly observable by the influence they produce on their host stars;  one possible form of the oft searched-for stellar SPIs (e.g., Shkolnik et al. 2003; 2005; 2017;  Lanza 2008, Lanza 2013).\nocite{shkolnik03,shkolnik05,lanza08,lanza13,shkolnik17}   The magnitude of this SPI, as measured by the energy dissipated in the stellar atmosphere, should depend on the strength of the stellar magnetic field, the planetary magnetic field, and the relative speed of the planet's orbital velocity compared to the stellar magnetic rotation rate (Lanza 2012).   While this technique does not provide a direct measure of the planetary magnetic field strength, it does allow for both the detection of exoplanetary magnetic fields and their influence on their host stars.

Tidal (gravitational) SPIs may alter the rotational evolution of the host star and the orbital evolution of the planet~\citep{poppenhaeger14}. In this way, tides may significantly affect the stellar activity level. This phenomenon should be particularly efficient for massive late-type stars, where the convective layers driving the stellar activity are thin and thus more easily affected by tides induced from the planet. Pillitteri et al. (2014) and Fossati et al. (2018) concluded that this is the case of the WASP-18 system, which contains a massive $\approx$10 M$_{J}$ planet orbiting a mid-F-type star with a period of $\approx$1 day.  X-ray and far-UV observations of WASP-18 indicate that the star has an anomalously low activity level for its young age, which Pillitteri et al. (2014) argued is driven by the tidal forces induced by the massive planet disrupting the $\alpha$-$\Omega$ hydromagnetic dynamo in the host star.

Using data from the MUSCLES survey of planet hosting M dwarfs~\citep{france16,youngblood16,loyd16}, we recently presented a  tentative detection of stellar SPI~\citep{france16}.   Because magnetic field strength increases with planetary mass in the solar system, one may expect that the most massive, closest-in planets in exoplanetary systems produce the largest signal on their host stars, therefore SPI signals could be expected to correlate with $M_{plan}$/$a_{plan}$ 
(or other proportionalities between the dissipated power and the star-planet system configuration, see Section 4.2), where $M_{plan}$ is the planetary mass and $a_{plan}$ is the semi-major axis (see, e.g., Miller et al. 2015).    The MUSCLES database allowed us to explore SPI as a function of emission line formation temperature.  Probing different temperature regimes was critical for the tentative detection of SPI in MUSCLES (described below), and can be used to constrain the possible location of magnetic field line reconnection and subsequent location of the plasma heating.  \citet{france16}~suggested that the systems with close-in, massive planets may indeed be generating enhanced transition region activity, as probed by $\sim$~(0.3~--~2)~$\times$~10$^{5}$ K gas.   Conversely, no correlations with the cooler gas emitting in the lower-chromosphere were observed (traced by \ion{Mg}{2} and \ion{Si}{2}, $T_{form}$~$\lesssim$~10$^{4}$ K).  

However, the small  sample size and low-significance  of the MUSCLES result ($\approx$~2-$\sigma$) compelled us to develop a larger sample with broader spectral type coverage.  Expanding the observational basis for understanding the environmental drivers of exoplanet atmospheres and refining the SPI study were the primary motivations for assembling the large sample of exoplanet host stars and non-planet hosting control group presented in this work.   

\subsection{A Survey of the Chromospheric and Transition Region Activity of Exoplanet Host Stars}

In this paper,  we present a new far-UV emission line survey of exoplanet host stars, including all of the available archival data from $HST$-STIS and COS (spectra from $IUE$ are largely too low quality for this work; France et al.  2016b)\nocite{france16}.  We acquired new $HST$-COS  observations of 45 host stars, and have assembled the largest UV spectroscopic exoplanet host star and non-planet hosting control sample to date (Tables 1 and 2).   For simplicity,  we refer to stars without known planetary systems as `non-planet hosts', but acknowledge that many of these stars likely have planetary systems not yet discovered (Section 2.2).   We use these data to compare UV activity levels  from a range of formation temperatures in the chromosphere and transition region ($T_{form}$~$\approx$~20,000~--~200,000 K) in F, G,  K, and M dwarfs with and without (known) planets.   Using our planet-hosting sample, we examine the correlations between stellar activity and a proposed parameterization of the SPI strength (\spi).   In Section 2,  we describe the stellar sample, target selection process, and the new $HST$ observations made in support of this work.  Section 3 describes the data reduction and spectral line analysis.   Section 4 presents an overview of the results on activity levels of planet-hosts, a new scaling to the EUV flux from these stars, the strength of the SPI signal in the data, and numerical techniques developed to compare the UV activity to stellar and planetary parameters.  We  present a brief summary of this work in Section 5.

\begin{figure}[!h]
\begin{center}
\vspace{0.0in}
\begin{tabular}{c}
\includegraphics[width=0.5\textwidth,angle=0,trim={.0in 0.0in 0.0in 0.0in},clip]{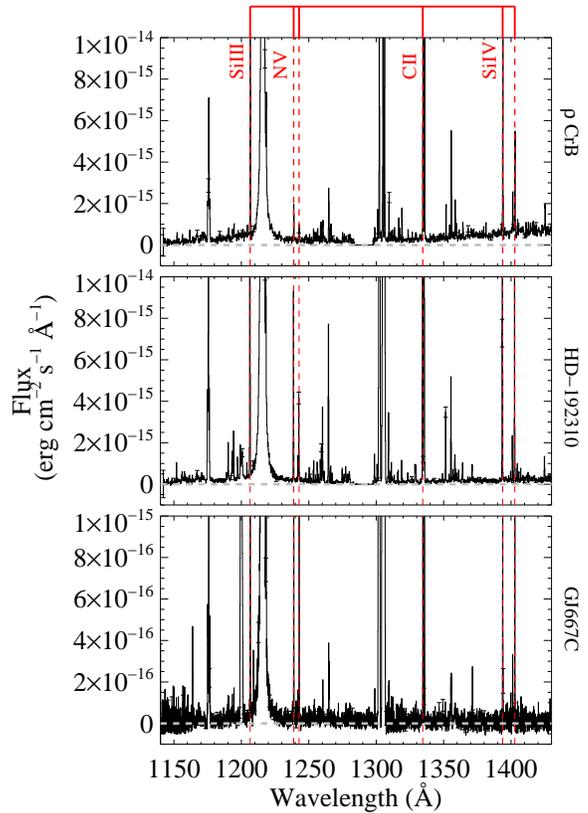}
\end{tabular}
\end{center}
\vspace{-0.3in}
\caption{Example FUV exoplanet host star spectra used in this work.  From top-to-bottom, representative G dwarf ($\rho$ CrB; G0V, $V$~=~5.39), K dwarf (HD 192310; K2V, $V$~=~5.72), and M dwarf (GJ 667 C; M2.5V, $V$~=~10.22) spectra obtained with $HST$-COS G130M.  Prominent hot gas lines studied here (\ion{Si}{3}~$\lambda$~1206~\AA; log$_{10}$$T_{form}$~=~4.7, \ion{N}{5}~$\lambda$~1240~\AA; log$_{10}$$T_{form}$~=~5.2, \ion{C}{2}~$\lambda$~1335~\AA; log$_{10}$$T_{form}$~=~4.5, and \ion{Si}{4}~$\lambda$~1400~\AA; log$_{10}$$T_{form}$~=~4.9; Dere et al. 2009) are marked with red dashed lines.  Strong emission lines at 1216 and 1304~\AA\ are mainly geocoronal emission from neutral hydrogen and oxygen in Earth's upper atmosphere.    }
\label{fig-proveit} 
\vspace{0.05in}
\end{figure}

\begin{figure}[]
\begin{center}
\vspace{0.0in}
\begin{tabular}{c}
\includegraphics[width=0.5\textwidth,angle=0,trim={.0in 0.0in 0.0in 0.0in},clip]{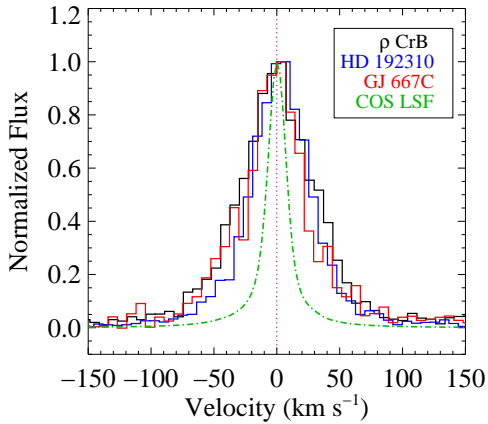}
\end{tabular}
\end{center}
\vspace{-2.3in}
\caption{Spectral blow-up of the \ion{Si}{3}~$\lambda$~1206~\AA\ (log$_{10}$$T_{form}$~=~4.7) upper chromospheric emission line for the three example stars shown in Figure 1 ($\rho$ CrB: G0V, $V$~=~5.39; HD 192310: K2V, $V$~=~5.72; GJ 667 C: M2.5V, $V$~=~10.22).  The 1206~\AA\ line-spread function of $HST$-COS is shown as the green dash-dot line, illustrating that the lines are spectrally resolved in all targets.  The spectra have been smoothed by 3 pixels (half of an $HST$-COS spectral resolution  element) for display.  }
\label{fig-proveit} 
\vspace{-0.0in}
\end{figure} 

\section{Stellar Targets and Observations}

In order to quantify the absolute UV irradiance levels incident on orbiting planets, we require direct observations of cool stars.  To date, very few stellar atmosphere codes  have incorporated complete spectral irradiance modeling that includes contributions from the chromosphere, transition region, and corona (although see, e.g.,  Fontenla et al. 2016).  Most models,  including the widely used PHEONIX~\citep{husser13} and  Kurucz stellar atmosphere models~\citep{castelli04}, only include emission from the stellar photosphere and thus underpredict the flux below $\sim$~2000~\AA\ for cool stars by orders of magnitude~\citep{shkolnik14,loyd16}. 

We also wish to understand how the UV activity levels of exoplanet host stars compare with similar stars without planets.   Therefore, we have assembled a sample of known exoplanet host stars and a ``control'' sample without known planets (or where the presence of massive, short period planets has been ruled out; see discussion below). Of course, $Kepler$ and RV surveys have shown us that most cool stars have planets, so our ``non-planet hosts'' may be stars for which planets have not yet been discovered, but are possible target candidates for current and future planet discovery missions like $TESS$~\citep{sullivan15} and $LUVOIR$ (Roberge et al. 2017).   In practice, when we refer to ``non-planet hosts'', we are referring to field stars that have been observed in previous $HST$ observing programs for other primary science objectives (e.g., solar twins, the sun in time, etc.).\nocite{roberge17}  

 In assembling this sample, we restricted ourselves to the use of observations from the broad wavelength coverage UV spectrographs aboard the {\it Hubble Space Telescope} (STIS and COS), as this allows us to preserve a quality control threshold for wavelength and flux calibration and ensures that ``optically inactive'' M dwarfs are included.  In the following two subsections, we briefly describe these samples.   

\subsection{Exoplanet Host Stars}  

As the original motivation for this work was the intriguing SPI signal found in the MUSCLES Treasury Survey dataset~\citep{france16}, we began assembling the list of known exoplanet host stars with archival $HST$-STIS and -COS observations.  Some stars, e.g., Proxima Cen,  moved from the non-planet host list to the planet host list during the course of this work~\citep{anglada16}.   This list is also populated with stars hosting transiting planets that have been observed at UV wavelengths for absorption spectroscopy during transit~\citep{linsky10,ehrenreich12, ballester13,loyd17}.    We note that transits impact the observed line fluxes by less than 5\% for the combined observations, therefore we do not attempt to phase-separate these data.  Combining the archival observations with the 45 new exoplanet host star observations presented in Section 2.3, we have assembled 1150~--~1450~\AA\ spectra of 71 stars hosting extrasolar planets.   

\subsection{Stars Without Known Planets: ``Non-planet Hosts'' }  

We have also assembled a sample of stars with no known exoplanets to compare against the list of planet-hosting stars (see Table \ref{nonplanethoststarprops}).  In order to obtain medium-to-high signal-to-noise FUV spectral observations of cool stars,  they must be brighter than roughly 10th magnitude in V-band (V $<$ 10; brighter for the SNAP observations of solar type stars, and somewhat fainter for M dwarfs).   This places the requirement that we select our sample from large RV surveys that target nearby stars (see, e.g., Valenti \& Fischer et al. 2005 and references therein).  As a result, what we refer to as a ``non-planet-host'' really means that a planet has not been detected down to the sensitivity of these surveys.  For example,  the sample of 1300 FGKM stars described by~\citet{marcy04} has radial velocity precision to 3 m s$^{-1}$ for FGK stars and 5 m s$^{-1}$ for M dwarfs.  This translates into a planetary mass limit of $M$~sin~$i$~of roughly 0.1~$M_{Jup}$ for planets with roughly 5~--~10 year orbital periods (semi-major axes~$\lesssim$~3 AU).  The HARPS survey has pushed to less than 1 m s$^{-1}$~\citep{pepe11}, enabling the detection  of Earth-mass planets with orbital periods up to tens of days around nearby M dwarfs~\citep{anglada16}.  Suffice to say, these caveats should be kept in mind as we describe differences between the planet-hosting and non-planet-hosting samples.   

The FUV observations for the bulk of the non-planet hosts were drawn from StarCat \citep{Ayres2010}, a database of ultraviolet stellar spectra from \emph{HST}-STIS. To prevent omission of targets that were observed after StarCat was assembled, we cross-referenced the catalogs of \citet{Valenti2005}, \citet{Neves2013}, \citet{Buchhave2015}, and \citet{Terrien2015} with the \emph{HST}-COS and \emph{HST}-STIS archives. These surveys include comparisons of metallicities of planet-hosting versus non-planet-hosting systems, so our search yielded several more stars with ultraviolet spectra that had previously been identified as non-planet hosts. 


\subsection{New Observations with $HST$-COS}

We carried out a SNAP program with the $HST$-COS instrument (HST GO 14633; PI~--~K. France) to fill out the sample of UV activity from exoplanet host stars.   We used {\tt exoplanets.org} to assemble a list of 151 confirmed planet hosting late F through K dwarfs within 50 pc.   From these, we eliminated duplicates from the list above and applied a brightness constraint, visual magnitude 5~$<$~V~$<$~8.5, to enable robust emission line flux fitting without compromising $HST$-COS instrument safety (see Table \ref{planethoststarprops}).  

In order to obtain a robust census of line formation temperatures in the upper atmospheres of cool stars, we selected spectral coverage from 1150~--~1450~\AA.  The G130M mode of COS provides the necessary wavelength coverage, the highest sensitivity of any spectral mode at these wavelengths aboard $HST$, and the spectral resolution ($R$~$\sim$~16,000) to cleanly separate and resolve the emission lines.  Our $HST$ SNAP observations with COS G130M provided access to a suite of spectral tracers, including neutrals:  \ion{N}{1} $\lambda$1200~\AA, \ion{C}{1} $\lambda$1275~\AA, \ion{O}{1} $\lambda$1304, 1356~\AA, \ion{S}{1} $\lambda$1425~\AA;  low-ionization metals and intermediate formation temperature species: \ion{Si}{3} $\lambda$1206~\AA, \ion{Si}{2} $\lambda$1260, 1264~\AA, \ion{C}{2} $\lambda$1335~\AA; and the high formation temperature lines \ion{C}{3} $\lambda$1175~\AA, \ion{O}{5} $\lambda$1218~\AA, \ion{N}{5} $\lambda$1239, 1243~\AA, \ion{Si}{4} $\lambda$1394, 1403~\AA.  While all of these ions were present in the highest S/N observations, only \ion{C}{2}, \ion{Si}{3}, \ion{Si}{4}, and \ion{N}{5} were detected at high significance in most of our target stars, and we consequently focus on these tracers in this work.   Figures 1 and 2 display the full spectra of a sample of stars used in this work, and a zoom in on the \ion{Si}{3} emission line, respectively.   

The COS G130M exposure times were between 1905~--~2020 seconds per star (the typical exposure time was 1920 s), in the CENWAVE 1291 setting.  The total exposures were  split between two focal plane offset positions (FP-POS) to mitigate both the long-term effects of Ly$\alpha$ gain sag on the detector and detector fixed pattern noise.   The observing program executed from 29 November 2016 through 17 February 2018, with 45 out of the original 80 SNAP targets (56\%) observed.   

\subsection{ {\it Extreme-Ultraviolet Explorer} Spectra} 

For our complete list of planet-hosting and non-planet-hosting stars, we identified 12 stars with observations in the $EUVE$ archive that were considered detections by~\citet{craig97}.  We assembled these datasets from the MAST $EUVE$ archive, and took neutral hydrogen column densities from~\citet{linsky14}.  The $EUVE$ overlap sample we analyzed included: Procyon, $\alpha$ Cen A, $\chi^{1}$ Ori, $\kappa$ Cet, $\xi$ Boo, 70 Oph, $\epsilon$ Eri, AU Mic, EV Lac, AD Leo, Proxima Cen, and YY Gem.   Analysis of the $EUVE$ data is presented in Section 3.2 and is presented in the context of our FUV activity survey in Section 4.1.2.  


\section{Analysis:  Emission Line Fluxes and Bolometric Luminosities} 

\subsection{FUV Emission Line Fluxes of \ion{Si}{3}, \ion{N}{5}, \ion{C}{2}, and \ion{Si}{4}; 1200~--~1420~\AA}

We quantify the FUV activity level from our planet-hosting and non-planet-host samples by defining the ``UV activity index'', $F_{ion}$/$F_{bolom}$, for the four primary ions studied in this work: \ion{Si}{3}~$\lambda$~1206~\AA; log$_{10}$$T_{form}$~=~4.7, \ion{N}{5}~$\lambda$~1240~\AA; log$_{10}$$T_{form}$~=~5.2, \ion{C}{2}~$\lambda$~1335~\AA; log$_{10}$$T_{form}$~=~4.5, and \ion{Si}{4}~$\lambda$~1400~\AA; log$_{10}$$T_{form}$~=~4.9.   The emission line luminosities, $L_{ion}$, are simply the wavelength-integrated fluxes scaled by the distance, $L_{ion}$~=~4$\pi$$d^{2}$$F_{ion}$, where $d$ is the distance to the star and $F_{ion}$ is the line flux in units of [erg cm$^{-2}$ s$^{-1}$], described in the next paragraph.   
The formation temperatures are taken from the CHIANTI database~\citep{chianti09}, however we note that different ions trace different atmospheric altitude, pressure, and temperature regimes as a function of stellar mass. 

Emission line fluxes from \ion{N}{5} ($\lambda$~1238.82~\AA, $\lambda$~1242.80~\AA), \ion{C}{2} ($\lambda$~1334.53~\AA, $\lambda$~1335.66~\AA, $\lambda$~1335.71~\AA), \ion{Si}{3} ($\lambda$~1206.49~\AA, $\lambda$~1206.55~\AA, $\lambda$~1207.51~\AA), and \ion{Si}{4} ($\lambda$~1393.75~\AA, $\lambda$~1402.76~\AA) were measured for both the planet-hosting and non-planet-hosting samples (see Tables \ref{planethostfluxes} and \ref{nonplanethostfluxes}). Since all targets are located within the Local Bubble, the dust reddening along the line of sight was assumed to be negligible.  However,  absorption from low-ionization gas in the local ISM, particularly in the \ion{C}{2} $\lambda$ 1334.53~\AA\ line, can lead to systematic underestimation of the intrinsic \ion{C}{2} emission strength~\citep{redfield04}. Many of the systems had faint emission lines with low S/N, making it difficult to fit line profiles to the data. For all sources, the fluxes were calculated as 
\begin{equation}
F_{ion} = \sum \limits_{\lambda = \lambda_0 - \delta \lambda}^{\lambda_0 + \delta \lambda} \Delta \lambda F_{\lambda} - \sum \limits_{\lambda = \lambda_0 - \delta \lambda}^{\lambda_0 + \delta \lambda} \Delta \lambda F_{cont} 
\end{equation} 
where $\Delta\lambda$ is the average spacing between adjacent data points ($\sim$~0.01 \AA) and $F_{cont}$ is the flux in the continuum, estimated from a linear interpolation across the emission line. $\delta\lambda$ was set to roughly 0.5~\AA\, with adjustments made as needed to accommodate wider features. 

$L_{bolom}$ is the bolometric luminosity, $L_{bolom}$~=~4$\pi$$d^{2}$$F_{bolom}$.  Bolometric fluxes, $F_{bolom}$, were calculated as 
\begin{equation}
F_{bolom}= \sigma T_{eff}^4 \left( \frac{R_{\ast}}{d} \right)^2
\end{equation}
using the stellar parameters for each target (Tables 1 and 2). Although \citet{loyd16} measured bolometric fluxes for each of the MUSCLES stars by incorporating $HST$ spectroscopy and Tycho photometry, the simpler calculations were adopted for all objects in the survey to preserve uniformity across the sample. Comparing with the MUSCLES luminosities, we find that this simple prescription differs by as much as $\sim$~30~\% for the cooler M dwarfs (e.g., GJ 1214) and less than 10~\% for the warmer stars in the MUSCLES sample.   Fractional luminosities, $F_{ion}$/$F_{bolom}$, were then obtained for each of the four measured ions by dividing the line flux by the bolometric flux. 

Figures 3 and 4 display the UV activity levels as a function of the various stellar parameters studied here.   Figure 3 displays the relationship between $F_{SiIV}$/$F_{bolom}$ and spectral slope ($\equiv$~$B$~--~$V$) and distance.  Figure 4 compares $F_{SiIV}$/$F_{bolom}$ and the stellar rotational period, for both the exoplanet host and non-planet-hosting samples.   For stars without published rotation periods, we display upper limits based on v sin $i$ measurements (these stars are noted with ** in Tables 1 and 2).  To avoid cluttering the body of the paper, we use \ion{Si}{4} as the representative example ion in  this section; plots for all four ions are presented in Appendix B.

\begin{figure*}[h!]
\begin{center}
\vspace{0.0in}
\begin{tabular}{c}
\hspace{-0.36in}
\includegraphics[width=0.55\textwidth,angle=0,trim={.0in 0.0in 0.0in 0.0in},clip]{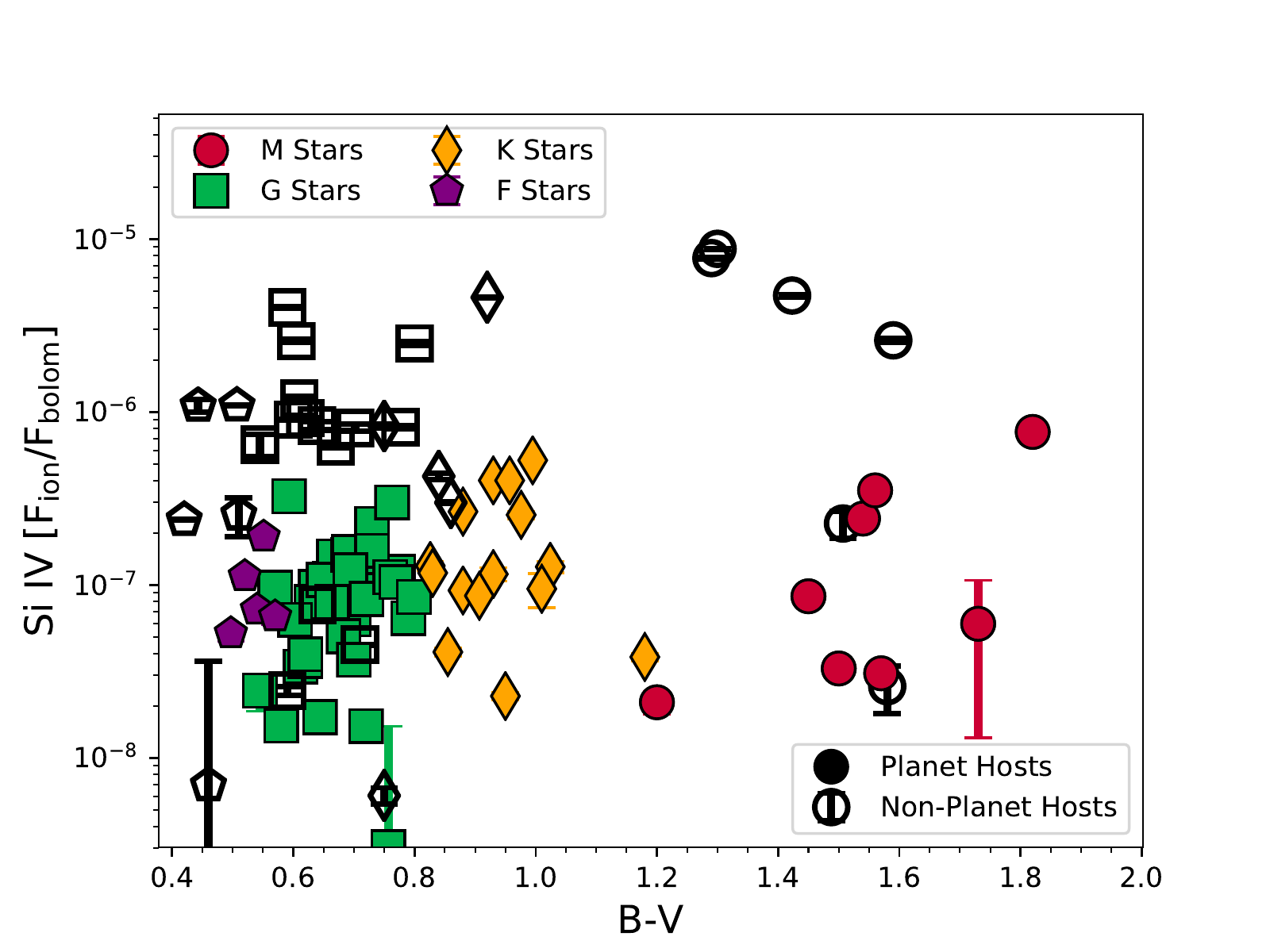}
\hspace{-0.36in}
\includegraphics[width=0.55\textwidth,angle=0,trim={.0in 0.0in 0.0in 0.0in},clip]{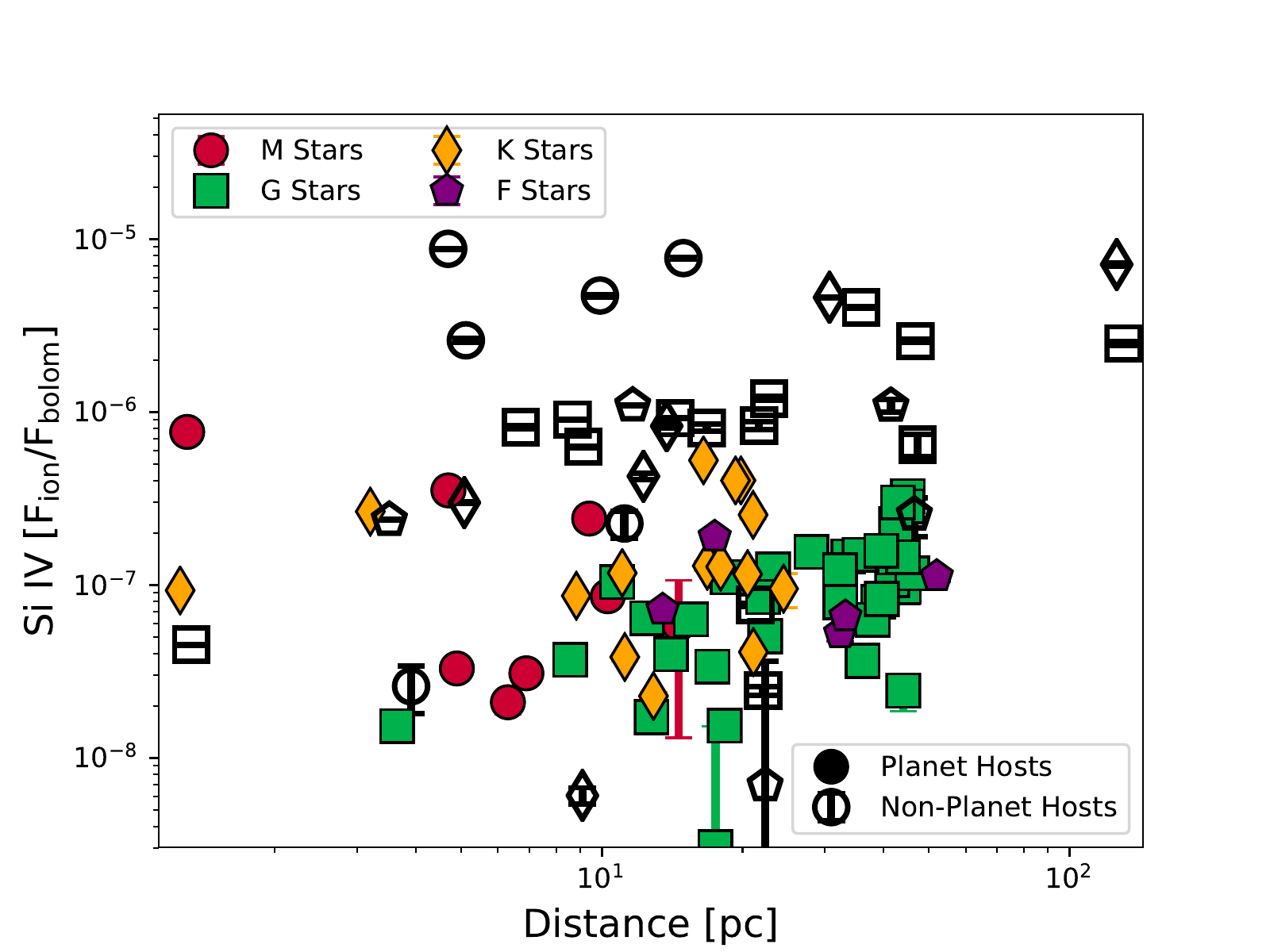} 
\end{tabular}
\end{center}
\vspace{-0.2in}
\caption{The full planet-hosting sample (in filled color symbols) and non-planet control sample (in open black symbols), showing \ion{Si}{4} fractional hot gas luminosity as a function of $B$~--~$V$ color ($left$, a proxy for effective surface temperature) and distance ($right$).  Spectral types are given by different symbols (circles: M dwarfs, diamonds: K dwarfs, squares: G dwarfs, pentagons: F dwarfs) as shown in the legend.  The non-planet hosting stars are shown to be systematically factors of 5~--~10 brighter in the high-temperature FUV lines.  The \ion{Si}{4} behavior is representative of the behavior of all 4 FUV activity indicators studied here; the full plot set is presented in Appendix B.}
\label{fig-proveit} 
\vspace{-0.15in}
\end{figure*}

\begin{figure}[!h]
\begin{center}
\vspace{0.0in}
\begin{tabular}{c}
\includegraphics[width=0.5\textwidth,angle=0,trim={.0in 0.0in 0.0in 0.0in},clip]{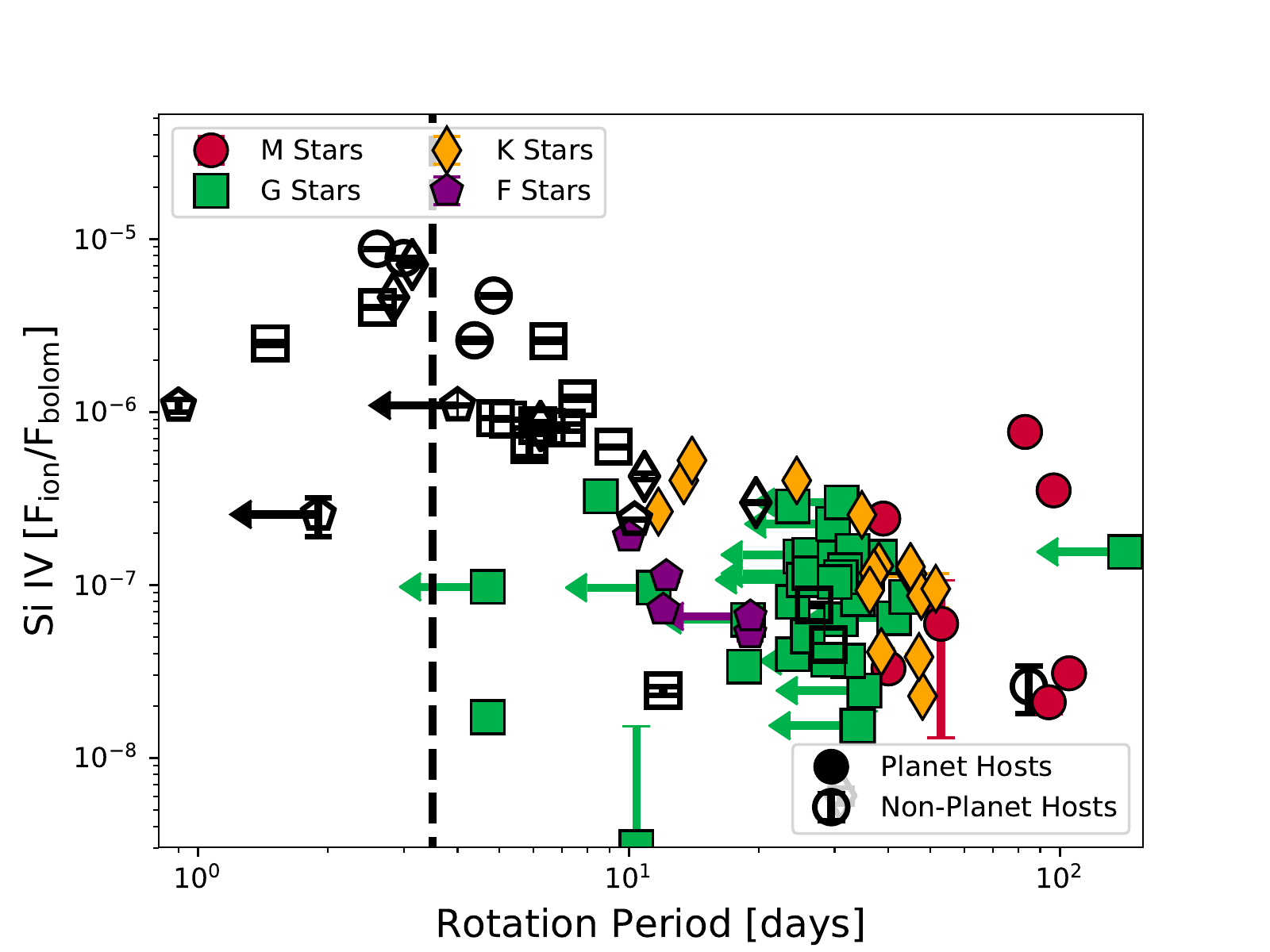}
\end{tabular}
\end{center}
\vspace{-0.2in}
\caption{The full planet-hosting sample (in filled color symbols) and non-planet control sample (in open black symbols), showing the \ion{Si}{4} activity level ($\propto$ fractional hot gas luminosity) as a function of the stellar rotation period  ($P_{rot}$).  Spectral types are given by different symbols (circles: M dwarfs, diamonds: K dwarfs, squares: G dwarfs, pentagons: F dwarfs) as shown in the legend.  This figure succinctly demonstrates the bimodal distribution of targets, with non-planet hosts typically having $P_{rot}$~$\lesssim$~20 days and the planet hosts having $P_{rot}$~$\gtrsim$~20 days.  This is a natural consequence of the selection bias for RV planet searches~\citep{marcy04}.  The saturated activity level is log$_{10}$$F_{SiIV}$/$F_{bolom}$ $\approx$~$-$5.6, and the power law slope beyond the $P_{orb}$~$\approx$~3.5 day break point is  $-$1.1~$\pm$~0.1 (Section 4.1).   
 }
\label{fig-proveit} 
\vspace{-0.05in}
\end{figure} 

\subsection{EUV Fluxes, 90~--~360~\AA}

For each $EUVE$ spectrum, we converted the data to flux density units (erg cm$^{-2}$ s$^{-1}$ \AA$^{-1}$) and integrated over the spectral region where most stars had appreciable flux (90~--~360 \AA, or 9~--~36 nm).  These raw integrated fluxes (erg cm$^{-2}$ s$^{-1}$) were first background corrected by subtracting the flux level of an $EUVE$ non-detection in this band ($\gamma$ Tau,  $F(EUV)_{back}$~$\approx$~1~$\times$~10$^{-12}$ erg cm$^{-2}$ s$^{-1}$).  YY Gem was dropped at this point because its post-subtraction integrated flux was less than 10\% of the background level.    The fluxes were then corrected for neutral hydrogen, neutral helium, and ionized helium attenuation by calculating optical depth spectra for the appropriate $N$(\ion{H}{1}) from the references collated by Linsky et al. (2014).  The ionization fraction of helium (0.6) and the neutral hydrogen to helium ratios (0.08) were taken from the observed local ISM values from~\citet{dupuis95}.       The  $N$(\ion{H}{1}) values were necessarily low (all less than 10$^{18.5}$ and 10/12 less than 10$^{18.1}$ cm$^{-2}$; see Table 3 of Linsky et al. 2014), the ISM transmission functions are relatively linear at these wavelengths and we calculated the average flux correction for the 90~--~360 \AA\ band.  These intrinsic EUV fluxes are compared with the FUV activity sample in Section 4.1.2.

\section{Results: UV Activity Levels of Exoplanet Host Stars}

Figure 4 shows the relationship between FUV activity index and the stellar rotation period.   The stars are identified by symbol type and separated into planet vs. non-planet-hosting by the use of color or black symbols, respectively.  
Comparing the FUV activity indices with the stellar rotation periods, we observe a ``saturated'' plateau followed by a  roughly continuous, power-law, decline in UV activity.  We classify the UV activity into two rough categories: {\it High UV-activity stars} with  $F_{SiIV}$/$F_{bolom}$~$>$~10$^{-6}$ and  {\it Intermediate-to-Low UV-activity stars} with $F_{SiIV}$/$F_{bolom}$~$<$~10$^{-6}$.   Very roughly, this transition occurs around rotation periods of 3.5 days.
There is some evidence that a low-activity plateau ($F_{SiIV}$/$F_{bolom}$~$<$~10$^{-7}$) is reached around a rotation period of 20 days, but larger samples of slowly rotating stars are needed to fill out this trend.   
We fitted the \ion{Si}{4} activity-rotation diagram with a power-law of the form: 
\begin{equation}
log_{10}F_{ion}/F_{bolom} = \begin{cases} R_{sat}, &  P_{rot} < P_{break}  \\  R_{sat} \times (P_{rot}/P_{break})^{\alpha},  &  P_{rot}~\geq~P_{break}  \end{cases} 
\end{equation}
where $R_{sat}$ is the logarithmic saturated activity level and $P_{break}$ is the turnover rotation period where the activity declines.   We used the MCMC sampler \textit{emcee} \citep{foreman13} to explore the posterior probability of the free parameters of this model ($R_{sat}$, $P_{break}$, $\alpha$), modeling the data scatter as Gaussian in log space with constant standard deviation that we treated as a fourth free parameter.   We applied a uniform prior of 1 day~$<$~$P_{break}$~$<$~10 days based on the clear visual trend in the data and treated $P_{rot} / \sin i$ upper limits derived from $v \sin i$ measurements as equivalent to $P_{rot}$ in the fits. 



With sparse coverage of stars with rotation $P_{rot}$~$<$~3 days, we are only able to place an upper limit on $P_{break}$ for the ions studied here, $P_{break}$~$\lesssim$~3.5 days.  For all four ions, $R_{sat}$ is between $-$5.5~--~$-$6.0.  For the \ion{Si}{4} plot shown in Figure 4, $\alpha$ = $-$1.1~$\pm$~0.1.    For \ion{C}{2}, $\alpha$ = $-$1.0~$\pm$~0.1.  For \ion{Si}{3}, $\alpha$ = $-$1.1~$\pm$~0.1.  For \ion{N}{5}, $\alpha$ =  $-$1.3~$\pm$~0.1.

The UV-activity-rotation diagram is qualitatively reminiscent of the H$\alpha$-rotation relationship for M dwarfs presented by~\citet{newton17}, as well as the X-ray-rotation relationship presented by~\citet{pizzolato03} for cool stars.  The transition to the low-activity UV state takes place at shorter rotation periods for cool stars as a whole, relative to M dwarf-only samples.  This indicates that warmer stars `turn over' to a lower activity level at shorter rotation periods than for M stars.  Due to the primary goals of the surveys that acquired our UV M dwarf observations, we have too few stars with intermediate rotation periods (10~--~30 days) to make a detailed comparison with the H$\alpha$ sample.     

Figure 4 also indicates outliers on the high- and low-activity ends of the distribution:  intermediate activity levels can be found out to rotation periods~$\approx$~100 days (Proxima Cen and GJ 876;  due in some measure to flare activity during their UV observations; Christian et al. 2004; Ribas et al. 2017; France et al. 2012; 2016)\nocite{christian04,ribas17,france12a,france16}  while anomalously low activity levels (HD 28033 and  HD 13931) may be reminiscent of planet-induced rotational modulation, as has been suggested for WASP-18~\citep{pillitteri14,fossati18}.

\subsection{Comparison with Non-planet Host Control Sample}

Figure 3 shows a clear bimodality of UV activity index of our sample.  The non-planet-hosting sample (open, black symbols) are factors of roughly 5~--~10 higher than the planet-hosting sample.  At first glance, these plots suggest that non-planet-hosting stars are more active then  their planet-hosting cousins, however Figure 4 shows that this is clearly an effect of the different rotation periods sampled in the two populations.    We can interpret the differences between the planet-hosting and non-planet-hosting samples as an age bias arising from the detection technique. The large RV surveys of the 1990s and 2000s made specific cuts on \ion{Ca}{2} activity indices to avoid excess stellar jitter, higher activity stars making the extraction of the radial velocity signal more challenging (although see also  Issacson \& Fisher 2010).  Therefore, these surveys are biased by self-selection for  ages~$\gtrsim$~2 Gyr for solar-type stars~\citep{marcy04,valenti05}; the exoplanet host star observations essentially give us a picture of the radiation environment at ages $\gtrsim$~2 Gyr.    On the other hand, observations of the control sample were originally acquired, in part, because some of these systems were interesting active stars, and therefore provide a better picture of the typical UV irradiance level experienced by orbiting planets during the initial  $\sim$~1.7 Gyr when life would be forming and evolving~\citep{jones10}.\nocite{issacson10}   

   \subsubsection{Individual ``Like-star'' Comparisons}
   
   A complementary approach to comparing the ensemble properties of planet-hosting and non-planet-hosting stars is to examine individual systems with very similar spectral type and rotation period.  The goal here is to find stars whose most obvious difference is the presence of a planetary system.  We note that due to the limited size of the survey, finding systems with like-stellar parameters and like-planetary systems was not possible (e.g., HD 189733, $\epsilon$ Eri, and HD 128311 below).  Using the \ion{Si}{4} activity index as representative of the behavior of the FUV emission from these stars, we identified the following ``case studies'' for comparison: 
   
  \begin{itemize}
  \item The $P_{rot}$~$\sim$~11 day K dwarfs (see Tables 1 and 2 for stellar parameter references):  Comparing the similar planet-hosting stars HD189733 (K0 V, $T_{eff}$ = 4880 K, $P_{rot}$ = 13.4 days), $\epsilon$ Eri (K2 V, $T_{eff}$ = 4900 K, $P_{rot}$ = 11.7 days), and   HD  128311 (K3 V, $T_{eff}$ = 4965 K, $P_{rot}$ = 14 days) with the non-planet hosting K dwarf HR 1925 (K1 V, $T_{eff}$ = 5309 K, $P_{rot}$ = 10.86 days), 
  we find the average $F_{SiIV}$/$F_{bolom}$ value for the planet hosting stars is 4.3 ($\pm$~0.2)~$\times$~10$^{-7}$, while HR 1925 displays the identical 4.3 ($\pm$~0.2)~$\times$~10$^{-7}$.
   
   \item The $P_{rot}$~$\sim$~28 day G dwarfs:  Comparing the similar planet-hosting stars  
  $\mu$~Ara (G3 V, $T_{eff}$ = 5800 K, $P_{rot}$ = 31 days), 16 Cyg B (G3 V, $T_{eff}$ = 5770 K, $P_{rot}$ = 29.1 days), and HD1461 (G3 V, $T_{eff}$ = 5765 K, $P_{rot}$ = 29 days) with the non-planet-hosting G dwarfs 16 Cyg A (G1.5 V, $T_{eff}$ = 5825 K, $P_{rot}$ = 26.9 days) and $\alpha$~Cen~A (G2 V, $T_{eff}$ = 5770 K, $P_{rot}$ = 29 days), 
  we find the average $F_{SiIV}$/$F_{bolom}$ value for the planet hosting stars is 9.8 ($\pm$~0.6)~$\times$~10$^{-8}$, while the non-planet hosting sample displays the somewhat lower 6.0 ($\pm$~0.2)~$\times$~10$^{-8}$.
   
   \item The $P_{rot}$~$\sim$~100 day M dwarfs: Comparing the planet-hosting star GJ 667C (M1.5 V, $T_{eff}$ = 3440 K, $P_{rot}$ = 105 days) with the non-planet hosting Kapteyn's Star (M1 V, $T_{eff}$ = 3527 K, $P_{rot}$ = 84.7 days), we see that  $F_{SiIV}$/$F_{bolom}$ for GJ 667C is 2.7 ($\pm$~0.2)~$\times$~10$^{-8}$, while Kapteyn's star displays an statistically indistinguishable 3.0 ($\pm$~1.0)~$\times$~10$^{-8}$.
   
      \end{itemize}
      
 The comparisons above show that other than a slightly higher \ion{Si}{4} activity level in the solar-type planet-hosting stars, there is essentially no discernable difference between the FUV activity levels of the planet-hosting and non-planet hosting samples.  This supports the assertion made above  that we are observing an age spread of a single stellar population as  opposed to two distinct planet-hosting and non-planet-hosting groups.

\subsubsection{FUV Activity Index as a Proxy for EUV Irradiance}

The stellar EUV energy budget contains contributions from both the transition region (Lyman continuum as well as helium and metal line emission in the 228~--~911~\AA\
bandpass) and corona. The FUV emission lines (\ion{N}{5} and \ion{Si}{4}) are required to estimate
the former (Fontenla et al. 2011; Linsky et al. 2014), while X-ray data provide constraints on the latter (e.g., Sanz-Forcada et al. 2011).

\begin{figure}[]
\hspace{-1.26in}
\includegraphics[width=0.7\textwidth,angle=0,trim={.0in 0.0in 0.0in 0.0in},clip]{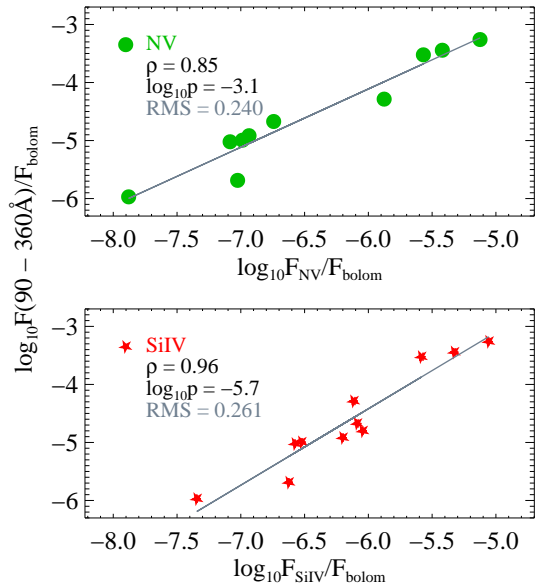}
\vspace{-0.2in}
\caption{Correlations between the \ion{N}{5} ($top$) and \ion{Si}{4} ($bottom$) activity index and the fractional 90~--~360~\AA\ flux (from archival $EUVE$ observations).  The $EUVE$ fluxes have been  corrected for interstellar \ion{H}{1} opacity.  The Spearman rank coefficient ($\rho$), the $p$-value, and the RMS scatter about the best fit line are shown in the legend.  The tight correlation argues that broadband EUV fluxes in this region can be estimated  to within a factor of $\sim$~2 from the FUV activity index.    }
\label{fig-proveit} 
\vspace{-0.05in}
\end{figure} 

We combine our large FUV data set with the smaller number  of overlapping $EUVE$ observations to:  {\bf 1)} evaluate if the UV transition region emission lines directly scale with the EUV flux and if so 
{\bf 2)} present a new method for  estimating the 90~--~360~\AA\ flux from cool stars.  Both of these topics are critical to modeling the atmospheric response of all types of planets, from rocky worlds~\citep{lammer09,wheatley17} to hot Jupiters~\citep{murray-clay09,Koskinen13}. 

We find that the FUV activity indices that we presented in Section 3 can  be correlated with the comparable EUV fractional luminosity to  develop scaling relations for the EUV flux that hold across spectral type and activity level.  These relations do not rely on Ly$\alpha$ flux reconstructions or scalings from other lines to estimate the Ly$\alpha$ flux\footnote{Ly$\alpha$ flux reconstructions and Ly$\alpha$ scaling relations have uncertainties that can range from 20\%~to factors-of-several depending on the signal-to-noise and spectral resolution of the observations~\citep{linsky14}}.  Other than local ISM absorption of the ground-state \ion{C}{2} 1334~\AA\ line, our FUV activity measurements are straightforward and do not suffer from any significant line-of-sight attenuation or uncertain intrinsic emission line  shapes (see, e.g., the discussion of the intrinsic Ly$\alpha$ emission line profiles of cool stars in Wood et al. 2005 and  Youngblood et al. 2016).   We parameterize the 90~--~360~\AA\ flux as a function of the UV activity indices presented above: 
\begin{equation}
log_{10} \left(F(90-360 \textup{\AA} ) / F_{bolom} \right) = m \times log_{10} \left(F_{ion} / F_{bolom} \right) + b
\end{equation}
(see Figure 5), where the wavelength range (90~--~360) is in \AA.  We computed the residuals for each ion (defined as the difference between the $F(90-360 \textup{\AA} )$/$F_{bolom}$ data and the best fit linear model), and unsurprisingly, the  highest temperature FUV lines showed the smallest residuals.   We therefore favor \ion{N}{5} and \ion{Si}{4} as the best proxies for the fractional EUV flux.  The RMS scatter on the residuals for these two ions are between factors of 1.7 and 1.8 in linear flux, even though both flux ratios span approximately two and half orders of magnitude in  activity level.   
The best-fit coefficients for the UV activity index-to-EUV activity for \ion{N}{5} are [$m$,$b$] = [1.0~($\pm$~0.1), 1.9~($\pm$~0.6)], and the coefficients for \ion{Si}{4} are [$m$,$b$] = [1.3~($\pm$~0.1), 3.5~($\pm$~0.8)]\footnote{We note that Proxima Cen is the only star in this sample with a planet inside 0.2 AU.  Excluding Proxima from the fits does not change the fit coefficients beyond their 1-$\sigma$ uncertainty ranges.}.

While we recommend transition region tracers (\ion{Si}{4} and \ion{N}{5}) because these lines are formed in plasma conditions closer to the EUV emission and do not suffer from ISM absorption effects, a correlation exists with the lower temperature chromospheric lines as well (\ion{C}{2}). The best-fit coefficients for the UV activity index-to-EUV activity 
for \ion{C}{2} are [$m$,$b$] = [1.4~($\pm$~0.2), 3.5~($\pm$~0.9)].  
The \ion{C}{2}--EUV correlation has larger scatter than those for \ion{Si}{4} and \ion{N}{5}; the higher ionization relationships should be used when possible.  

Based on the above analysis, we determine that {\bf 1)} the EUV fluxes follow a power-law relationship with the FUV transition region activity indices over a wide range of spectral types and rotation periods 
and {\bf 2)} with an estimate of the star's bolometric luminosity and a measurement of one of the higher temperature FUV emission lines, the stellar EUV flux in the 90~--~360~\AA\ band can be estimated to roughly a factor of two.    Using the above relationship for \ion{N}{5},  we calculated the  90~--~360~\AA\ flux for all stars in the survey and these are presented in Tables 3 and 4.  The computed EUV fluxes are plotted as a function of stellar rotation period in Figure 6.  The largest uncertainties on calculated $F(90-360 \textup{\AA} )$ comes from the uncertainties on the linear fit parameters, which correspond to approximately a factor of 2.3 uncertainty on $F(90-360 \textup{\AA} )$ when using the \ion{N}{5}--EUV relations.

A rough estimate of the total EUV irradiance can be computed for the quiet Sun~\citep{woods09} and an inactive M dwarf (GJ 832) using the model spectra of~\citet{fontenla16}.  $F_{\star}$(90~--~911 \AA) = $F(90-360 \textup{\AA} )$ + $F(360-911 \textup{\AA} )$. For the quiet Sun, 
\begin{equation}
F_{G2V}(90-911 \textup{\AA} ) =  F(90-360 \textup{\AA} ) + [0.57~\times~F(90-360 \textup{\AA} )]. 
\end{equation} 
For a quiescent M1V star, 
\begin{equation}
F_{M1V}(90-911 \textup{\AA} ) =  F(90-360 \textup{\AA} ) + [1.12~\times~F(90-360 \textup{\AA} )]
\end{equation}
where $F(90-360 \textup{\AA} )$ is the computed EUV flux described above.  We note that because the \ion{Si}{4} and \ion{N}{5} formation temperatures are an order of magnitude (or more) less than the typical  coronal temperature of these stars, we do not suggest extending these relations to the X-ray wavelengths (5~--~100~\AA).  We refer the reader to~\citet{poppenhaeger10},~\citet{forcada11}, and~\citet{loyd16} for a discussion about the X-ray properties of planet hosting stars of various spectral types.     

\begin{figure}[]
\begin{center}
\vspace{0.0in}
\begin{tabular}{c}
\includegraphics[width=0.5\textwidth,angle=0,trim={.0in 0.0in 0.0in 0.0in},clip]{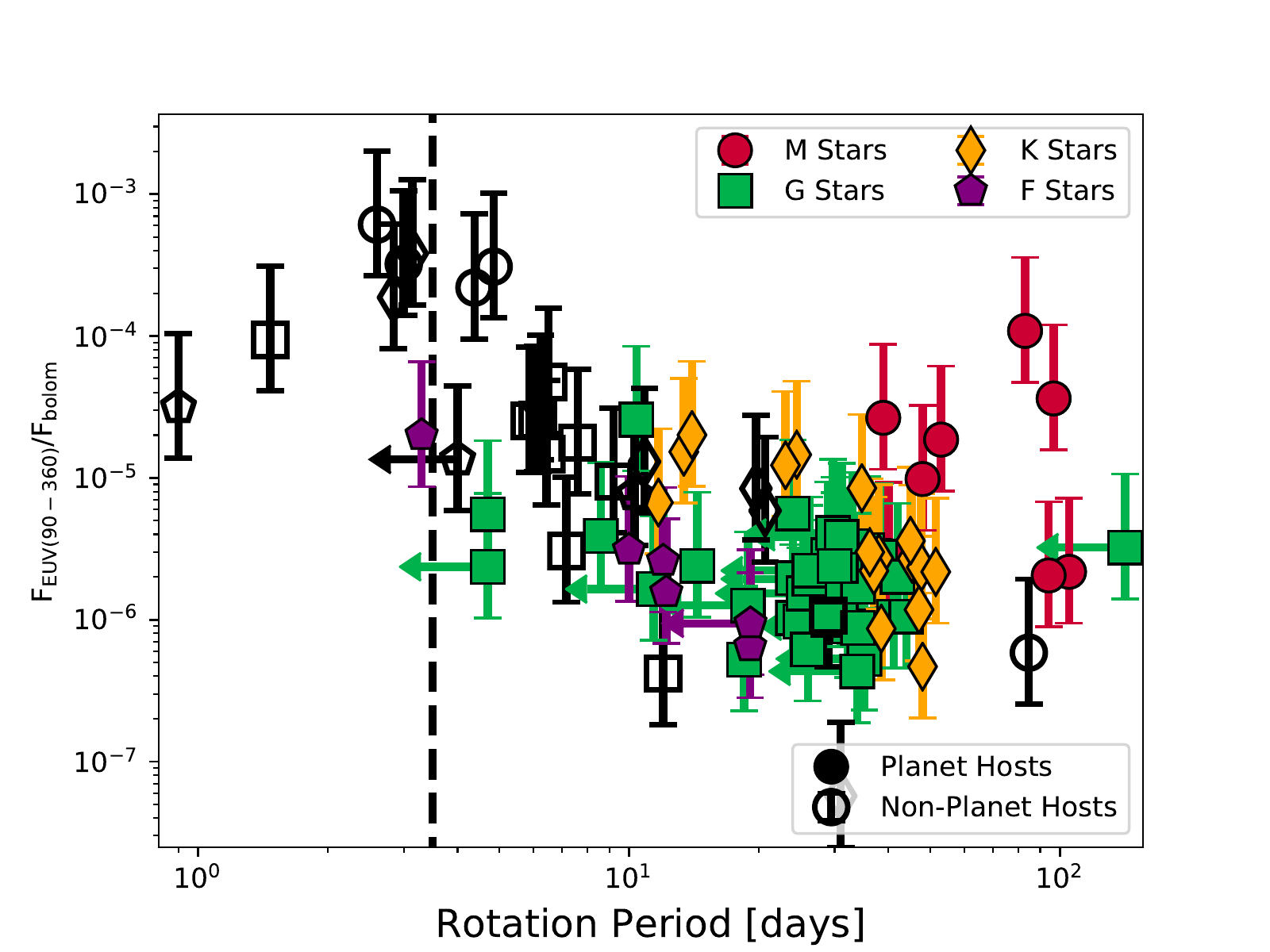}
\end{tabular}
\end{center}
\vspace{-0.2in}
\caption{Converting the \ion{N}{5} UV activity index to the relative EUV flux in the 90~--~360~\AA\ band (Equation 4), we calculate  the ISM-corrected $F$(90~--~360\AA) / $F_{bolom}$ flux ratio for all stars in our sample with \ion{N}{5} measurements.  EUV error estimates are propagated from the uncertainty on the best fit parameters to Equation 4.  One observes a $\sim$two-order-of-magnitude decline in the EUV emission strength as cool stars move from the saturated activity regime at rotation periods $P_{rot}$~$\lesssim$~3.5 days to the presumably older population at  $P_{rot}$~$\gtrsim$~20 days. }
\label{fig-proveit} 
\vspace{-0.0in}
\end{figure} 

These results argue that the EUV evolution from younger to older stars (shorter to longer rotation periods) is similar to that from the chromospheric/transition region emission.   X-ray+EUV evolution studies for solar-type stars~\citep{ribas05} find comparable decrease ($\sim$~10~--~20 in the 20~--~360~\AA\ band) for solar type stars from ~$\sim$~0.6 Gyr to ~$\sim$~4 Gyr.   The two results suggest a common picture where the overall XUV + FUV (5~--~1800~\AA) flux decreases by one-to-two orders of magnitude as the stars age from $\sim$~0.5~--~5~Gyr.\nocite{fontenla11,linsky14,forcada11}  This result is consistent with the relative FUV flux decline in the $GALEX$ sample of early M dwarfs presented by~\citet{schneider18}.   

What are the potential impacts of this flux evolution on orbiting planets?  For terrestrial atmospheres, increasing the EUV flux to levels estimated for the young Sun (~$\sim$ 1 Gyr; Ayres 1997) can increase the temperature of the thermosphere by a factor of $\gtrsim$ 10 (Tian et al. 2008), potentially causing significant and rapid atmospheric mass-loss.\nocite{ayres97,tian08,lammer18}  The issue of increased EUV irradiance and the atmospheric stability of rocky planets (see, e.g., Lammer et al. 2018) is even greater for M dwarfs, where the EUV irradiance levels of even field-age stars (ages~$\sim$~2~--~6 Gyr) are predicted to drive runaway oxidation as many Earth oceans worth of hydrogen are lost (e.g.,  Ribas et al. 2017; Wheatley et al. 2017).\nocite{ribas17,wheatley17}  Our results provide an estimate of the enhancement level of the total EUV + FUV radiation environment around F through M stars, anchored by direct observations.


\subsection{UV Activity Diagnostics and Star-Planet Interactions}

Figure 7 shows the UV activity indices versus the SPI parameter $\left(M_{planet} / a_{planet} \right)$, assuming the mass and semi-major axis of the most massive planet in multi-planet systems, for the sample of planet-hosting stars. The results are quantitatively similar when calculating the correlations with the closest planet.  We find a significant linear correlation between the fractional luminosities for all four ions and the SPI parameter, suggesting that stars with more massive and close-in planets emit more ultraviolet photons from their chromospheres and transition regions relative to their bolometric luminosity.  In Figure 7, 
Spearman $\rho$ and $p$-values are calculated for the log$_{10}$SPI vs. log$_{10}$$F_{ion}$/$F_{bolom}$ relations\footnote{The $p$-value is a measure of the ability of the distribution to be consistent with a null correlation, i.e., an uncorrelated scatter plot.  A $p$-value of 1 is a perfect scatter plot and $p$-values of less than 0.05 typically indicate a strong correlation for samples sizes larger than a few tens of data points.}.  
We find that for \ion{N}{5}, [$\rho_{NV}$,$p_{NV}$] = [0.303, 0.016], for \ion{C}{2}, [$\rho_{CII}$,$p_{CII}$] = [0.296, 0.019], for \ion{Si}{3}, [$\rho_{SiIII}$,$p_{SiIII}$] = [0.343, 0.005], and for \ion{Si}{4}, [$\rho_{SiIV}$,$p_{SiIV}$] = [0.315, 0.015].  In addition to the $p$-values all being at statistically significant levels, Spearman coefficients near 0.3 for samples sizes between 60 and 80 (our sample has 71 stars) represent a statistically significant correlation at the $\sim$~99\% confidence level.   

\begin{figure*}[!h]
\begin{center}
\vspace{0.0in}
\begin{tabular}{cc}
\includegraphics[width=0.5\textwidth,angle=0,trim={.0in 0.0in 0.0in 0.0in},clip]{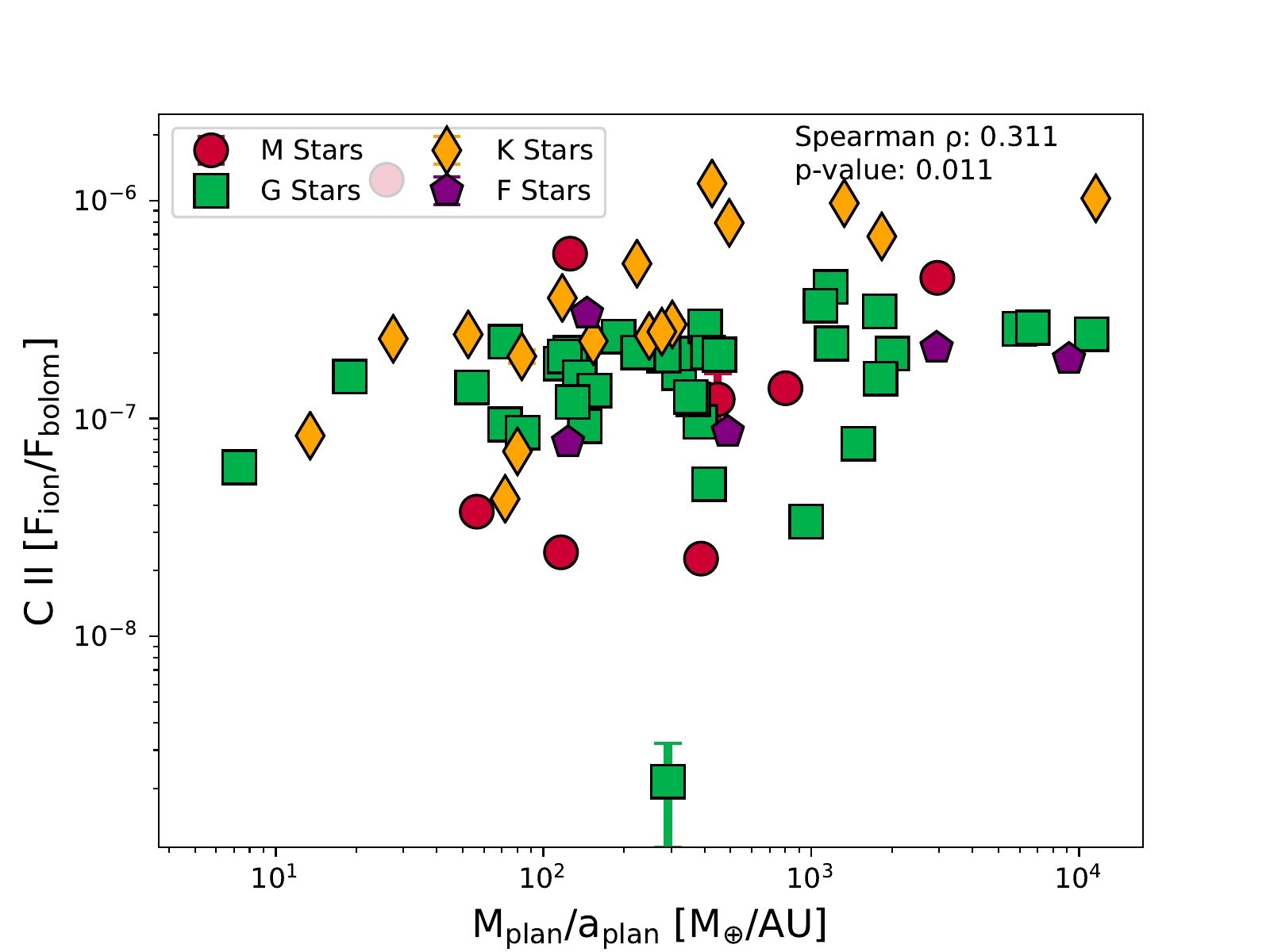} & 
\includegraphics[width=0.5\textwidth,angle=0,trim={.0in 0.0in 0.0in 0.0in},clip]{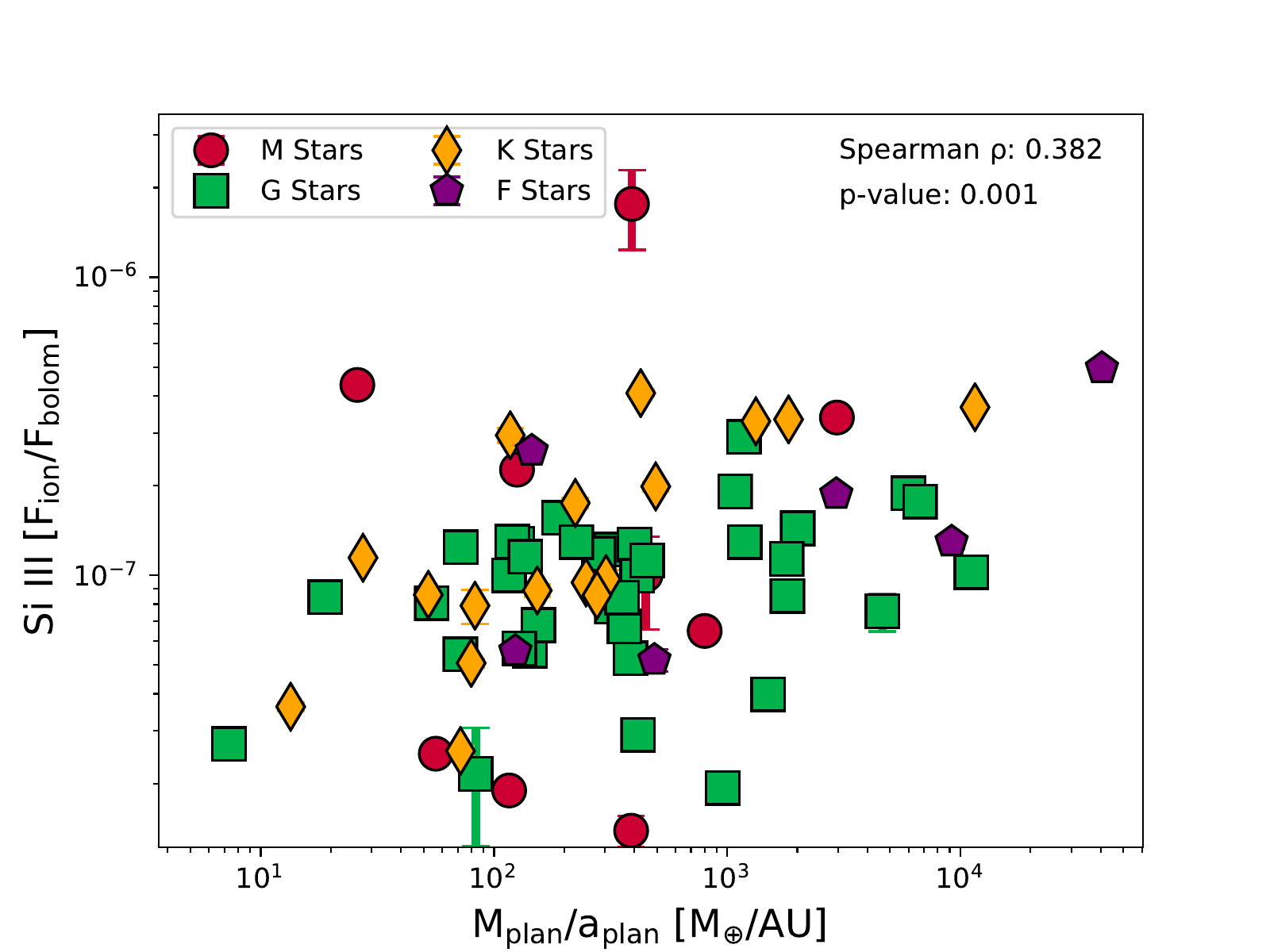} \\
\includegraphics[width=0.5\textwidth,angle=0,trim={.0in 0.0in 0.0in 0.0in},clip]{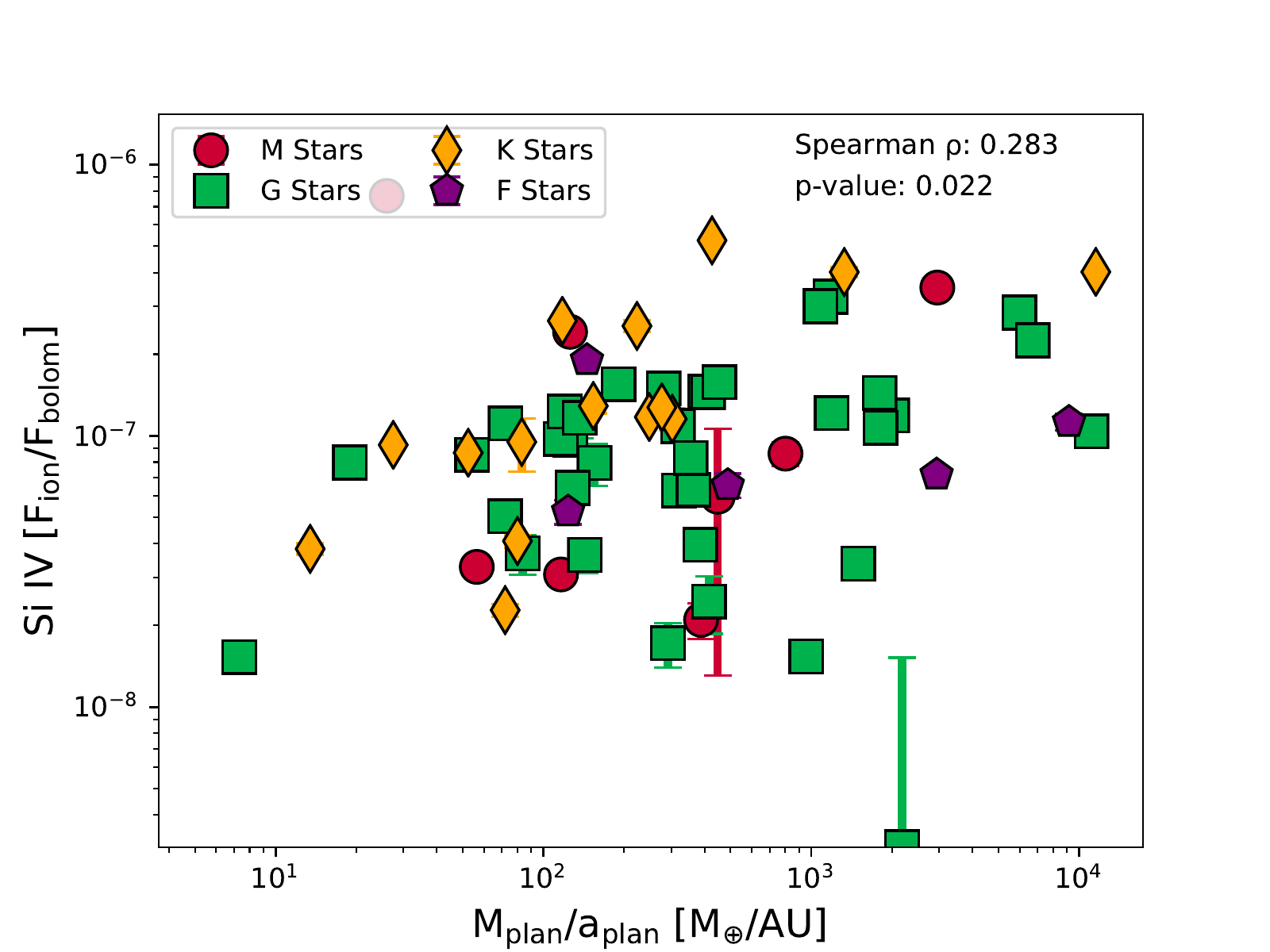} & 
\includegraphics[width=0.5\textwidth,angle=0,trim={.0in 0.0in 0.0in 0.0in},clip]{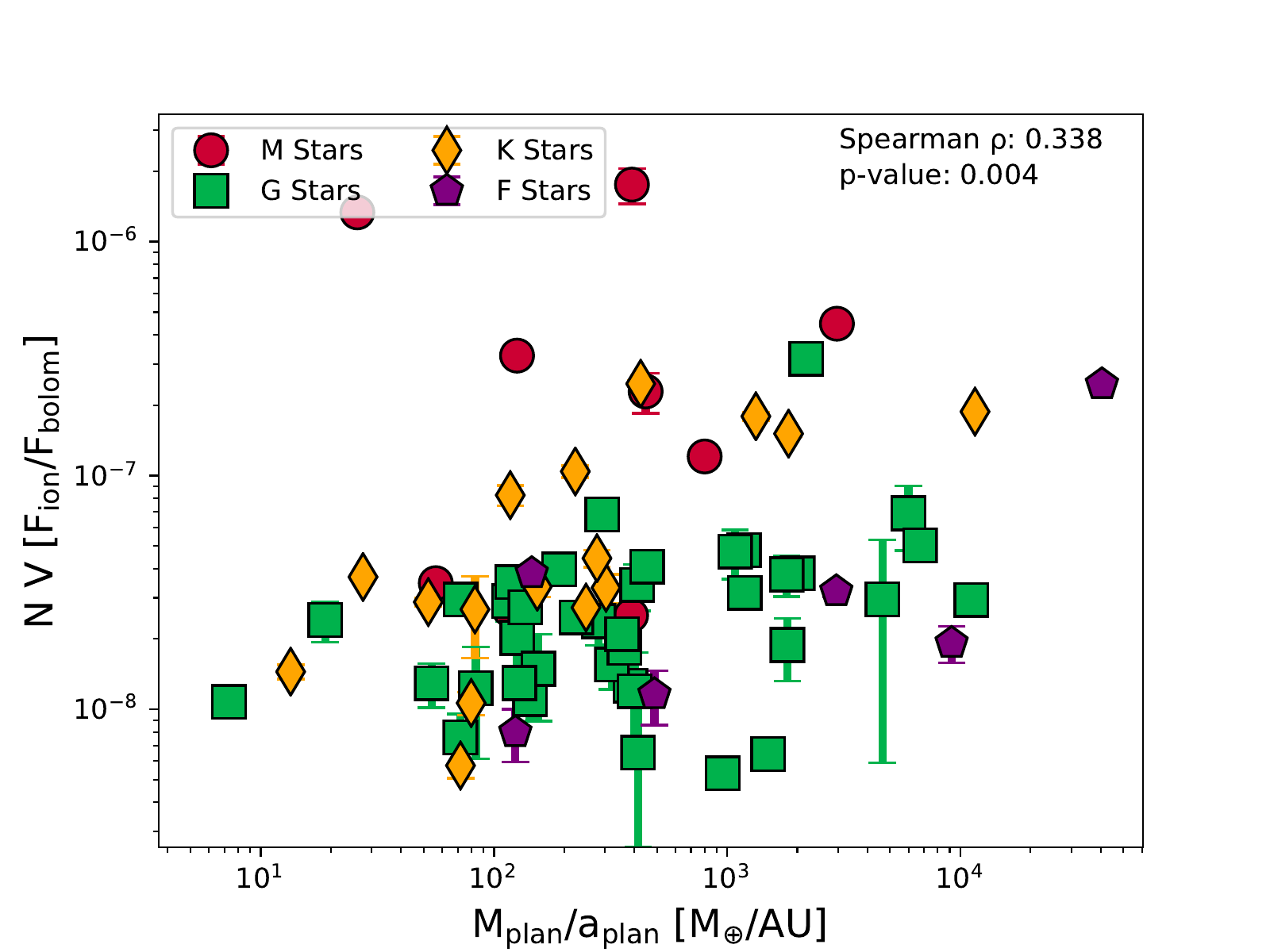}
\end{tabular}
\end{center}
\vspace{-0.1in}
\caption{UV activity levels as a function of the ``SPI parameter'' ($M_{plan}$/$a_{plan}$) for (top left to lower right)  \ion{Si}{3}~$\lambda$~1206~\AA, \ion{N}{5}~$\lambda$~1240~\AA, \ion{C}{2}~$\lambda$~1335~\AA, and \ion{Si}{4}~$\lambda$~1400~\AA.  Spectral types are given by different symbols (circles: M dwarfs, diamonds: K dwarfs, squares: G dwarfs, pentagons: F dwarfs) as shown in the legend.  Best fit linear models in log-log space are shown as the diagonal overplotted line.  All UV activity vs SPI parameter correlations have Spearman rank coefficients between 0.28 and 0.38 with $p$-values between 1~$\times$~10$^{-3}$ and 2~$\times$~10$^{-2}$.  While the UV activity vs SPI parameter correlations are all statically significant, underlying correlations with the stellar parameters driven by population selection biases are also present (see Section 4.2.2).  Section 4.2 describes the analysis of alternative SPI proportionalities. }
\label{fig-proveit} 
\vspace{-0.15in}
\end{figure*}

This result confirms the general trend between log$_{10}$SPI vs. log$_{10}$$F_{ion}$/$F_{bolom}$ identified for M  dwarfs by~\citet{france16}, with the caveat that our larger sample identifies significant stellar and observational biases that may drive this result (see Section 4.2.1).  Our spectroscopic line sample does not include lower formation-temperature species like \ion{Si}{2} and \ion{Mg}{2}, so we are unable to test the fall-off of this correlation with atmospheric emitting region temperature.  We confirm that over the roughly 20,000~--~200,000 K temperature range spanned by our four target ions, this trend holds.   Care should be taken in parsing this sample up into sub-categories, as individual flare events, stellar activity cycles, or specific star-planet systems will more strongly influence the results.  With that caveat in mind,  we note that the correlation coefficients of the full sample are quite a bit lower than found by France et al. for the MUSCLES stars ($\rho$~$\gtrsim$~0.6).  The  log$_{10}$SPI vs. log$_{10}$$F_{ion}$/$F_{bolom}$ correlations are much stronger in our (albeit small) sample of K dwarfs compared to full F~--~M sample.     The Spearman $\rho$ coefficients and $p$-values for the K dwarf sample are between 0.77~--~0.79 and $p$~$<$~0.002, respectively, for all four ions.  

We also considered other potential proportionalities between the UV power deposition and the star-planet system architecture that may provide clues about the physical mechanism responsible for enhanced atmospheric heating.  Specifically, we explored $1)$ magnetic reconnection between stellar and planetary magnetic fields~\citep{lanza12}, with  $F_{ion}$/$F_{bolom}$ $\propto$ $M_{plan}^{2/3}$ ($a_{plan}$/$R_{\star}$)$^{-4}$ $a_{plan}^{-1/2}$, $2)$  magnetic loop stresses between the stellar and planetary fields~\citep{lanza13}, $F_{ion}$/$F_{bolom}$ $\propto$ $M_{plan}$$^{2}$ $a_{plan}^{-1/2}$, and $3)$ tidal torques (e.g., Zahn 2008), with $F_{ion}$/$F_{bolom}$ $\propto$ ($M_{plan}$/$M_{\star}$)$^{2}$ ($a_{plan}$/$R_{\star}$)$^{-6}$\nocite{zahn08}.  Unlike the case of the ``simple'' SPI parameter, $M_{plan}$/$a_{plan}$, we did not find strong and consistent evidence for correlations between any of these alternative SPI metrics and the fractional UV.  The \ion{N}{5}--SPI correlation was significant for both the magnetic reconnection and tidal torque scenario, but this did not hold across the other ions.    Table 5 presents the Spearman rank coefficients and $p$-values for each of these cases.   

\subsubsection{Underlying Stellar Correlations and Planet-hosting Sample Bias}

\begin{figure*}[t]
\begin{center}
\vspace{0.0in}
\begin{tabular}{c}
\hspace{-0.4in}
\includegraphics[width=0.36\textwidth,angle=0,trim={.0in 0.0in 0.0in 0.0in},clip]{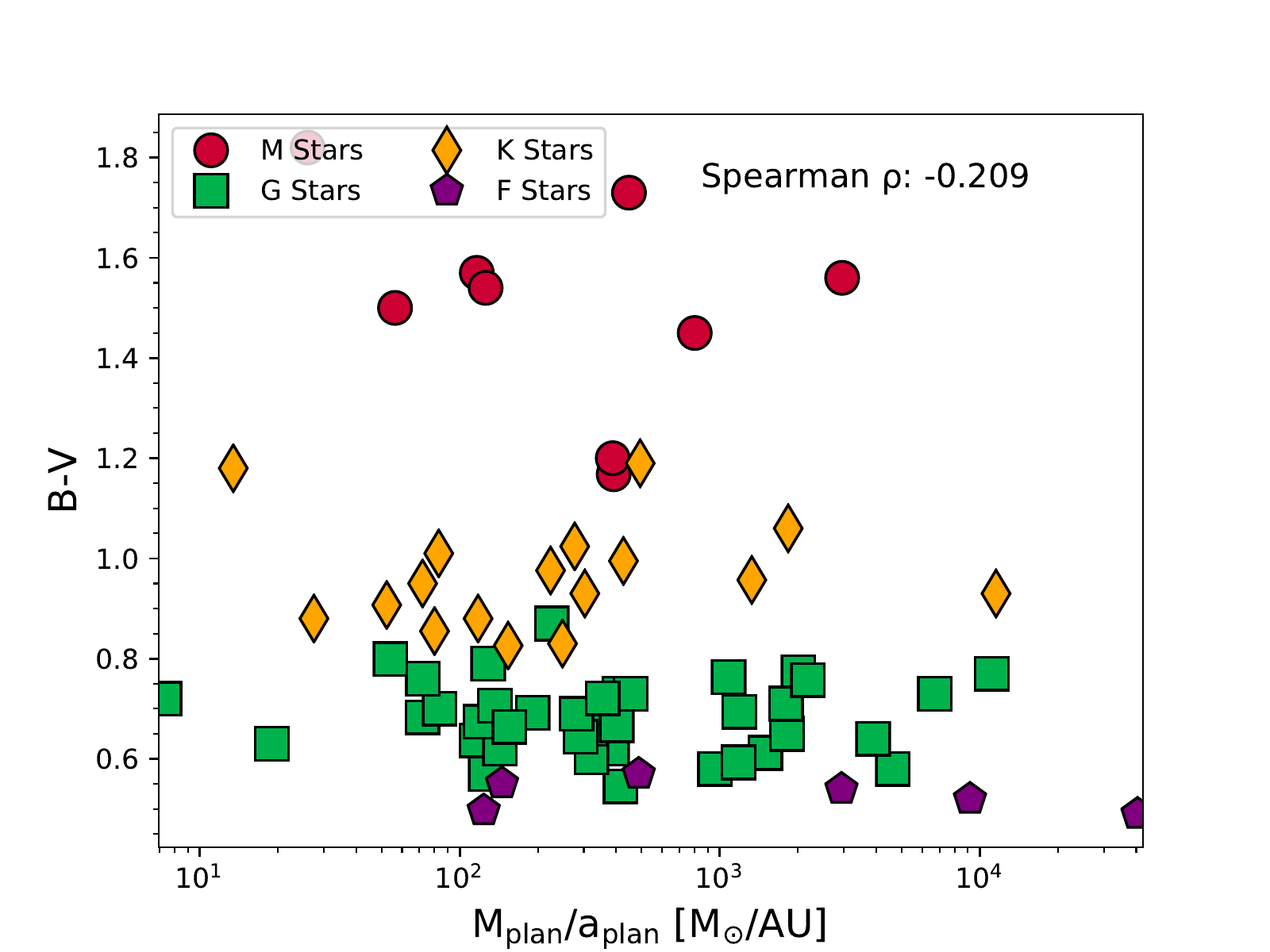}
\hspace{-0.1in}
\includegraphics[width=0.36\textwidth,angle=0,trim={.0in 0.0in 0.0in 0.0in},clip]{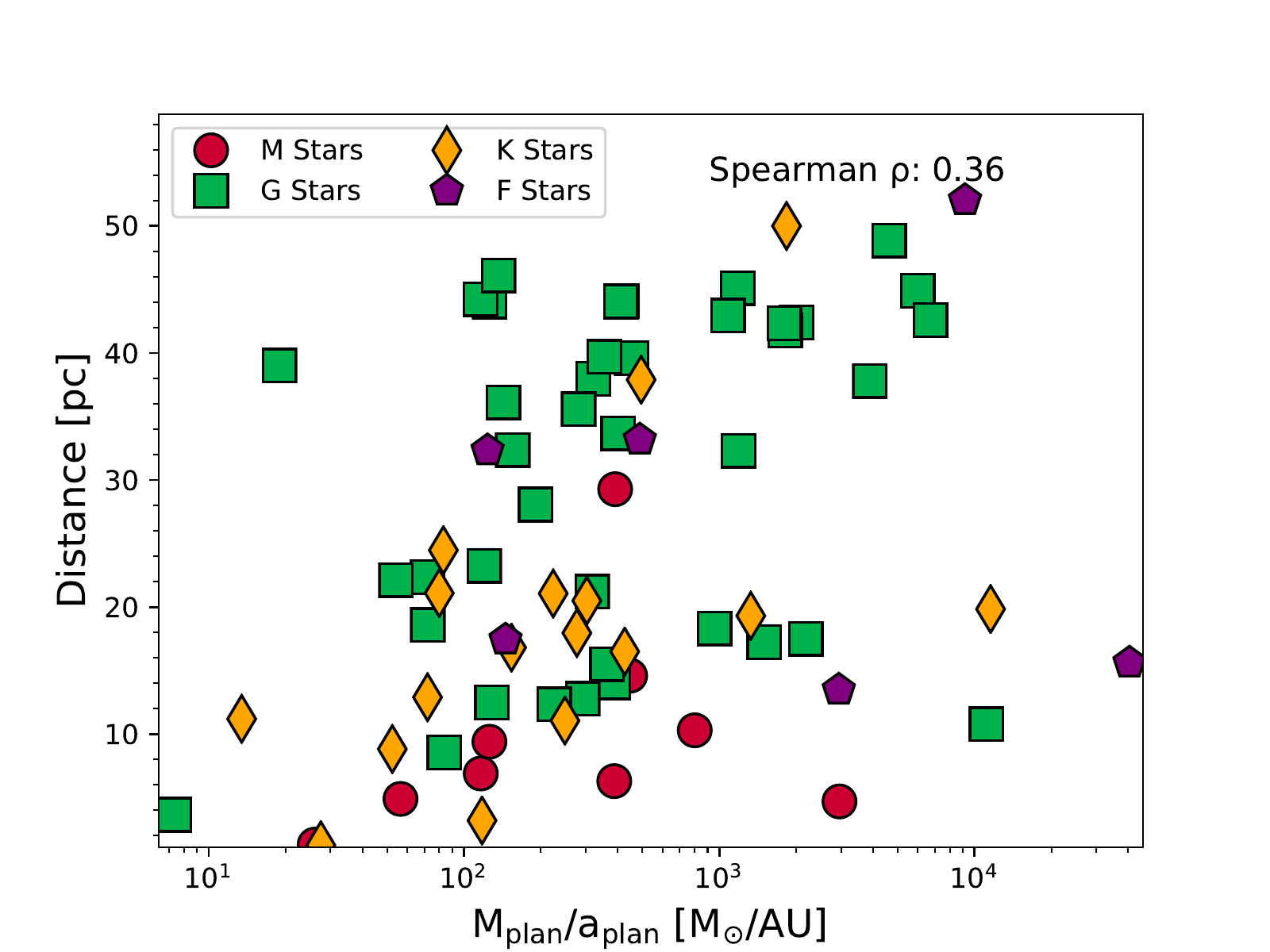}
\hspace{-0.1in}
\includegraphics[width=0.36\textwidth,angle=0,trim={.0in 0.0in 0.0in 0.0in},clip]{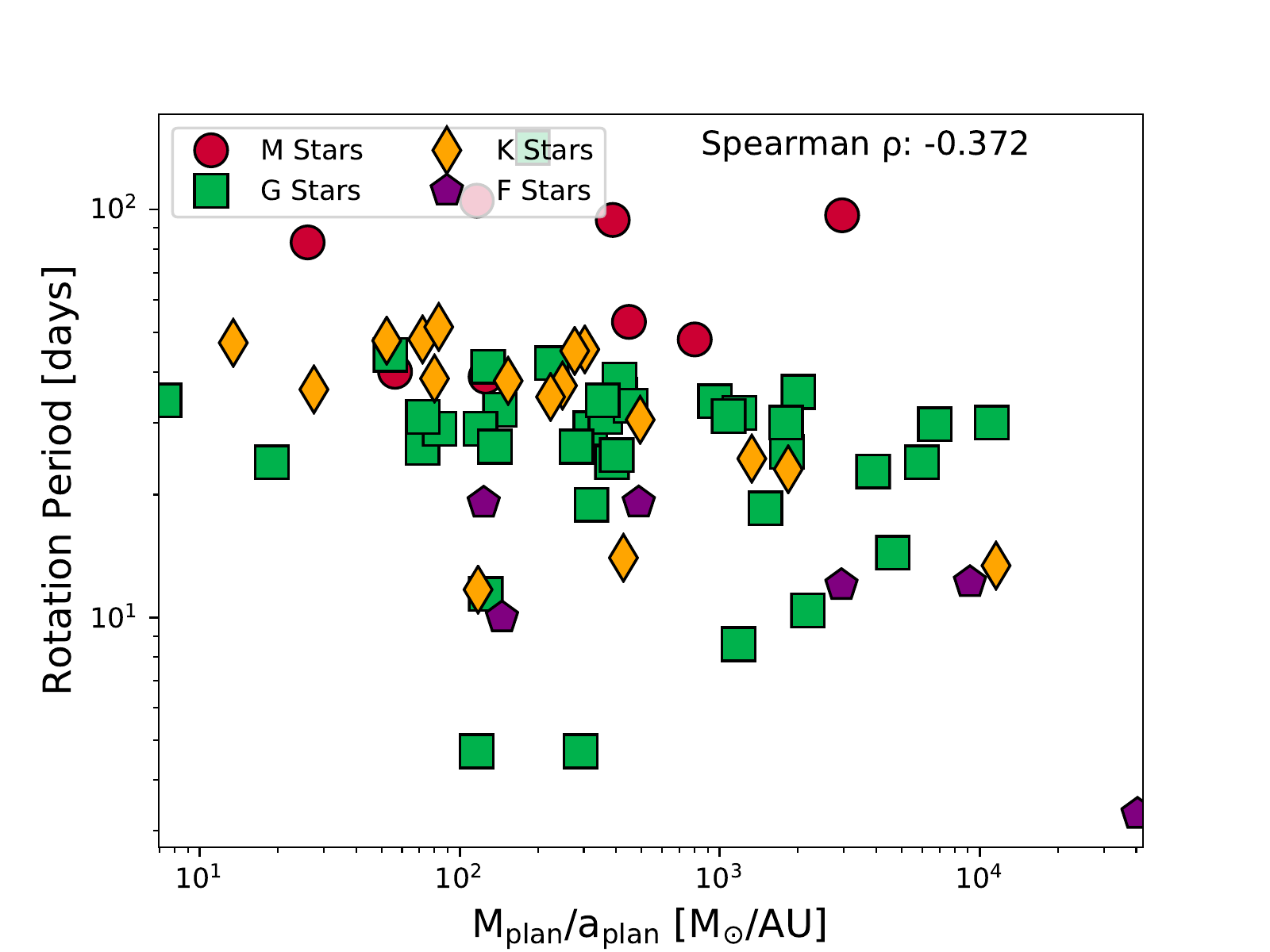} 
\end{tabular}
\end{center}
\vspace{-0.1in}
\caption{We investigate underlying stellar correlations by comparing $B$~--~$V$ ($left$), distance ($center$), and $P_{rot}$ ($right$) vs. the SPI parameter.  All plotting symbols are the same as Figure 7 (and shown in the legend) and the Spearman $\rho$ coefficients are listed in the upper right hand corner of each figure. }
\label{fig-proveit} 
\vspace{-0.15in}
\end{figure*}

Although there appears to be a significant power-law relationship between the SPI parameter and the fractional luminosities in \ion{N}{5}, \ion{C}{2}, \ion{Si}{3}, and \ion{Si}{4}, it is possible that the trend is produced by observational biases within the sample of planet hosts.   Stellar sample biases will serve to limit the detectable bounds of the SPI that can be confidently claimed. To investigate this effect, Spearman rank coefficients were calculated between the SPI parameter and stellar rotation period $\left(P_{rot} \right)$, effective temperature $\left(T_{eff} \right)$, distance $\left(d \right)$, $V$, and $B-V$ (see Table 6). We find that the SPI parameter is correlated with $P_{rot}$ $\left( \rho = -0.371 \right)$ and $d$ $\left(\rho = 0.351 \right)$, at the same level as the SPI parameter is correlated with the UV activity indices. 

The underlying  dependencies on the stellar parameters can be understood by considering stellar and observational biases towards detecting certain types of planets.  First, we find an inverse correlation between the SPI parameter and the rotation period (Figure 8, right).  The RV detection method is less sensitive to lower mass planets around more active stars because the stellar activity adds noise to the RV signal, therefore, only the most massive short period planets are found around stars with small rotation periods.  Second,  we see a correlation of SPI with distance, which we also attribute to an observational bias: for fainter stars, only large RV signals are able to be clearly detected above the photon shot noise.  Given a sample with similar stellar properties (e.g., K and G dwarfs), RV searches will only be sensitive to massive short period planets as the stars become fainter, that is to say that only planets with large $M_{plan}$/$a_{plan}$ will be readily detected at large distances.  This finding is similar to conclusions from previous X-ray SPI analyses demonstrating that the correlation between planet mass and X-ray luminosity was driven by distance effects and stellar sample biases~\citep{poppenhaeger11}.

Because of interdependency of the stellar parameters, an increase in the value of the SPI parameter cannot be directly associated with an enhancement in fractional luminosity. However, the impact of the SPI parameter can still be investigated by properly accounting for the stellar properties in our analysis, discussed in the following subsection.

\subsubsection{Statistical Analysis of Star-Planet Interaction Signal }

One method of incorporating the stellar parameters is to assume that each, like the SPI parameter, 
contributes linearly to the UV activity index.     In a multiple linear regression model, a coefficient $\beta$ represents the amount added by the  corresponding parameter. However, the stellar properties themselves are correlated (e.g. $B-V \propto T_{eff}$; see Table \ref{stellarparam_correlations}), which complicates the standard interpretation. To remove this bias, we first conduct a principal component analysis (PCA) to map the stellar properties and the SPI parameter into a new basis. 

The purpose of the PCA is to transform the multiple linear regression model into a domain where the ``predictor variables," or principal components, are independent and orthogonal to each other \citep{Pearson1901}. Each principal component is constructed as a linear combination of the original variables, which in our case are the stellar parameters and the SPI parameter. A full description is presented in Appendix C.  We reduce the problem to a set of three principal components (PCs), with $PC_1$ being most strongly correlated with $P_{rot}$, $B$~--~$V$, distance, and $T_{eff}$.   The SPI parameter contributes more strongly to $PC_2$ and $PC_3$.  

Equations C.3 and C.5 list the coefficients $\left( \beta \right)$ of the multiple linear regression analysis in the principal components. The results show that three of the principal components contribute significantly to the observed linear relationship with the \ion{N}{5} fractional luminosities (Figure 9, $left$). However, only $PC_2$ and $PC_3$ add to the \ion{Si}{3} and \ion{Si}{4} fractional luminosities, and \ion{C}{2} is only significantly dependent on $PC_3$. When we calculate the Spearman rank coefficients between the multiple linear regression models and fractional luminosities for each ion (Figure C.1), we find that the correlations appear to decrease with formation temperature. \ion{N}{5}, with the highest formation temperature, has $\rho = 0.58$, while \ion{C}{2}, with the lowest formation temperature, has $\rho = 0.33$. \ion{Si}{3} and \ion{Si}{4} fall closer to \ion{C}{2}, with $\rho = 0.32$ and $\rho = 0.33$ for both \ion{Si}{3} and \ion{Si}{4},  respectively. 

While these Spearman coefficients suggest a stronger correlation with the highest ionization emission line, one needs to evaluate the statistical significance  of the importance of the SPI parameter to the observed UV activity levels.   We do this by computing the Bayesian Information Criterion (BIC; Schwarz 1978) in the $F_{NV}$/$F_{bolom}$ versus linear regression plots with and without the SPI term. \nocite{schwarz78, liddle07} We find that the BIC does not change appreciably with the inclusion of the SPI term in the principal components of the linear regression.  Figure 9 ($right$) shows the same PCA analysis for \ion{N}{5} with the SPI information excluded from the regression models.  We conclude that the SPI does not play an explicit role in shaping the distribution of UV activity indices in our sample.  Section C.1 in the Appendix describes a comparable analysis of the non-planet-hosting stars.     

Do these results mean that SPIs are not enhancing the FUV activity indices?  Not necessarily.   Models of both magnetic and tidal SPI indicate that one influence of the planet would be to spin-up the host star, disrupting nominal gyrochronological relationships~\citep{lanza10,maxted15,brown14}.   Indeed, we observe a correlation between the SPI parameter and the stellar rotation period (Figure 8, right).  This may indicate that what we observe as a ``stellar interdependence'' may in fact be a planet-induced effect whereby the interaction with the planetary system is altering the underlying stellar population.  However, if we assume that the rotation period is strongly correlated with the UV activity level, an open question is why the exoplanet host stars as a group are not spun-up to the level of the non-planet-hosting sample.  

\begin{figure*}[h]
\begin{center}
\vspace{0.0in}
\begin{tabular}{c}
\includegraphics[width=0.45\textwidth,angle=0,trim={.0in 0.0in 0.0in 0.0in},clip]{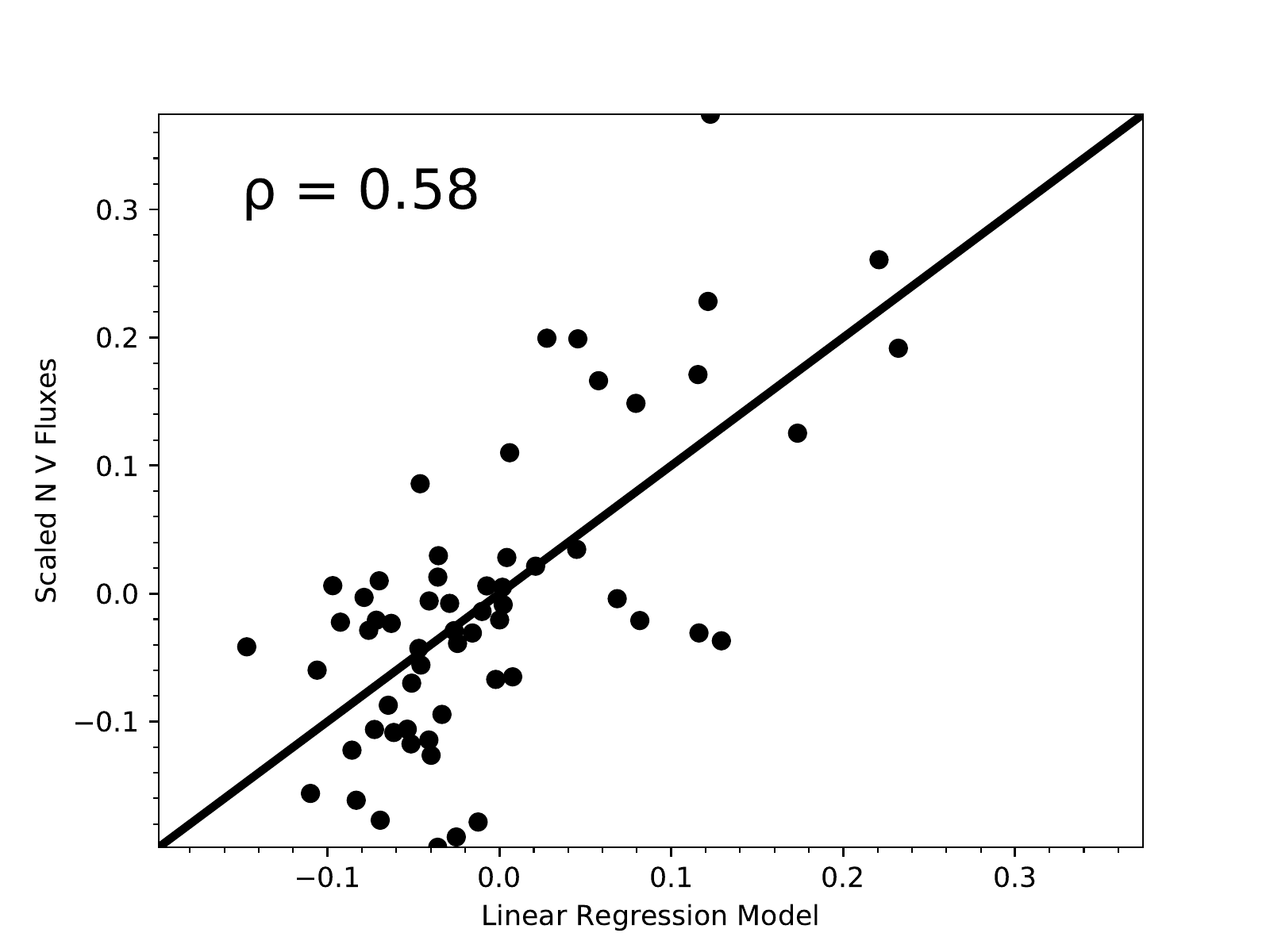}
\includegraphics[width=0.45\textwidth,angle=0,trim={.0in 0.0in 0.0in 0.0in},clip]{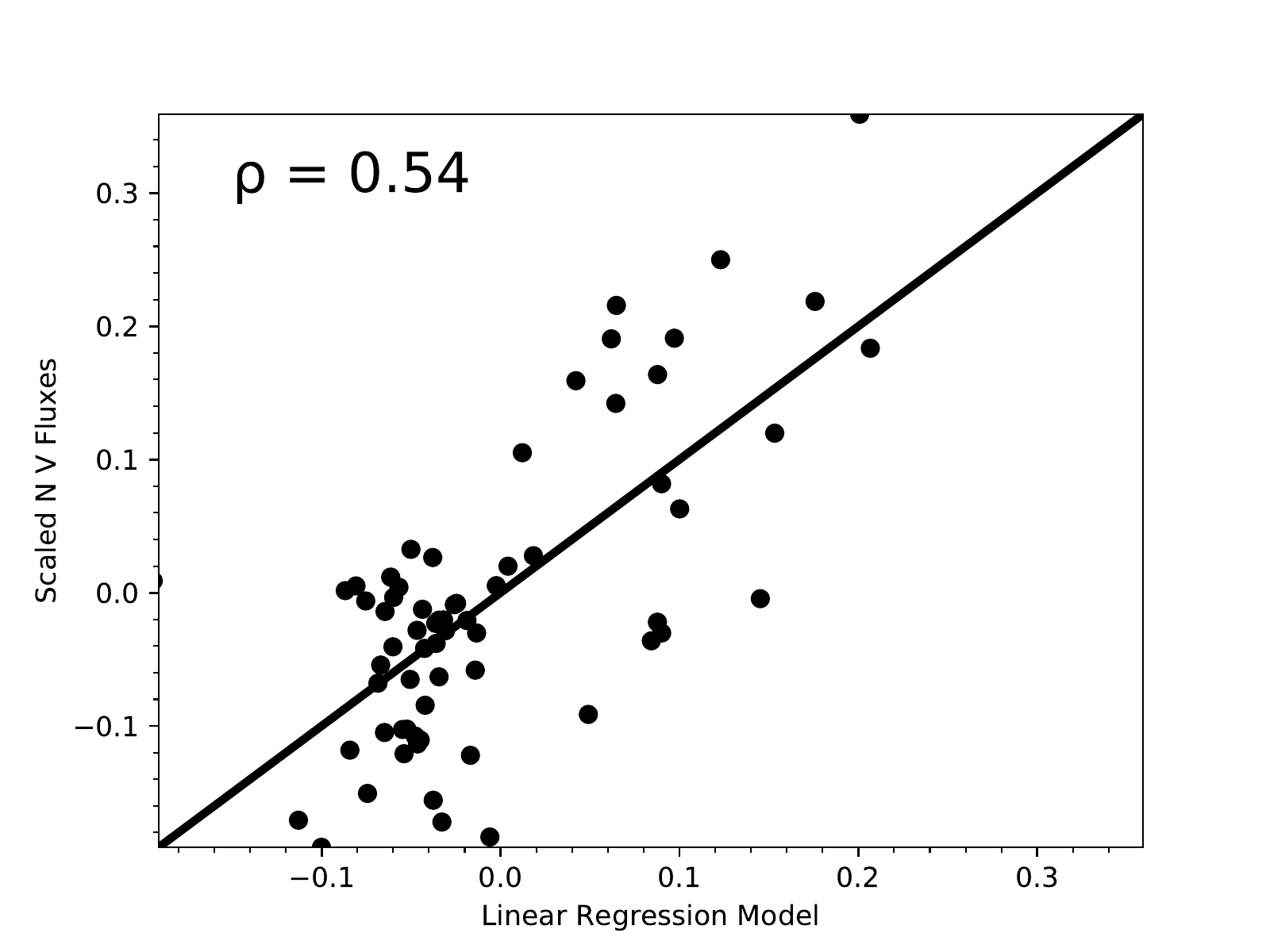}
\end{tabular}
\end{center}
\vspace{-0.2in}
\caption{
($left$) Comparison of the observed \ion{N}{5} fluxes versus those fluxes predicted from a PCA analysis incorporating stellar and planetary properties.   ($right$) The same PCA analysis without the SPI parameter included; the two analyses do not show statistically significant differences (Section 4.2.2).  
The Spearman correlation coefficient $\rho$ describes the agreement between the multivariate linear model and our observations, where we expect $|\rho| = 1$ for a perfect model. Linear regression plots including the stellar and planetary properties for all four ions are displayed in Appendix C. }
\label{fig-proveit} 
\vspace{-0.15in}
\end{figure*}

\section{Summary}

We have presented a survey of UV emission line activity indices in F, G, K,  and M dwarf exoplanet host stars.   We analyzed the largest FUV spectroscopic data set of planet hosting stars (71) assembled to date.  This was complemented by a control sample of 33 stars not currently known to host planets.  These observations were taken from a combination of archival and new programs with $HST$-COS and -STIS, targeting the chromospheric and transition region emission lines of \ion{Si}{3}~$\lambda$~1206~\AA; log$_{10}$$T_{form}$~=~4.7, \ion{N}{5}~$\lambda$~1240~\AA; log$_{10}$$T_{form}$~=~5.2, \ion{C}{2}~$\lambda$~1335~\AA; log$_{10}$$T_{form}$~=~4.5, and \ion{Si}{4}~$\lambda$~1400~\AA; log$_{10}$$T_{form}$~=~4.9.   We studied this data set to compare the UV activity properties of planet-hosting and non-planet-hosting systems, assess the connection between the FUV and EUV irradiance levels incident on orbiting planets, and to search for enhanced stellar activity that may result from the interaction of the planet and the host star.   

  The main results of this work are: 
\begin{enumerate}
	\item The planet-hosting and non-planet-hosting samples display a bimodal distribution in FUV activity level, with the planet-hosting stars factors of 5~--~10 $fainter$ in high-energy emission lines than the non-planet hosts.  This can be explained by a sample bias:  exoplanet host stars bright enough to obtain UV observations largely come from radial-velocity surveys that specifically select for low-activity stars.  Conversely,  previous observations of stars in the solar neighborhood often were originally targeted specifically because of their high levels of activity.  Thus, we are largely seeing the difference between a field population of young (shorter rotation period) non-planet-hosting stars and an  older (longer rotation period) exoplanet host star population.  While this result is straightforward,  it does present a note of caution for researchers modeling exoplanetary atmospheres: by selecting stellar irradiance levels based solely on samples of exoplanet host stars, 
	one is underestimating the flux levels seen earlier in that planet's evolution by an order of magnitude or more.

	\item We have compared the FUV activity indices measured in this work with a sample of overlapping stars with moderate-to-high quality EUV spectra in the 90~--~360~\AA\ range from $EUVE$.   We use these samples to derive a tight relationship between the fractional FUV emission line luminosity and the fractional EUV luminosity.  We present a new relationship for estimating ISM-corrected EUV irradiance in the 90~--~360~\AA\ band, accurate to approximately a factor of 2, for low-mass stars with \ion{N}{5} or \ion{Si}{4} spectra.    EUV fluxes for each of our sample stars are given in Tables 3 and 4.
	
	\item Comparing the FUV activity indices with a star-planet-interaction parameter ($M_{plan}$/$a_{plan}$), we found a significant correlation ($\sim$~99\% confidence) between the presence of massive, short-period planets and stellar activity as indicated by enhanced FUV line emission.   However, observational and astrophysical biases complicate the direct connection of the enhanced UV activity with the planetary system.  We mitigated these interdependencies by creating a principal component analysis treatment of the linear regression problem, finding that fits including SPI do not present a statistically better description of the observations.  On the other hand, our observations do not conclusively rule out the influence of SPIs.  Tides raised on the star by the orbiting planets could influence the stellar rotation period variations, but we do not observe correlations between the UV activity and tidal SPI strength proportionalities. 
\end{enumerate}

\acknowledgments
The data presented here were obtained as part of the $HST$ Guest Observing programs \#12464, \#13650, and \#14633.  N.A. and K.F. thank Sebastian Pineda for enjoyable discussion about the statistical analysis of this data set.  We also acknowledge valuable discussions with Jeffrey Linsky.  This work was supported by STScI grant HST-GO-14633.01.  N.A. is supported by a NASA Earth and Space Sciences Fellowship (NESSF; 80NSSC17K0531) to the University of Colorado at Boulder.   K.F. acknowledges the hospitality of the Reagan Test Site on Kwajalein Atoll, Republic of the Marshall Islands, where a portion of this work was carried out.


\LongTables
\begin{deluxetable*}{cccccccccc}
\tablecaption{Stellar Properties of Planet-Hosts \label{planethoststarprops}
}
\tablewidth{0 pt}
\tabletypesize{\scriptsize}
\tablehead{
\colhead{Name} & \colhead{SpT} & \colhead{$V$} & \colhead{$B-V$} & \colhead{$T_{eff}$} & \colhead{$R_{\ast}$} & \colhead{$d$} & \colhead{$M_{planet}$} & \colhead{$a_{planet}$} & \colhead{$P_{rot}$} \\
\colhead{} & \colhead{} & \colhead{} & \colhead{} & \colhead{[K]} & \colhead{$\left[ R_{\odot} \right]$} & \colhead{[pc]} & \colhead{$\left[M_{\oplus} \right]$} & \colhead{[au]} & \colhead{[days]} \\
}
\startdata
HD 120136 & F7V & 4.49 & 0.49 & 6310\tablenotemark{1} & 1.33\tablenotemark{2} & 15.6 & 1860 & 0.046 & 3.3\tablenotemark{3} \\
HD 197037 & F7V & 6.813 & 0.497 & 6150\tablenotemark{4} & 1.15\tablenotemark{*} & 32.3 & 256.6 & 2.07 & 19.1\tablenotemark{4} \\
HD 136118 & F7V & 6.94 & 0.52 & 6003\tablenotemark{5} & 1.58\tablenotemark{5} & 52.0 & 13300 & 1.45 & 12.2\tablenotemark{5} \\
HD 9826 & F9V & 4.1 & 0.54 & 6210\tablenotemark{2} & 1.63\tablenotemark{2} & 13.5 & 7494 & 2.55 & 12\tablenotemark{6} \\
HD 10647 & F9V & 5.52 & 0.551 & 6039\tablenotemark{1} & 1.1\tablenotemark{7} & 17.4 & 294 & 2.02 & 10\tablenotemark{8} \\
HD 23079 & F9V & 7.11 & 0.57 & 5848\tablenotemark{1} & 1.13\tablenotemark{7} & 33.2 & 779 & 1.596 & 19.1\tablenotemark{**} \\
HD 155358 & G0V & 7.28 & 0.545 & 5900\tablenotemark{4} & 1.39\tablenotemark{9} & 44.1 & 260 & 0.63 & 35.2\tablenotemark{**} \\
$\rho$ CrB & G0V & 5.39 & 0.612 & 5627\tablenotemark{10} & 1.362\tablenotemark{10} & 17.2 & 338 & 0.23 & 18.5\tablenotemark{10} \\
HD 39091 & G0V & 5.67 & 0.58 & 5888\tablenotemark{1} & 2.1\tablenotemark{*} & 18.3 & 3206 & 3.3 & 33.9\tablenotemark{**} \\
HD 187085 & G0V & 7.21 & 0.57 & 6075\tablenotemark{11} & 1.15\tablenotemark{*} & 44.0 & 255 & 2.0 & 11.4\tablenotemark{**} \\
HD 106252 & G0V & 7.36 & 0.64 & 5750\tablenotemark{1} & 1.09\tablenotemark{7} & 37.8 & 10500 & 2.7 & 22.8\tablenotemark{5} \\
HD 209458 & G0V & 7.63 & 0.58 & 6090\tablenotemark{12} & 1.20\tablenotemark{13} & 48.9 & 220 & 0.05 & 14.4\tablenotemark{14} \\
HD 114729 A & G0V & 6.69 & 0.62 & 5662\tablenotemark{1} & 1.46\tablenotemark{7} & 36.1 & 300 & 2.1 & 32.3\tablenotemark{**} \\
HD 13931 & G0V & 7.6 & 0.637 & 5900\tablenotemark{16} & 1.17\tablenotemark{16} & 44.2 & 598 & 5.15 & 4.7\tablenotemark{**} \\
47 UMa & G1V & 5.04 & 0.62 & 5892\tablenotemark{15} & 1.24\tablenotemark{15} & 14.1 & 809 & 2.1 & 24\tablenotemark{5} \\
HD 10180 & G1V & 7.32 & 0.63 & 5911\tablenotemark{17} & 1.11\tablenotemark{*} & 39.0 & 64.4 & 3.4 & 24\tablenotemark{17} \\
HD 117618 & G2V & 7.17 & 0.603 & 5855\tablenotemark{18} & 1.19\tablenotemark{7} & 38.0 & 56.1 & 0.17 & 18.9\tablenotemark{**} \\
HD 121504 & G2V & 7.54 & 0.593 & 6075\tablenotemark{1} & 1.096\tablenotemark{*} & 45.1 & 388 & 0.33 & 8.6\tablenotemark{19} \\
$\mu$ Ara & G3V & 5.15 & 0.7 & 5800\tablenotemark{20} & 1.245\tablenotemark{*} & 15.5 & 555 & 1.5 & 31\tablenotemark{20} \\
16 Cyg B & G3V & 6.2 & 0.66 & 5770\tablenotemark{21} & 0.98\tablenotemark{2} & 21.2 & 534 & 1.68 & 29.1\tablenotemark{22} \\
HD 1461 & G3V & 6.6 & 0.674 & 5765\tablenotemark{23} & 1.095\tablenotemark{23} & 23.2 & 7.6 & 0.06 & 29\tablenotemark{24} \\
HD 38529 & G4V & 5.924 & 0.773 & 5600\tablenotemark{25} & 2.82\tablenotemark{25} & 42.4 & 255 & 0.13 & 35.7\tablenotemark{25} \\
HD 37124 & G4IV-V & 7.68 & 0.667 & 5763\tablenotemark{27} & 0.82\tablenotemark{*} & 33.7 & 214 & 0.5 & 25\tablenotemark{26} \\
HD 147513 & G5V & 5.376 & 0.644 & 5700\tablenotemark{1} & 1.0\tablenotemark{*} & 12.8 & 385 & 1.3 & 4.7\tablenotemark{19} \\ 
HD 222582 & G5V & 7.69 & 0.65 & 5662\tablenotemark{1} & 1.15\tablenotemark{7} & 41.8 & 2425 & 1.3 & 25.4\tablenotemark{**} \\
HD 28185 & G5V & 7.81 & 0.71 & 5705\tablenotemark{28} & 1.03\tablenotemark{7} & 42.3 & 1842 & 1.02 & 30\tablenotemark{28} \\
HD 4113 & G5V & 7.88 & 0.73 & 5688\tablenotemark{29} & 1.036\tablenotemark{*} & 44.0 & 524 & 1.3 & 38.3\tablenotemark{**} \\
HD 65216 & G5V & 7.96 & 0.69 & 5718\tablenotemark{27} & 1.036\tablenotemark{*} & 35.6 & 387 & 1.4 & 26.2\tablenotemark{**} \\
HD 178911 B & G5V & 7.98 & 0.73 & 5667\tablenotemark{7} & 1.14\tablenotemark{7} & 42.6 & 2317 & 0.3 & 29.7\tablenotemark{**} \\
HD 79498 & G5V & 8.02 & 0.706 & 5740\tablenotemark{4} & 1.036\tablenotemark{*} & 46.1 & 428 & 3.1 & 26.2\tablenotemark{**} \\
HIP 91258 & G5V & 8.65 & 0.01 & 5519\tablenotemark{30} & 1.036\tablenotemark{*} & 44.9 & 339 & 0.06 & 24\tablenotemark{30} \\
HD 90156 & G5V & 6.92 & 0.683 & 5599\tablenotemark{31} & 1.036\tablenotemark{*} & 22.4 & 18 & 0.2 & 26\tablenotemark{31} \\
HD 115617 & G6.5V & 4.74 & 0.7 & 5530\tablenotemark{7} & 0.94\tablenotemark{7} & 8.6 & 18.2 & 0.2 & 29\tablenotemark{32} \\
HD 70642 & G6V & 7.17 & 0.692 & 5670\tablenotemark{1} & 0.97\tablenotemark{27} & 28.1 & 607 & 3.2 & 142\tablenotemark{**} \\
HD 47186 & G6V & 7.63 & 0.73 & 5675\tablenotemark{33} & 1.017\tablenotemark{*} & 39.6 & 22.6 & 0.05 & 33\tablenotemark{33} \\
HD 92788 & G6V & 7.3 & 0.694 & 5754\tablenotemark{34} & 1.05\tablenotemark{34} & 32.3 & 1133 & 0.95 & 31.7\tablenotemark{34} \\
HD 102117 & G6V & 7.47 & 0.721 & 5672\tablenotemark{35} & 1.27\tablenotemark{7} & 39.7 & 54 & 0.15 & 34\tablenotemark{35} \\
HD 4208 & G7V & 7.78 & 0.664 & 5571\tablenotemark{1} & 0.85\tablenotemark{*} & 32.4 & 257 & 1.7 & 0\tablenotemark{7} \\
HD 10700 & G8V & 3.5 & 0.72 & 5340\tablenotemark{36} & 0.793\tablenotemark{36} & 3.7 & 3.94 & 0.538 & 34\tablenotemark{36} \\
HD 69830 & G8V & 5.95 & 0.79 & 5385\tablenotemark{37} & 0.895\tablenotemark{*} & 12.5 & 10 & 0.08 & 41.2\tablenotemark{**} \\
55 Cnc & G8V & 5.95 & 0.87 & 5200\tablenotemark{38} & 0.943\tablenotemark{38} & 12.3 & 1230 & 5.4 & 42\tablenotemark{34} \\
HD 1237 & G8V & 6.578 & 0.757 & 5417\tablenotemark{39} & 0.9\tablenotemark{*} & 17.5 & 1070 & 0.49 & 10.4\tablenotemark{39} \\
HD 154345 & G8V & 6.74 & 0.76 & 5468\tablenotemark{40} & 0.94\tablenotemark{40} & 18.6 & 304 & 4.2 & 31\tablenotemark{40} \\
GJ 86 & G9V & 6.17 & 0.77 & 5350\tablenotemark{41} & 0.855\tablenotemark{*} & 10.8 & 1272 & 0.1 & 30\tablenotemark{42} \\
HD 147018 & G9V & 8.3 & 0.763 & 5441\tablenotemark{43} & 0.96\tablenotemark{*} & 43.0 & 2080 & 1.9 & 31.1\tablenotemark{**} \\
HD 164922 & G9V & 7.01 & 0.799 & 5293\tablenotemark{10} & 0.999\tablenotemark{10} & 22.1 & 114 & 2.1 & 44\tablenotemark{45} \\
HD 189733 & K0V+M4V & 7.648 & 0.93 & 4880\tablenotemark{2} & 0.805\tablenotemark{2} & 19.8 & 363 & 0.03 & 13.4\tablenotemark{47} \\
HD 7924 & K0.5V & 7.185 & 0.826 & 5177\tablenotemark{46} & 0.78\tablenotemark{46} & 16.8 & 8.7 & 0.06 & 38\tablenotemark{46} \\
HD 3651 & K0.5V & 5.88 & 0.83 & 5270\tablenotemark{16} & 0.88\tablenotemark{48} & 11.1 & 73.3 & 0.295 & 37\tablenotemark{49} \\
HD 128621 & K1V & 1.33 & 0.88 & 5336\tablenotemark{50} & 0.863\tablenotemark{50} & 1.25 & 1.1 & 0.04 & 36.2\tablenotemark{50} \\
HD 114783 & K1V & 7.56 & 0.93 & 5105\tablenotemark{1} & 0.78\tablenotemark{7} & 20.5 & 351 & 1.2 & 45.4\tablenotemark{**} \\
HD 97658 & K1V & 7.714 & 0.855 & 5050\tablenotemark{51} & 0.908\tablenotemark{51} & 21.1 & 6.4 & 0.08 & 38.5\tablenotemark{51} \\
HD 40307 & K2.5V & 7.147 & 0.95 & 4750\tablenotemark{51} & 0.856\tablenotemark{51} & 12.9 & 9.5 & 0.132 & 48\tablenotemark{51} \\
HD 192263 & K2.5V & 7.767 & 0.957 & 4965\tablenotemark{1} & 0.75\tablenotemark{7} & 19.3 & 203 & 0.2 & 24.5\tablenotemark{52} \\
$\epsilon$ Eri & K2V & 3.73 & 0.88 & 4900\tablenotemark{51} & 0.882\tablenotemark{51} & 3.2 & 400 & 3.4 & 11.7\tablenotemark{51} \\
HD 192310 & K2V & 5.723 & 0.907 & 5166\tablenotemark{53} & 0.8\tablenotemark{27} & 8.8 & 16.9 & 0.3 & 47.67\tablenotemark{53} \\
HD 99492 & K2V & 7.53 & 1.024 & 4740\tablenotemark{54} & 0.96\tablenotemark{54} & 18.0 & 33.7 & 0.1 & 45\tablenotemark{54} \\
HD 128311 & K3V & 7.446 & 0.995 & 4965\tablenotemark{7} & 0.73\tablenotemark{7} & 16.5 & 463 & 1.1 & 14\tablenotemark{26} \\
HD 104067 & K3V & 7.921 & 0.976 & 4969\tablenotemark{55} & 0.856\tablenotemark{*} & 21.1 & 59 & 0.3 & 34.7\tablenotemark{55} \\
HD 156668 & K3V & 8.42 & 1.01 & 4850\tablenotemark{56} & 0.72\tablenotemark{56} & 24.5 & 4.2 & 0.05 & 51.5\tablenotemark{56} \\
HAT P 11 & K4V & 9.47 & 1.19 & 4780\tablenotemark{57} & 0.75\tablenotemark{57} & 37.9 & 26.2 & 0.053 & 30.5\tablenotemark{58} \\
WASP 69 & K5V & 9.87 & 1.06 & 4720\tablenotemark{59} & 0.813\tablenotemark{59} & 50 & 83 & 0.05 & 23.07\tablenotemark{59} \\
HD 85512 & K6V & 7.651 & 1.18 & 4300\tablenotemark{51} & 0.778\tablenotemark{51} & 11.2 & 3.5 & 0.26 & 47.1\tablenotemark{51} \\
GJ 832 & M1.5V & 8.672 & 1.5 & 3816\tablenotemark{51} & 0.631\tablenotemark{51} & 4.9 & 203 & 3.6 & 40\tablenotemark{51} \\
GJ 667 C & M1.5V & 10.22 & 1.57 & 3440\tablenotemark{51} & 0.562\tablenotemark{51} & 6.9 & 5.7 & 0.049 & 105\tablenotemark{51} \\
GJ 3470 & M2V & 12.332 & 1.168 & 3600\tablenotemark{44} & 0.550\tablenotemark{44} & 29.3 & 13.9 & 0.04 & \nodata \\
GJ 176 & M2.5V & 9.951 & 1.54 & 3310\tablenotemark{51} & 0.493\tablenotemark{51} & 9.4 & 8.3 & 0.066 & 38.9\tablenotemark{51} \\
GJ 436 & M3.5V & 10.613 & 1.45 & 3310\tablenotemark{51} & 0.493\tablenotemark{51} & 10.3 & 23 & 0.03 & 48\tablenotemark{51} \\
GJ 1214 & M4.5V & 14.67 & 1.73 & 2920\tablenotemark{51} & 0.286\tablenotemark{51} & 14.6 & 6.4 & 0.01 & 53\tablenotemark{51} \\
GJ 876 & M5V & 10.192 & 1.56 & 3180\tablenotemark{51} & 0.424\tablenotemark{51} & 4.7 & 615 & 0.208 & 96.7\tablenotemark{51} \\
GJ 581 & M5V & 10.56 & 1.2 & 3310\tablenotemark{51} & 0.493\tablenotemark{51} & 6.3 & 15.9 & 0.04 & 94.2\tablenotemark{51} \\
Proxima Centauri & M5.5V & 11.13 & 1.82 & 3050\tablenotemark{60} & 0.14\tablenotemark{60} & 1.299 & 1.3\tablenotemark{60} & 0.05\tablenotemark{60} & 83\tablenotemark{60} 
\enddata
\tablenotetext{*}{$R_{\ast}$ estimated based on spectral type}
\tablenotetext{**}{Rotation periods are upper limits (calculated from $v \sin i$)}
\tablerefs{\scriptsize (1) \citet{Nordstrom2004}; (2) \citet{Baines2008}; (3) \citet{Baliunas1997}; (4) \citet{Robertson2012}; (5) \citet{Fischer2002}; (6) \citet{Butler1999}; (7) \citet{Valenti2005}; (8) \citet{Marmier2013}; (9) \citet{Fuhrmann2008}; (10) \citet{Fulton2016}; (11) \citet{Jones2006}; (12) \citet{Schuler2011}; (13) \citet{Brown2001}; (14) \citet{Mazeh2000}; (15) \citet{Fuhrmann1997}; (16) \citet{Wittrock2017}; (17) \citet{Lovis2011}; (18) \citet{Tinney2005}; (19) \citet{Mayor2004}; (20) \citet{Santos2004}; (21) \citet{Fuhrmann1998}; (22) \citet{Hale1994}; (23) \citet{Rivera2010}; (24) \citet{Wright2004}; (25) \citet{Fischer2003}; (26) \citet{Vogt2005}; (27) \citet{Bonfanti2015}; (28) \citet{Santos2001}; (29) \citet{Tamuz2008}; (30) \citet{Moutou2014}; (31) \citet{Mordasini2011}; (32) \citet{Baliunas1996}; (33) \citet{Bouchy2009}; (34) \citet{Fischer2001}; (35) \citet{Lovis2005}; (36) \citet{Tuomi2013}; (37) \citet{Lovis2006}; (38) \citet{vonBraun2011}; (39) \citet{Naef2000}; (40) \citet{Wright2008}; (41) \citet{Queloz1999}; (42) \citet{Saar1997}; (43) \citet{Segransan2010}; (44) \citet{Bonfils2012}; (45) \citet{Isaacson2010}; (46) \citet{Howard2009}; (47) \citet{Knutson2007}; (48) \citet{See2017}; (49) \citet{Olspert2017}; (50) \citet{DeWarf2010}; (51) \citet{France2016_MUSCLESI}; (52) \citet{Henry2002}; (53) \citet{Pepe2011}; (54) \citet{Meschiari2011}; (55) \citet{Segransan2011}; (56) \citet{Howard2011}; (57) \citet{Bakos2010}; (58) \citet{SO2011}; (59) \citet{Anderson2014}; (60) \citet{AngladaEscude2016}}
\tablecomments{$M_{planet}$ and $a_{planet}$ are for the most massive planet in each system, as listed in The Extrasolar Planets Encyclopedia. Spectral types, $V$, $B-V$, and distances were taken from Simbad.}
\end{deluxetable*}
\clearpage

\begin{deluxetable*}{cccccccc}
\tablecaption{Stellar Properties of Non-Planet Hosts \label{nonplanethoststarprops}
}
\tablewidth{0 pt}
\tabletypesize{\scriptsize}
\tablehead{
\colhead{Name} & \colhead{SpT} & \colhead{$V$} & \colhead{$B-V$} & \colhead{$T_{eff}$} & \colhead{$R_{\ast}$} & \colhead{$d$} & \colhead{$P_{rot}$} \\
\colhead{} & \colhead{} & \colhead{} & \colhead{} & \colhead{[K]} & \colhead{$\left[ R_{\odot} \right]$} & \colhead{[pc]} & \colhead{[days]} \\
}
\startdata
HD 28568 & F2V & 6.484 & 0.443 & 6656\tablenotemark{1} & 1.35\tablenotemark{2} & 41.5 & 0.9\tablenotemark{53} \\
HD 28033 & F8V & 7.35 & 0.51 & 6167\tablenotemark{*} & 1.2\tablenotemark{2} & 46.6 & 1.9\tablenotemark{**} \\
HD 33262 & F9V & 4.708 & 0.507 & 6158\tablenotemark{4} & 0.96\tablenotemark{2} & 11.6 & 4\tablenotemark{3**} \\
HD 106516 & F9V & 6.11 & 0.46 & 6327\tablenotemark{5} & 1.15\tablenotemark{*} & 22.4 & \nodata \\
HD 28205 & G0V & 7.404 & 0.545 & 6306\tablenotemark{6} & 1.15\tablenotemark{*} & 47.3 & 5.87\tablenotemark{6} \\
HD 25825 & G0V & 7.811 & 0.605 & 6097\tablenotemark{7} & 1\tablenotemark{8} & 46.9 & 6.5\tablenotemark{8} \\
HD 97334 & G0V & 6.41 & 0.61 & 5898\tablenotemark{9} & 1.01\tablenotemark{3} & 22.8 & 7.6\tablenotemark{10} \\
HD 39587 & G0V & 4.4 & 0.6 & 5890\tablenotemark{11} & 0.96\tablenotemark{11} & 8.7 & 5.24\tablenotemark{11} \\
HII 314 & G1V & 10.4 & 0.8 & 5845\tablenotemark{12} & 0.99\tablenotemark{12} & 130.5 & 1.47\tablenotemark{12} \\
16 Cyg A & G1.5V & 5.95 & 0.64 & 5825\tablenotemark{13} & 1.22\tablenotemark{13} & 21.3 & 26.9\tablenotemark{14} \\
HD 72905 & G1.5V & 5.64 & 0.62 & 5850\tablenotemark{11} & 0.95\tablenotemark{11} & 14.4 & 4.9\tablenotemark{11} \\
HD 129333 & G1.5V & 7.61 & 0.59 & 5853\tablenotemark{15} & 1\tablenotemark{16} & 35.8 & 2.606\tablenotemark{16} \\
HD 199288 & G2V & 6.52 & 0.59 & 5757\tablenotemark{17} & 0.969\tablenotemark{17} & 22.1 & 12\tablenotemark{17} \\
$\alpha$ Cen A & G2V & 0.01 & 0.71 & 5770\tablenotemark{18} & 1.22\tablenotemark{19} & 1.3 & 29\tablenotemark{20} \\
HD 59967 & G3V & 6.635 & 0.639 & 5847\tablenotemark{21} & 0.89\tablenotemark{3} & 21.7 & 6.14\tablenotemark{3} \\
HD 20630 & G5V & 4.85 & 0.67 & 5776\tablenotemark{15} & 0.93\tablenotemark{11} & 9.1 & 9.2\tablenotemark{22} \\
HD 43162 & G6.5V & 6.366 & 0.702 & 5473\tablenotemark{23} & 0.901\tablenotemark{23} & 16.8 & 7.158\tablenotemark{24} \\
HD 131156 & G8V & 4.593 & 0.777 & 5550\tablenotemark{25} & 0.8\tablenotemark{27} & 6.7 & 6.43\tablenotemark{26} \\
KIC 11560431 & K0V & 9.5 & \nodata & 5094\tablenotemark{28} & 0.892\tablenotemark{28} & \nodata & 3.14\tablenotemark{29} \\
HD 166 & K0V & 6.13 & 0.75 & 5509\tablenotemark{15} & 0.9\tablenotemark{2} & 13.8 & 6.23\tablenotemark{30} \\
HD 165341 & K0V & 4.03 & 0.86 & 5407\tablenotemark{31} & 0.85\tablenotemark{2} & 5.1 & 19.7\tablenotemark{32} \\
HD 103095 & K1V & 6.45 & 0.75 & 5033\tablenotemark{*} & 0.93\tablenotemark{*} & 9.1 & 31\tablenotemark{22} \\
HR 1925 & K1V & 6.23 & 0.84 & 5309\tablenotemark{15} & 0.93\tablenotemark{*} & 12.3 & 10.86\tablenotemark{33} \\
HD 22468 & K2V & 5.71 & 0.92 & 4867\tablenotemark{*} & 3.9\tablenotemark{34} & 30.7 & 2.84\tablenotemark{34} \\
HD 155886 & K2V & 5.08 & 0.85 & 4867\tablenotemark{*} & 0.69\tablenotemark{35} & 5.5 & 20.69\tablenotemark{36} \\
LTT 2050 & M1V & 10.331 & 1.507 & 3438\tablenotemark{41} & 0.4\tablenotemark{*} & 11.2 & \nodata \\
HD 197481 & M1V & 8.627 & 1.423 & 3600\tablenotemark{42} & 0.61\tablenotemark{2} & 9.9 & 4.85\tablenotemark{10} \\
Kapteyn's Star & M1V & 8.853 & 1.58 & 3527\tablenotemark{43} & 0.341\tablenotemark{43} & 3.9 & 84.7\tablenotemark{44} \\
LP 415-1619 & M2V & 13.338 & 1.482 & 3420\tablenotemark{48} & 0.58\tablenotemark{48} & 46.3 & \nodata \\
AD Leo & M4V & 9.52 & 1.3 & 3130\tablenotemark{*} & 0.38\tablenotemark{2} & 4.7 & 2.6\tablenotemark{10} \\
LHS-26 & M4V & 10.977 & 0.077 & 3130\tablenotemark{*} & 0.301\tablenotemark{46} & 5.6 & 87.1\tablenotemark{54} \\
Procyon & F5V & 0.37 & 0.42 & 6530\tablenotemark{49} & 2.03\tablenotemark{49} & 3.5 & 10.3\tablenotemark{50} \\
EV Lac & M4V & 10.26 & 1.59 & 3130\tablenotemark{*} & 0.38\tablenotemark{*} & 5.1 & 4.378\tablenotemark{51} \\
YY Gem & M0.5V & 9.27 & 1.29 & 3820\tablenotemark{52} & 0.6191\tablenotemark{52} & 14.9\tablenotemark{52} & 3\tablenotemark{***} 
\enddata
\tablenotetext{*}{$R_{\ast}$ or $T_{eff}$ estimated based on spectral type}
\tablenotetext{**}{Rotation periods are upper limits (calculated from $v \sin i$)}
\tablenotetext{***}{Rotation period estimated from age of Castor system~\citep{chabrier95,torres02}}
\tablerefs{\scriptsize (1) \citet{Boesgaard2016}; (2) \citet{Wood2005}; (3) \citet{Linsky2012}; (4) \citet{AvE2012}; (5) \citet{Ge2016}; (6) \citet{Ramirez2017}; (7) \citet{daSilva2015}; (8) \citet{Linsky2012b}; (9) \citet{Eisenbeiss2013}; (10) \citet{Hempelmann1995}; (11) \citet{Fichtinger2017}; (12) \citet{Rice2001}; (13) \citet{Booth2017}; (14) \citet{Hale1994}; (15) \citet{Rich2017}; (16) \citet{Berdyugina2005}; (17) \citet{Loyd2014}; (18) \citet{Zhao2018}; (19) \citet{Kervella2003}; (20) \citet{Hallam1991}; (21) \citet{Reddy2017}; (22) \citet{Brandenburg2017}; (23) \citet{Gaidos2002}; (24) \citet{Kajatkari2015}; (25) \citet{Gray1994}; (26) \citet{Toner1988}; (27) \citet{Petit2005}; (28) \citet{Brown2011}; (29) \citet{Balona2012}; (30) \citet{Gaidos2000}; (31) \citet{Huang2015}; (32) \citet{Noyes1984}; (33) \citet{Zhang2011}; (34) \citet{Fekel1983}; (35) \citet{Wood2012}; (36) \citet{Donahue1996}; (37) \citet{Messina2010}; (38) \citet{Maldonado2017}; (39) \citet{Malo2014}; (40) \citet{Woolf2005}; (41) \citet{Tuomi2014}; (42) \citet{Pawellek2014}; (43) \citet{Houdebine2010}; (44) \citet{Guinan2016}; (45) \citet{Pecaut2013}; (46) \citet{Newton2017}; (47) \citet{Kiraga2012}; (48) \citet{Mann2015}; (49) \citet{Yildiz2016}; (50) \citet{Arentoft2008}; (51) \citet{Pettersen1980}; (52) \citet{Torres2002}; (53) \citet{Stepien88}; (54) \citet{Newton16}}
\tablecomments{Spectral types, $V$, $B-V$, and distances were taken from Simbad.}
\end{deluxetable*}
\clearpage

\appendix

\section{\ion{Si}{3}, \ion{N}{5}, \ion{C}{2}, and \ion{Si}{4} Emission Line Measurements from the Planet-hosting and Non-planet-hosting Samples}

In  Tables 3 (planet-hosting stars) and 4 (non-planet hosts), we display the full emission line measurement lists for both samples studied in this work. 

\begin{deluxetable}{ccccccc}
\tablecaption{Planet Host Flux Measurements \label{planethostfluxes}
}
\tablewidth{0 pt}
\tabletypesize{\scriptsize}
\tablehead{
\colhead{Name} & \colhead{$F_{bol}$\tablenotemark{a}} & \colhead{N V} & \colhead{C II} & \colhead{Si III} & \colhead{Si IV} & \colhead{$F(90-360\AA)$} \\
\colhead{} & \colhead{[$10^{-7}$} & \colhead{[$10^{-15}$} & \colhead{[$10^{-15}$} & \colhead{[$10^{-15}$} & \colhead{[$10^{-15}$} & \colhead{[$10^{-14}$} \\
\colhead{} & \colhead{erg s$^{-1}$ cm$^{-2}$]} & \colhead{erg s$^{-1}$ cm$^{-2}$ \AA$^{-1}$]} & \colhead{erg s$^{-1}$ cm$^{-2}$ \AA$^{-1}$]} & \colhead{erg s$^{-1}$ cm$^{-2}$ \AA$^{-1}$]} & \colhead{erg s$^{-1}$ cm$^{-2}$ \AA$^{-1}$]} & \colhead{erg s$^{-1}$ cm$^{-2}$]} \\
}
\startdata
HD 120136 & 3.3 & $81.4 \pm 0.9$ & \nodata & $164 \pm 2$ & \nodata & 662 \\
HD 197037 & 0.5 & $0.4 \pm 0.1$ & $4.1 \pm 0.2$ & $2.9 \pm 0.1$ & $2.7 \pm 0.3$ & 3.38 \\
HD 136118 & 0.3 & $0.7 \pm 0.1$ & $6.5 \pm 0.2$ & $4.5 \pm 0.2$ & $3.9 \pm 0.3$ & 5.39 \\
HD 9826 & 6.3 & $20 \pm 1$ & $132 \pm 2$ & $117 \pm 4$ & $45 \pm 2$ & 163 \\
HD 10647 & 1.5 & $5.8 \pm 0.2$ & $46.2 \pm 0.6$ & $39.9 \pm 0.6$ & $29.0 \pm 0.7$ & 47.4 \\
HD 23079 & 0.4 & $0.5 \pm 0.1$ & $3.4 \pm 0.2$ & $2.0 \pm 0.2$ & $2.6 \pm 0.3$ & 3.68 \\
HD 155358 & 0.3 & $0.2 \pm 0.1$ & $1.7 \pm 0.2$ & $1.0 \pm 0.1$ & $0.8 \pm 0.2$ & 1.84 \\
$\rho$ CrB & 1.8 & $1.2 \pm 0.1$ & $13.8 \pm 0.3$ & $7.2 \pm 0.3$ & $6.1 \pm 0.3$ & 9.43 \\
HD 39091 & 4.6 & $2.4 \pm 0.2$ & $15.3 \pm 0.3$ & $8.8 \pm 0.2$ & $7.0 \pm 0.3$ & 19.7 \\
HD 187085 & 0.3 & $0.5 \pm 0.1$ & $5.4 \pm 0.2$ & $3.4 \pm 0.2$ & $2.6 \pm 0.2$ & 4.40 \\
HD 209458 & 0.2 & $0.7 \pm 0.6$ & \nodata & $1.8 \pm 0.3$ & \nodata & 5.74 \\
HD 114729 A & 0.5 & $0.5 \pm 0.1$ & $4.5 \pm 0.2$ & $2.7 \pm 0.2$ & $1.8 \pm 0.2$ & 4.35 \\
HD 13931 & 0.2 & $0.7 \pm 0.1$ & $4.4 \pm 0.2$ & $2.4 \pm 0.2$ & $2.4 \pm 0.3$ & 5.76 \\
47 UMa & 2.7 & $3.4 \pm 0.2$ & $26.0 \pm 0.5$ & $14.2 \pm 0.4$ & $10.7 \pm 0.4$ & 27.7 \\
HD 10180 & 0.3 & $0.7 \pm 0.1$ & $4.4 \pm 0.2$ & $2.4 \pm 0.2$ & $2.3 \pm 0.2$  & 5.58 \\
HD 117618 & 0.3 & $0.5 \pm 0.1$ & $5.4 \pm 0.2$ & $2.6 \pm 0.1$ & $2.1 \pm 0.2$  & 4.19 \\
HD 121504 & 0.2 & $1.1 \pm 0.2$ & $9.3 \pm 0.3$ & $6.7 \pm 0.2$ & $7.6 \pm 0.3$ & 9.02 \\
$\mu$ Ara & 1.9 & $3.6 \pm 0.2$ & $24.0 \pm 0.4$ & $13.2 \pm 0.3$ & $12.4 \pm 0.4$ & 29.2 \\
16 Cyg B & 0.7 & \nodata & $13.5 \pm 0.2$ & $8.3 \pm 0.1$ & $7.4 \pm 0.2$ & \nodata \\
HD 1461 & 0.7 & $2.5 \pm 0.1$ & $13.6 \pm 0.3$ & $9.1 \pm 0.2$ & $8.7 \pm 0.4$ & 20.1 \\
HD 38529 & 1.3 & $4.8 \pm 0.2$ & $24.9 \pm 0.4$ & $18.0 \pm 0.4$ & $14.9 \pm 0.4$ & 38.9 \\
HD 37124 & 0.2 & $0.2 \pm 0.1$ & $5.0 \pm 0.2$ & $2.4 \pm 0.2$ & $2.8 \pm 0.3$ & 1.83 \\
HD 147513 & 1.9 & $12.7 \pm 0.6$ & $0.4 \pm 0.2$ & \nodata & $3.2 \pm 0.6$ & 103 \\
HD 222582 & 0.2 & $0.4 \pm 0.1$ & $3.4 \pm 0.2$ & $1.9 \pm 0.1$ & $2.4 \pm 0.3$ & 3.44 \\
HD 28185 & 0.2 & $0.7 \pm 0.1$ & $5.6 \pm 0.2$ & $2.1 \pm 0.1$ & $2.6 \pm 0.3$ & 5.54 \\
HD 4113 & 0.2 & $0.6 \pm 0.1$ & $3.4 \pm 0.2$ & $1.7 \pm 0.2$ & $2.4 \pm 0.3$ & 4.62 \\
HD 65216 & 0.3 & $0.6 \pm 0.1$ & $5.1 \pm 0.3$ & $3.1 \pm 0.2$ & $3.9 \pm 0.3$ & 5.06 \\
HD 178911 B & 0.2 & $1.1 \pm 0.1$ & $5.6 \pm 0.2$ & $3.8 \pm 0.2$ & $4.8 \pm 0.4$ & 8.70 \\
HD 79498 & 0.2 & $0.4 \pm 0.1$ & $2.4 \pm 0.2$ & $1.8 \pm 0.1$ & $1.8 \pm 0.3$ & 3.50 \\
HIP 91258 & 0.1 & $1.0 \pm 0.3$ & $3.7 \pm 0.2$ & $2.7 \pm 0.2$ & $4.1 \pm 0.4$ & 7.97 \\
HD 90156 & 0.6 & $0.5 \pm 0.1$ & $5.7 \pm 0.2$ & $3.3 \pm 0.2$ & $3.1 \pm 0.3$ & 3.75 \\
HD 115617 & 3.3 & $4 \pm 2$ & $28 \pm 1$ & $7 \pm 3$ & $12 \pm 2$ & 32.5 \\
HD 70642 & 0.4 & $1.4 \pm 0.1$ & $8.5 \pm 0.2$ & $5.5 \pm 0.2$ & $5.5 \pm 0.3$ & 11.5 \\
HD 47186 & 0.2 & $0.8 \pm 0.1$ & $3.9 \pm 0.2$ & $2.2 \pm 0.2$ & $3.1 \pm 0.3$ & 6.54 \\
HD 92788 & 0.3 & $1.1 \pm 0.1$ & $7.4 \pm 0.3$ & $4.3 \pm 0.2$ & $4.1 \pm 0.3$ & 8.53 \\
HD 102117 & 0.3 & $0.6 \pm 0.1$ & $3.8 \pm 0.2$ & $2.6 \pm 0.2$ & $2.5 \pm 0.3$ & 5.21 \\
HD 4208 & 0.2 & $0.3 \pm 0.1$ & $2.6 \pm 0.2$ & $1.3 \pm 0.1$ & $1.5 \pm 0.3$ & 2.32 \\
HD 10700 & 11.1 & $11.9 \pm 0.7$ & $66.0 \pm 0.8$ & $30 \pm 2$ & $16.9 \pm 0.7$ & 96.7 \\
HD 69830 & 1.2 & $1.6 \pm 0.1$ & $14.8 \pm 0.3$ & $7.1 \pm 0.2$ & $8.0 \pm 0.4$ & 13.1 \\
55 Cnc & 1.2 & $3.04 \pm 0.4$ & $24.7 \pm 0.1$ & $15.9 \pm 0.1$ & \nodata & 24.7 \\
HD 1237 & 0.7 & $20.8 \pm 0.7$ & \nodata & \nodata & $0.2 \pm 0.8$ & 169 \\
HD 154345 & 0.7 & $1.9 \pm 0.1$ & $14.8 \pm 0.3$ & $8.2 \pm 0.2$ & $7.3 \pm 0.4$ & 15.8 \\
GJ 86 & 1.5 & $4.4 \pm 0.2$ & $36.2 \pm 0.5$ & $15.2 \pm 0.6$ & $15.4 \pm 0.5$ & 35.4 \\
HD 147018 & 0.1 & $0.6 \pm 0.1$ & $4.2 \pm 0.2$ & $2.4 \pm 0.2$ & $3.8 \pm 0.4$ & 4.84 \\
HD 164922 & 0.5 & $0.6 \pm 0.1$ & $6.4 \pm 0.3$ & $3.7 \pm 0.2$ & $3.9 \pm 0.3$ & 4.84 \\
HD 189733 & 0.3 & $5.8 \pm 0.1$ & $31.6 \pm 0.2$ & $11.3 \pm 0.1$ & $12.4 \pm 0.2$ & 47.1 \\
HD 7924 & 0.4 & $1.5 \pm 0.1$ & $10.1 \pm 0.3$ & $4.0 \pm 0.2$ & $5.7 \pm 0.3$ & 12.1 \\
HD 3651 & 1.4 & $3.8 \pm 0.2$ & $33.6 \pm 0.6$ & $13.3 \pm 0.3$ & $16.5 \pm 0.5$ & 31.1 \\
HD 128621 & 110.8 & $408 \pm 2$ & $2572 \pm 3$ & $1268 \pm 4$ & $1026 \pm 2$ & 3320 \\
HD 114783 & 0.3 & $0.9 \pm 0.1$ & $7.7 \pm 0.3$ & $2.8 \pm 0.2$ & $3.3 \pm 0.3$ & 7.62 \\
HD 97658 & 0.3 & $0.37 \pm 0.04$ & $2.45 \pm 0.07$ & $1.76 \pm 0.06$ & $1.42 \pm 0.07$ & 3.00 \\
HD 40307 & 0.6 & $0.37 \pm 0.04$ & $2.76 \pm 0.07$ & $1.66 \pm 0.06$ & $1.47 \pm 0.08$ & 3.02 \\
HD 192263 & 0.3 & $4.7 \pm 0.2$ & $25.7 \pm 0.5$ & $8.7 \pm 0.2$ & $10.6 \pm 0.4$ & 38.4 \\
$\epsilon$ Eri & 12.6 & $104 \pm 1$ & $451 \pm 7$ & $372 \pm 2$ & $335 \pm 6$ & 1200\tablenotemark{b} \\
HD 192310 & 1.7 & $4.9 \pm 0.2$ & $41.0 \pm 0.7$ & $14.5 \pm 0.3$ & $14.6 \pm 0.5$ & 39.4 \\
HD 99492 & 0.4 & $1.8 \pm 0.2$ & $10.4 \pm 0.3$ & $3.6 \pm 0.2$ & $5.3 \pm 0.4$ & 15.0 \\
HD 128311 & 0.3 & $8.5 \pm 0.3$ & $41.1 \pm 0.6$ & $14.0 \pm 0.3$ & $18.0 \pm 0.5$ & 68.8 \\
HD 104067 & 0.3 & $3.0 \pm 0.2$ & $14.9 \pm 0.4$ & $5.1 \pm 0.2$ & $7.4 \pm 0.4$ & 24.5 \\
HD 156668 & 0.1 & $0.4 \pm 0.1$ & $2.7 \pm 0.2$ & $1.1 \pm 0.1$ & $1.3 \pm 0.3$ & 3.00 \\
HAT P 11 & 0.06 & \nodata & $4.66 \pm 0.09$ & $1.17 \pm 0.04$ & \nodata & \nodata \\
WASP 69 & 0.04 & $0.57 \pm 0.02$ & $2.59 \pm 0.03$ & $1.26 \pm 0.2$ & \nodata & 4.64 \\
HD 85512 & 0.5 & $0.69 \pm 0.05$ & $3.96 \pm 0.08$ & $1.72 \pm 0.06$ & $1.82 \pm 0.09$ & 5.59 \\
GJ 832 & 1.0 & $3.51 \pm 0.08$ & $3.78 \pm 0.08$ & $2.55 \pm 0.06$ & $3.33 \pm 0.09$ & 28.5 \\
GJ 667 C & 0.3 & $0.72 \pm 0.05$ & $0.65 \pm 0.05$ & $0.51 \pm 0.04$ & $0.83 \pm 0.07$ & 5.82 \\
GJ 3470 & 0.02 & $3.0 \pm 0.5$ & \nodata & $3.0 \pm 0.9$ & \nodata & 24.4 \\
GJ 176 & 0.09 & $3.10 \pm 0.08$ & $5.4 \pm 0.1$ & $2.15 \pm 0.06$ & $2.30 \pm 0.09$ & 25.2 \\
GJ 436 & 0.08 & $0.96 \pm 0.05$ & $1.09 \pm 0.06$ & $0.52 \pm 0.04$ & $0.68 \pm 0.07$ & 7.77 \\
GJ 1214 & 0.008 & $0.18 \pm 0.04$ & $0.09 \pm 0.03$ & $0.08 \pm 0.03$ & $0.05 \pm 0.04$ & 1.50 \\
GJ 876 & 0.2 & $10.7 \pm 0.1$ & $10.6 \pm 0.1$ & $8.1 \pm 0.1$ & $8.4 \pm 0.1$ & 87.0 \\
GJ 581 & 0.2 & $0.53 \pm 0.04$ & $0.48 \pm 0.04$ & $0.29 \pm 0.04$ & $0.44 \pm 0.07$ & 4.34 \\
Proxima Centauri & 0.29 & $38.7 \pm 0.6$ & $36.1 \pm 0.4$ & $12.6 \pm 0.9$ & $22.2 \pm 0.5$ & 150\tablenotemark{b} 
\enddata
\tablenotetext{a}{$F_{bol}= \sigma T_{eff}^4 \left( R_{\ast} / d \right)^2$}
\tablenotetext{b}{Direct measurement of the 90~--~360~\AA\ flux from $EUVE$, corrected for interstellar hydrogen and helium attenuation.}
\end{deluxetable}

\clearpage

\clearpage

\begin{deluxetable}{ccccccc}
\tablecaption{Non-Planet Host Flux Measurements \label{nonplanethostfluxes}
}
\tablewidth{0 pt}
\tabletypesize{\scriptsize}
\tablehead{
\colhead{Name} & \colhead{$F_{bol}$\tablenotemark{a}} & \colhead{N V} & \colhead{C II} & \colhead{Si III} & \colhead{Si IV} & \colhead{$F(90-360\AA)$} \\
\colhead{} & \colhead{[$10^{-7}$} & \colhead{[$10^{-15}$} & \colhead{[$10^{-15}$} & \colhead{[$10^{-15}$} & \colhead{[$10^{-15}$} & \colhead{[$10^{-14}$} \\
\colhead{} & \colhead{erg s$^{-1}$ cm$^{-2}$]} & \colhead{erg s$^{-1}$ cm$^{-2}$ \AA$^{-1}$]} & \colhead{erg s$^{-1}$ cm$^{-2}$ \AA$^{-1}$]} & \colhead{erg s$^{-1}$ cm$^{-2}$ \AA$^{-1}$]} & \colhead{erg s$^{-1}$ cm$^{-2}$ \AA$^{-1}$]} & \colhead{erg s$^{-1}$ cm$^{-2}$ ]} \\
}
\startdata
HD 28568 & 0.6 & $23 \pm 5$ & $120 \pm 4$ & $82 \pm 8$ & $65 \pm 5$ & 189 \\
HD 28033 & 0.3 & \nodata & $4.8 \pm 0.9$ & $1 \pm 3$ & $7 \pm 2$ & \nodata \\
HD 33262 & 2.8 & $47 \pm 2$ & $287 \pm 2$ & $319 \pm 4$ & $307 \pm 3$ & 380 \\
HD 106516 & 1.2 & $4 \pm 3$ & $22 \pm 2$ & $4 \pm 5$ & $1 \pm 4$ & 32.6 \\
HD 28205 & 0.3 & $8 \pm 2$ & $27 \pm 2$ & $22 \pm 4$ & $17 \pm 3$ & 67.8 \\
HD 25825 & 0.2 & $10.6 \pm 0.5$ & $36.7 \pm 0.7$ & $40.9 \pm 0.8$ & $47 \pm 1$ & 86.2 \\
HD 97334 & 0.7 & $15 \pm 2$ & $86 \pm 2$ & $48 \pm 4$ & $82 \pm 2$ & 121 \\
HD 39587 & 4.3 & \nodata & $454 \pm 2$ & \nodata & $385 \pm 2$ & 680\tablenotemark{b}  \\
HII 314 & 0.02 & $2.24 \pm 0.08$ & $6.8 \pm 0.1$ & $3.70 \pm 0.9$ & $4.8 \pm 0.1$ & 18.2 \\
16 Cyg A & 1.1 & \nodata & $17.7 \pm 0.2$ & $10.8 \pm 0.2$ & $8.3 \pm 0.2$ & \nodata \\
HD 72905 & 1.5 & \nodata & $184 \pm 1$ & \nodata & $136 \pm 1$ & \nodata \\
HD 129333 & 0.3 & \nodata & $124 \pm 1$ & \nodata & $106 \pm 1$ & \nodata \\
HD 199288 & 0.6 & $0.31 \pm 0.06$ & $3.77 \pm 0.07$ & $1.79 \pm 0.08$ & $1.48 \pm 0.09$ & 2.53 \\
$\alpha$ Cen A & 271 & $356 \pm 5$ & $2690 \pm 8$ & $1590 \pm 12$ & $1220 \pm 7$ & 2900\tablenotemark{b}  \\
HD 59967 & 0.6 & $18 \pm 2$ & $60 \pm 2$ & $71 \pm 3$ & $47 \pm 2$ & 145 \\
HD 20630 & 3.3 & $39 \pm 1$ & $234 \pm 2$ & $216 \pm 4$ & $209 \pm 2$ & 40\tablenotemark{b}  \\
HD 43162 & 0.7 & $3 \pm 2$ & $73 \pm 2$ & $48 \pm 5$ & $61 \pm 3$ & 22.8 \\
HD 131156 & 3.9 & $70 \pm 4$ & $451 \pm 6$ & $281 \pm 12$ & $317 \pm 5$ & 830\tablenotemark{b}  \\
KIC 11560431 & 0.0002 & $4.6 \pm 1$ & $16.4 \pm 0.2$ & $5.7 \pm 0.1$ & $6.9 \pm 0.1$ & 37.0 \\
HD 166 & 1.1 & $43 \pm 2$ & $126 \pm 3$ & $96 \pm 5$ & $94 \pm 3$ & 350 \\
HD 165341 & 6.9 & $71 \pm 2$ & $437 \pm 2$ & $249 \pm 5$ & $206 \pm 2$ & 700\tablenotemark{b}  \\
HD 103095 & 1.9 & $0.1 \pm 0.1$ & $1.1 \pm 0.1$ & $0.56 \pm 0.09$ & $1.2 \pm 0.1$ & 1.11 \\
HR 1925 & 1.3 & $21 \pm 2$ & $85 \pm 2$ & $63 \pm 4$ & $56 \pm 2$ & 171 \\
HD 22468 & 2.6 & $600 \pm 3$ & $2550 \pm 5$ & $1280 \pm 12$ & $1200 \pm 4$ & 4880 \\
HD 155886 & 2.6 & $19 \pm 1$ & $131 \pm 1$ & $66 \pm 2$ & \nodata & 151 \\
LTT 2050 & 0.06 & $1.7 \pm 0.1$ & $2.0 \pm 0.2$ & $0.58 \pm 0.09$ & $1.4 \pm 0.2$ & 13.6 \\
HD 197481 & 0.2 & $70 \pm 1$ & $206 \pm 2$ & $788 \pm 2$ & $864 \pm 1$ & 720\tablenotemark{b}  \\
Kapteyn's Star & 0.3 & $0.2 \pm 0.1$ & $0.9 \pm 0.1$ & $0.8 \pm 0.1$ & $0.9 \pm 0.3$ & 1.98 \\
LP 415-1619 & 0.006 & \nodata & $2.97 \pm 0.07$ & $0.77 \pm 0.04$ & \nodata & \nodata \\
AD Leo & 0.2 & $136 \pm 1$ & $219 \pm 1$ & $142 \pm 1$ & $159 \pm 1$ & 1100\tablenotemark{b}  \\
LHS-26 & 0.08 & \nodata & \nodata & \nodata & \nodata & \nodata \\
Procyon & 170 & $1650 \pm 5$ & $8720 \pm 9$ & $4940 \pm 10$ & $4160 \pm 20$ & 3500\tablenotemark{b}  \\
EV Lac & 0.2 & $41 \pm 1$ & $51 \pm 10$ & $35 \pm 2$ & $39 \pm 1$ & 450\tablenotemark{b}  \\
YY Gem & 0.1 & $41.7 \pm 0.7$ & $223 \pm 1$ & $34 \pm 1$ & $82 \pm 1$ & 339 
\enddata
\tablenotetext{a}{$F_{bol}= \sigma T_{eff}^4 \left( R_{\ast} / d \right)^2$}
\tablenotetext{b}{Direct measurement of the 90~--~360~\AA\ flux from $EUVE$, corrected for interstellar \ion{H}{1} attenuation.}
\end{deluxetable}

\clearpage

\clearpage
\begin{deluxetable}{cccccc}
\tablecaption{Correlations between SPI parameters and $F_{ion}$/$F_{bolom}$ \label{stellarparam_correlations}
}
\tablewidth{0 pt}
\tabletypesize{\scriptsize}
\tablehead{
\colhead{ion} & \colhead{SPI parameter}& \colhead{$\rho$} & \colhead{$p$-value} &  \colhead{$\Delta$BIC} & \colhead{Improved BIC}  \\
}
\startdata
\ion{N}{5}   &     $M_{plan}$/$a_{plan}$   &      0.338   &    0.004   &   0.0172   &  with SPI \\
\ion{Si}{3}   &     $M_{plan}$/$a_{plan}$   &      0.382   &    0.001   &  \nodata   &  \nodata \\
\ion{Si}{4}   &     $M_{plan}$/$a_{plan}$   &      0.283   &    0.022   &  \nodata   &  \nodata \\
\ion{C}{2}   &     $M_{plan}$/$a_{plan}$   &      0.311   &    0.011   &  \nodata   &  \nodata \\
\tableline
\ion{N}{5}   &     $M_{plan}^{2/3}$ ($a_{plan}$/$R_{\star}$)$^{-4}$ $a_{plan}^{-1/2}$   &      0.300   &    0.017   &   4.089\tablenotemark{a}   &  without SPI \\
\ion{Si}{3}   &     $M_{plan}^{2/3}$ ($a_{plan}$/$R_{\star}$)$^{-4}$ $a_{plan}^{-1/2}$   &      0.108   &    0.392   &  \nodata   &  \nodata \\
\ion{Si}{4}   &     $M_{plan}^{2/3}$ ($a_{plan}$/$R_{\star}$)$^{-4}$ $a_{plan}^{-1/2}$   &      0.047   &    0.725   &  \nodata   &  \nodata \\
\ion{C}{2}   &     $M_{plan}^{2/3}$ ($a_{plan}$/$R_{\star}$)$^{-4}$ $a_{plan}^{-1/2}$   &      0.087   &    0.503   &  \nodata   &  \nodata \\
\tableline
\ion{N}{5}   &     $M_{plan}$$^{2}$ $a_{plan}^{-1/2}$   &      0.053   &    0.679   &   4.056\tablenotemark{a}   &  without SPI \\
\ion{Si}{3}   &     $M_{plan}$$^{2}$ $a_{plan}^{-1/2}$   &      0.264   &    0.034   &  \nodata   &  \nodata \\
\ion{Si}{4}   &     $M_{plan}$$^{2}$ $a_{plan}^{-1/2}$   &      0.277   &    0.034   &  \nodata   &  \nodata \\
\ion{C}{2}   &     $M_{plan}$$^{2}$ $a_{plan}^{-1/2}$   &      0.210   &    0.101   &  \nodata   &  \nodata \\
\tableline
\ion{N}{5}   &     ($M_{plan}$/$M_{\star}$)$^{2}$ ($a_{plan}$/$R_{\star}$)$^{-6}$   &      0.319   &    0.011   &   4.081\tablenotemark{a}   &  without SPI \\
\ion{Si}{3}   &     ($M_{plan}$/$M_{\star}$)$^{2}$ ($a_{plan}$/$R_{\star}$)$^{-6}$   &      0.146   &    0.246   &  \nodata   &  \nodata \\
\ion{Si}{4}   &     ($M_{plan}$/$M_{\star}$)$^{2}$ ($a_{plan}$/$R_{\star}$)$^{-6}$   &      0.089   &    0.505   &  \nodata   &  \nodata \\
\ion{C}{2}   &     ($M_{plan}$/$M_{\star}$)$^{2}$ ($a_{plan}$/$R_{\star}$)$^{-6}$   &      0.115   &    0.376   &  \nodata   &  \nodata \\
\enddata
\tablecomments{$\rho$ is the Spearman rank coefficient, where $| \rho | = 1$ for a perfect correlation and $\rho = 0$ when the data are uncorrelated. The $p$-value indicates the likelihood of obtaining $|\rho|$ closer to 1, under the assumption that there is no correlation between the two parameters (i.e. the data are randomly distributed). }
\tablenotetext{a}{The larger $\Delta$BIC values in the latter three SPI parameters is the result of 4 PCs being required to fit the $F_{ion}$/$F_{bolom}$ distribution as opposed to 3 PCs for the SPI parameter set to $M_{plan}$/$a_{plan}$ (see Section 4.2 and Appendix C).}
\end{deluxetable}
\clearpage

\clearpage

\begin{deluxetable}{ccccccccccccc}
\tablecaption{Correlations between Stellar Parameters: Planet-Hosts \label{stellarparam_correlations}
}
\tablewidth{0 pt}
\tabletypesize{\scriptsize}
\tablehead{
\colhead{} & \rlap{\kern\tabcolsep$P_{rot}$} & \colhead{} & \rlap{\kern\tabcolsep$T_{eff}$} & \colhead{} & \rlap{\kern\tabcolsep$d$} & \colhead{} & \rlap{\kern\tabcolsep$V$} & \colhead{} & \rlap{\kern\tabcolsep$B-V$} & \colhead{} & \rlap{\kern\tabcolsep$\log \left( SPI \right)$} & \colhead{} \\
\colhead{} & \colhead{$\rho$} & \colhead{$p$-value} & \colhead{$\rho$} & \colhead{$p$-value} & \colhead{$\rho$} & \colhead{$p$-value} & \colhead{$\rho$} & \colhead{$p$-value} & \colhead{$\rho$} & \colhead{$p$-value} & \colhead{$\rho$} & \colhead{$p$-value} \\
}
\startdata
$P_{rot}$ & \nodata & \nodata & -0.612 & $1 \times 10^{-8}$ & -0.338 & $1 \times 10^{-3}$ & 0.298 & $1 \times 10^{-2}$ & 0.648 & $1 \times 10^{-9}$ & -0.371 & $1 \times 10^{-3}$ \\
$T_{eff}$ & -0.612 & $1 \times 10^{-8}$ & \nodata & \nodata & 0.506 & $6 \times 10^{-6}$ & -0.475 & $2 \times 10^{-5}$ & -0.931 & $3 \times 10^{-32}$ & 0.174 & 0.1 \\
$d$ & -0.338 & $1 \times 10^{-3}$ & 0.506 & $6 \times 10^{-6}$ & \nodata & \nodata & 0.305 & $9 \times 10^{-3}$ & -0.473 & $3 \times 10^{-5}$ & 0.351 & $3 \times 10^{-3}$ \\
$V$ & 0.298 & 0.01 & -0.475 & $2 \times 10^{-5}$ & 0.305 & $9 \times 10^{-3}$ & \nodata & \nodata & 0.481 & $2 \times 10^{-5}$ & 0.191 & 0.1 \\
$B-V$ & 0.648 & $1 \times 10^{-9}$ & -0.931 & $3 \times 10^{-32}$ & -0.473 & $3 \times 10^{-5}$ & 0.481 & $2 \times 10^{-5}$ & \nodata & \nodata & -0.194 & 0.1 \\
\tableline
$\log \left( SPI \right)$ & \bf{-0.371} & $1 \times 10^{-3}$ & 0.174 & 0.1 & \bf{0.351} & $3 \times 10^{-3}$ & 0.191 & 0.1 & -0.194 & 0.1 & \nodata & \nodata
\enddata
\tablecomments{$\rho$ is the Spearman rank coefficient, where $| \rho | = 1$ for a perfect correlation and $\rho = 0$ when the data are uncorrelated. The $p$-value indicates the likelihood of obtaining $|\rho|$ closer to 1, under the assumption that there is no correlation between the two parameters (i.e. the data are randomly distributed). Here we define ``significant" correlations here as having $| \rho | > 0.3$ (marked in bold for the correlations with SPI), see Section 4.2.}
\end{deluxetable}

\clearpage

\section{\ion{Si}{3}, \ion{N}{5}, \ion{C}{2}, and \ion{Si}{4} Activity Levels vs. $B$~--~$V$,  Distance, and Rotation Period }

In Figures B.1~--~B.3 we show the correlation plots between the UV activity indices and $B$~--~$V$, $d$, and $P_{rot}$.   

\begin{figure}[!h]
\figurenum{B.1}
\begin{center}
\vspace{0.0in}
\begin{tabular}{cc}
\includegraphics[width=0.45\textwidth,angle=0,trim={.0in 0.0in 0.0in 0.0in},clip]{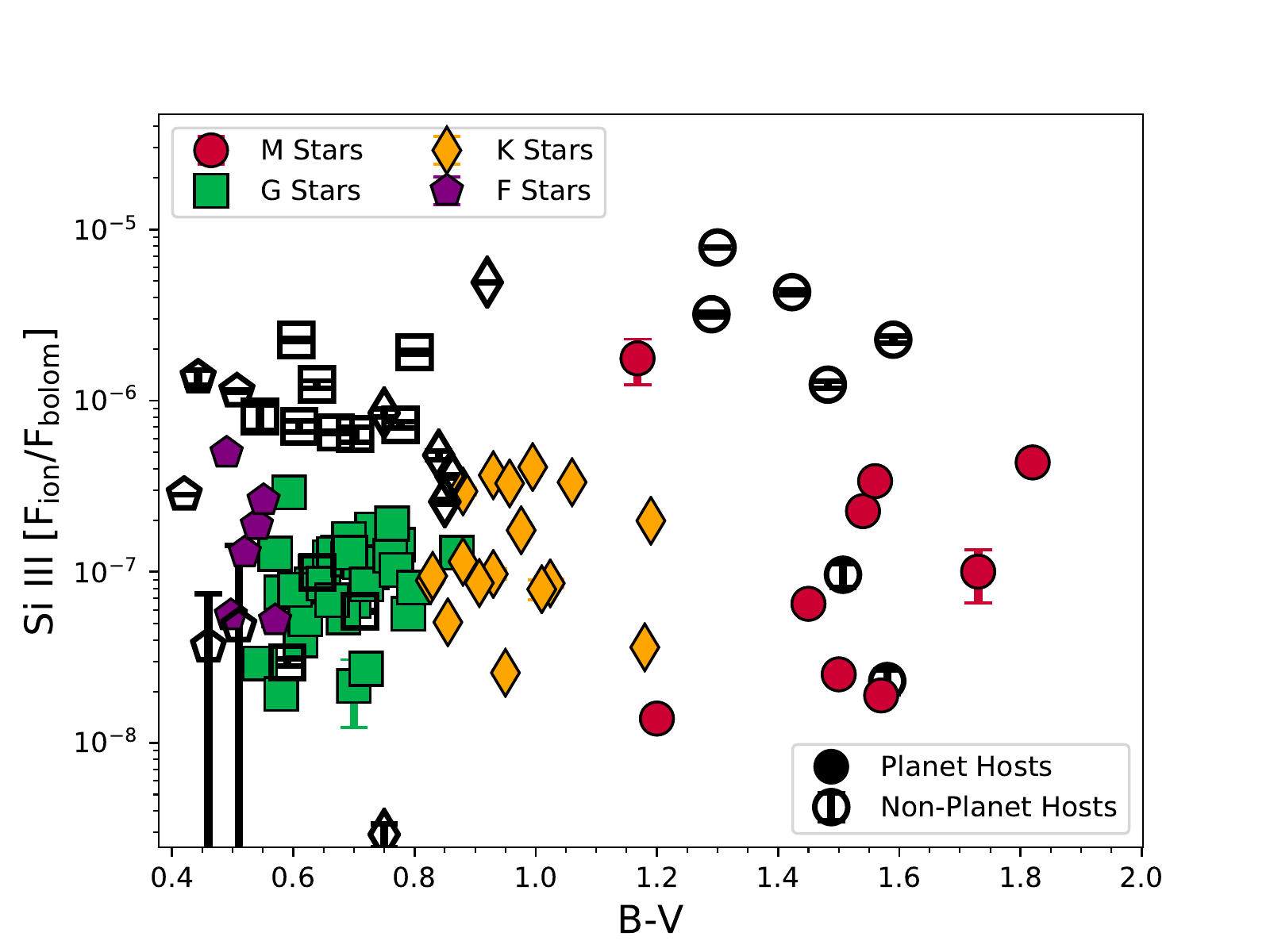} & 
\includegraphics[width=0.45\textwidth,angle=0,trim={.0in 0.0in 0.0in 0.0in},clip]{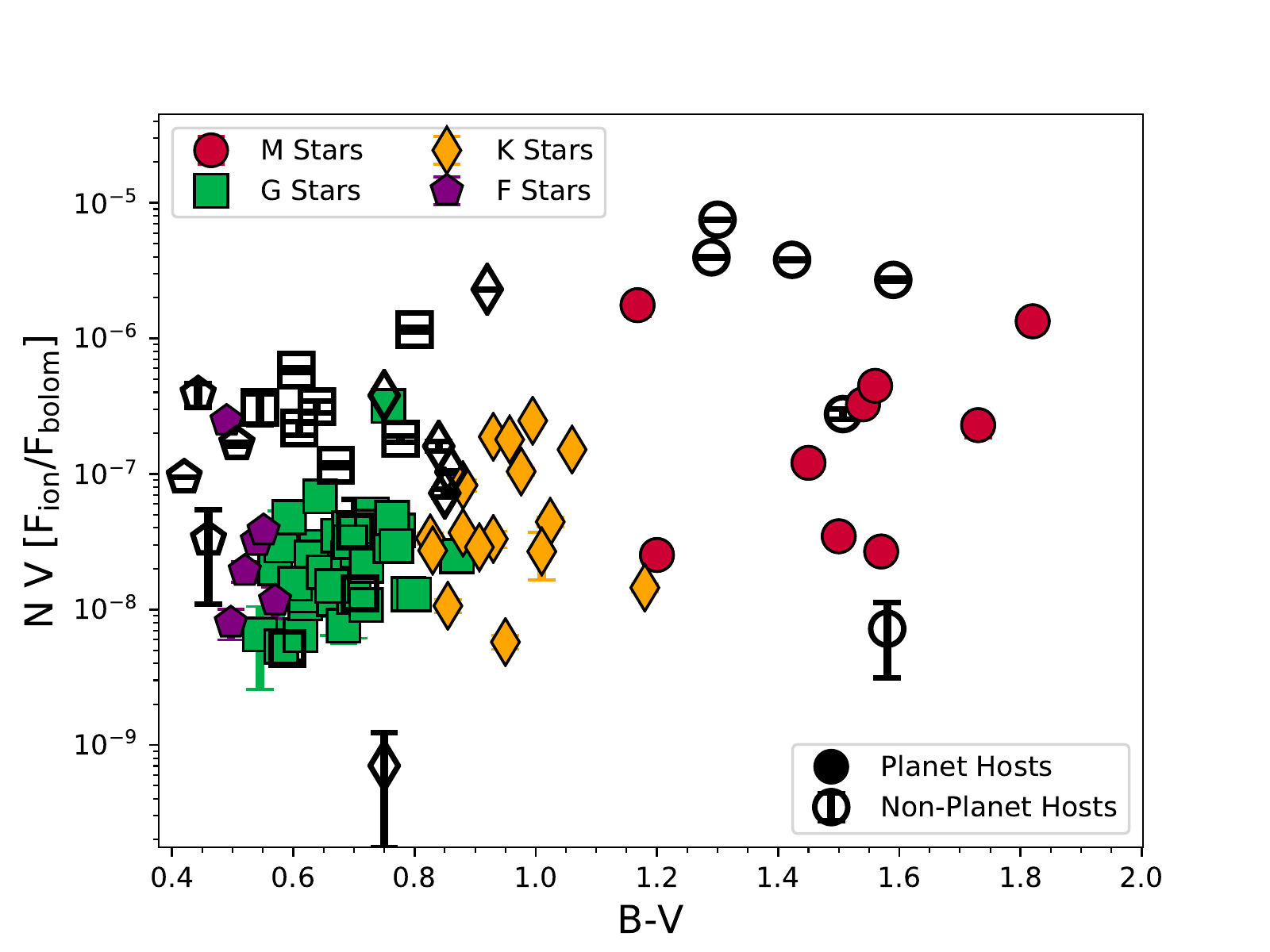} \\
\includegraphics[width=0.45\textwidth,angle=0,trim={.0in 0.0in 0.0in 0.0in},clip]{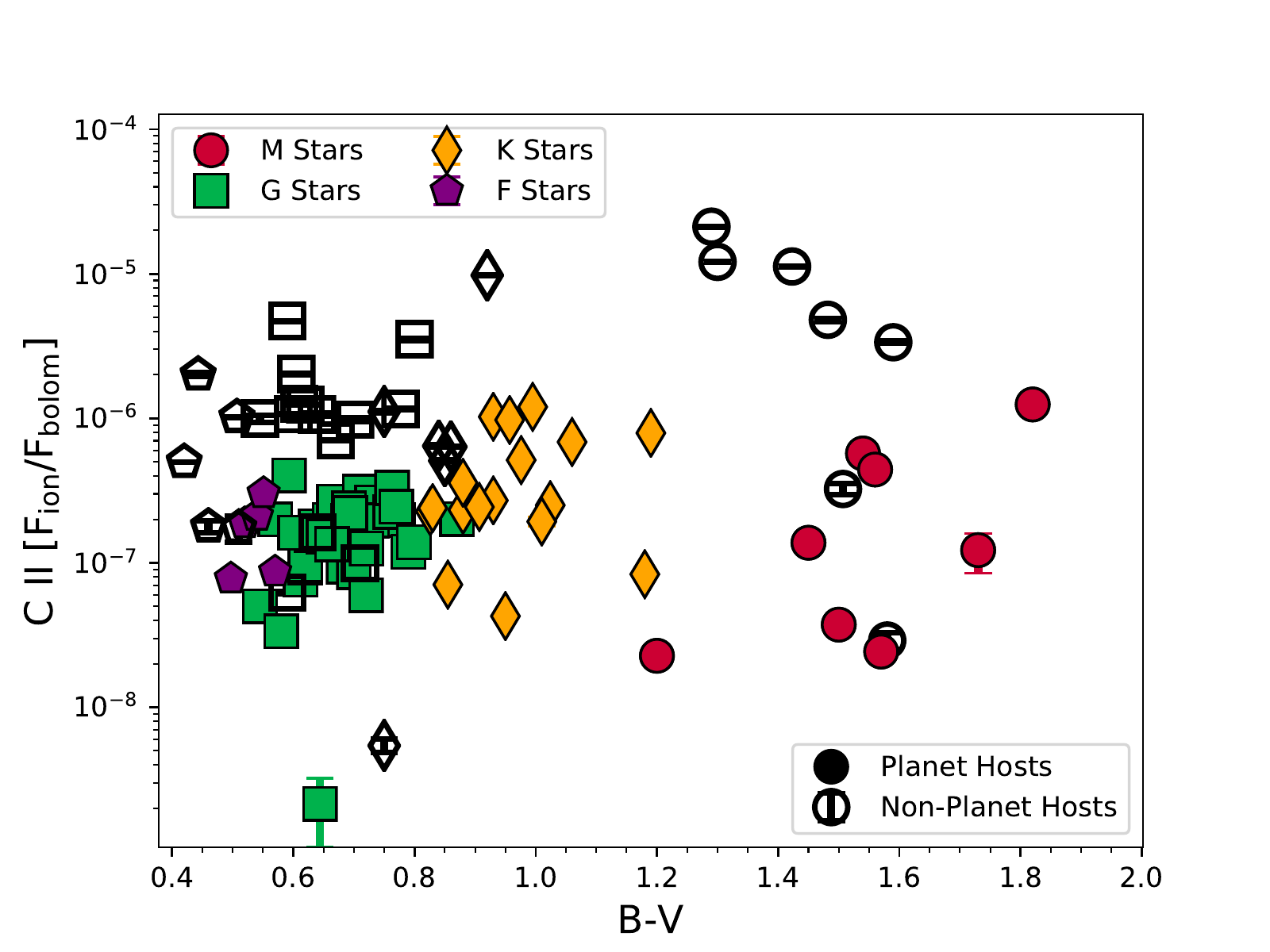} & 
\includegraphics[width=0.45\textwidth,angle=0,trim={.0in 0.0in 0.0in 0.0in},clip]{SiIV_fraclumvsB-V.pdf}
\end{tabular}
\end{center}
\vspace{-0.2in}
\caption{UV activity levels as a function of the spectral slope ($B$~--~$V$) for (top left to lower right) \ion{Si}{3}~$\lambda$~1206~\AA, \ion{N}{5}~$\lambda$~1240~\AA, \ion{C}{2}~$\lambda$~1335~\AA, and \ion{Si}{4}~$\lambda$~1400~\AA.  Spectral types are given by different symbols (circles: M dwarfs, diamonds: K dwarfs, squares: G dwarfs, pentagons: F dwarfs) as shown in the legend.   }
\label{fig-proveit} 
\vspace{-0.15in}
\end{figure} 

\begin{figure}[!h]
\figurenum{B.2}
\begin{center}
\vspace{0.0in}
\begin{tabular}{cc}
\includegraphics[width=0.45\textwidth,angle=0,trim={.0in 0.0in 0.0in 0.0in},clip]{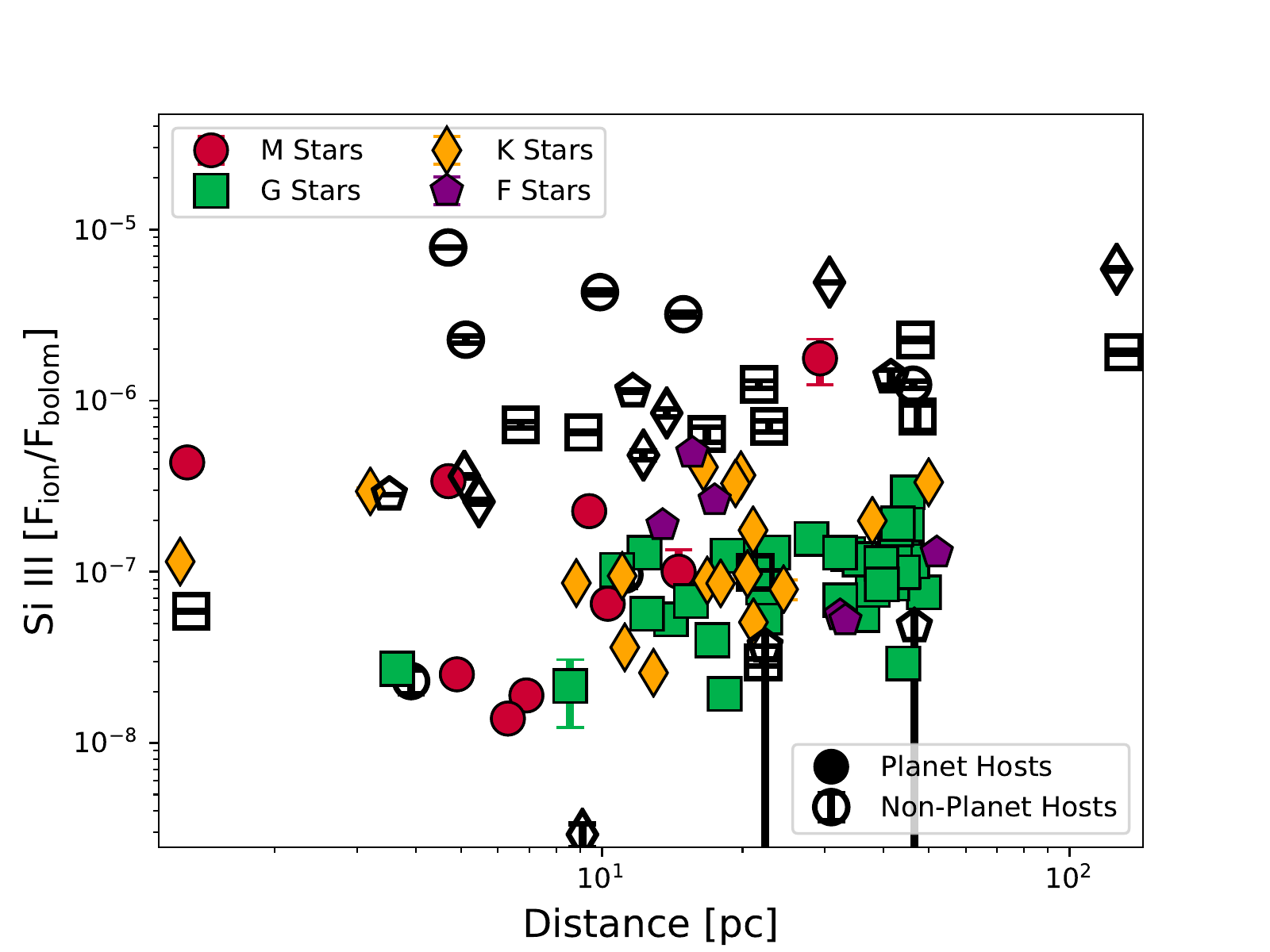} & 
\includegraphics[width=0.45\textwidth,angle=0,trim={.0in 0.0in 0.0in 0.0in},clip]{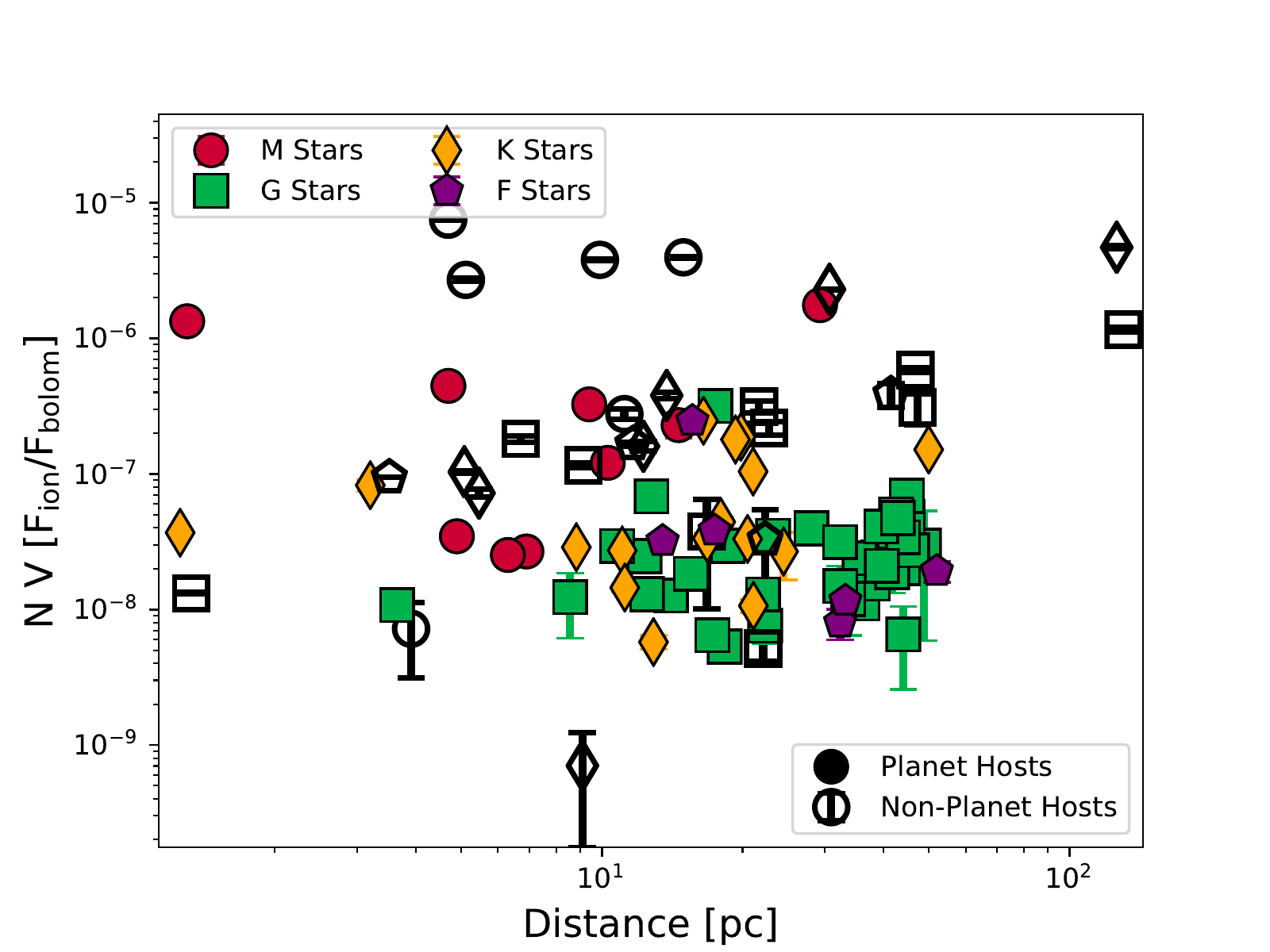} \\
\includegraphics[width=0.45\textwidth,angle=0,trim={.0in 0.0in 0.0in 0.0in},clip]{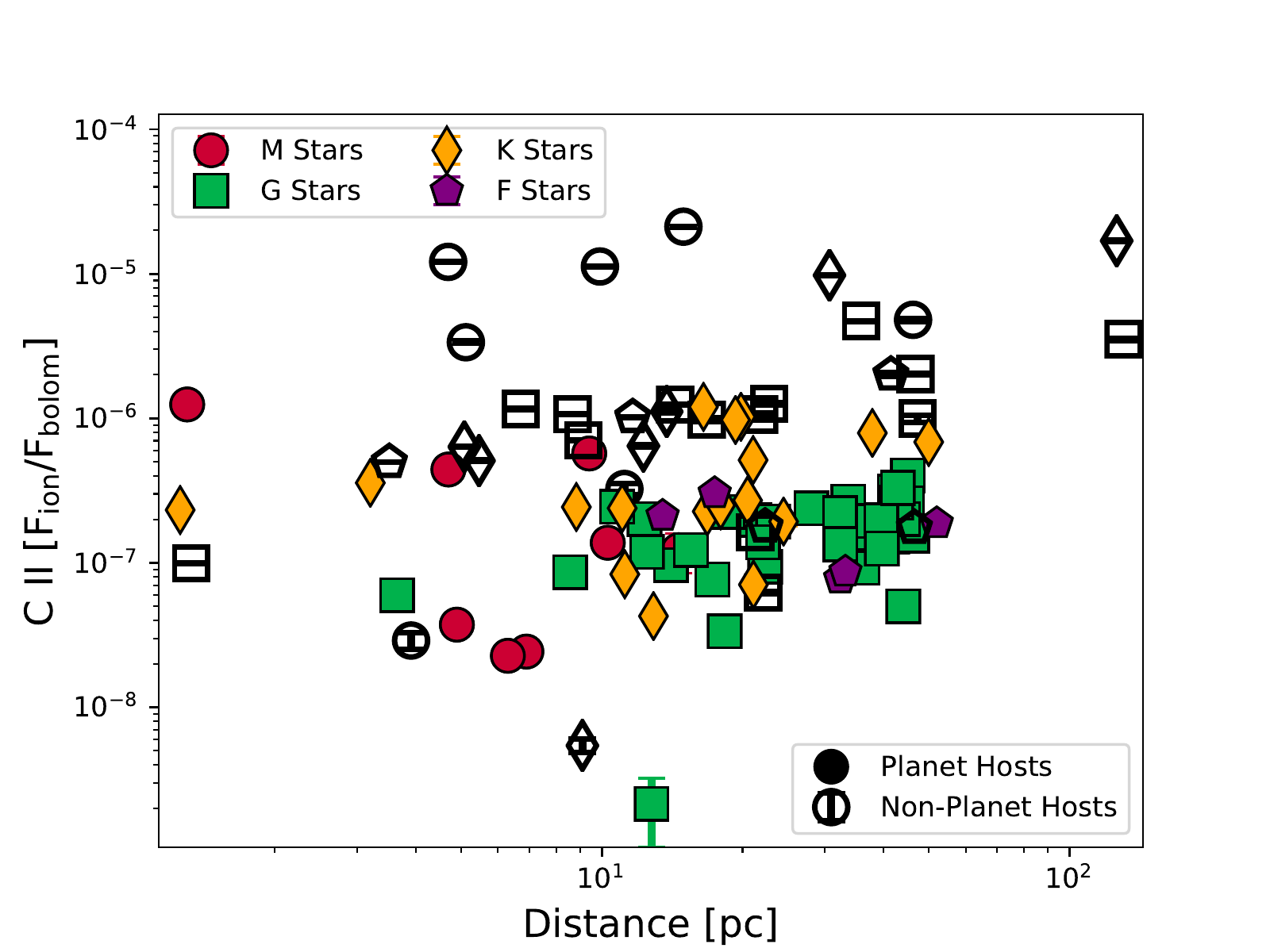} & 
\includegraphics[width=0.45\textwidth,angle=0,trim={.0in 0.0in 0.0in 0.0in},clip]{SiIV_fraclumvsDistance.pdf}
\end{tabular}
\end{center}
\vspace{-0.2in}
\caption{UV activity levels as a function of distance for (top left to lower right) \ion{Si}{3}~$\lambda$~1206~\AA, \ion{N}{5}~$\lambda$~1240~\AA, \ion{C}{2}~$\lambda$~1335~\AA, and \ion{Si}{4}~$\lambda$~1400~\AA.  Spectral types are given by different symbols (circles: M dwarfs, diamonds: K dwarfs, squares: G dwarfs, pentagons: F dwarfs) as shown in the legend.   }
\label{fig-proveit} 
\vspace{-0.15in}
\end{figure}

\begin{figure}[!h]
\figurenum{B.3}
\begin{center}
\vspace{0.0in}
\begin{tabular}{cc}
\includegraphics[width=0.45\textwidth,angle=0,trim={.0in 0.0in 0.0in 0.0in},clip]{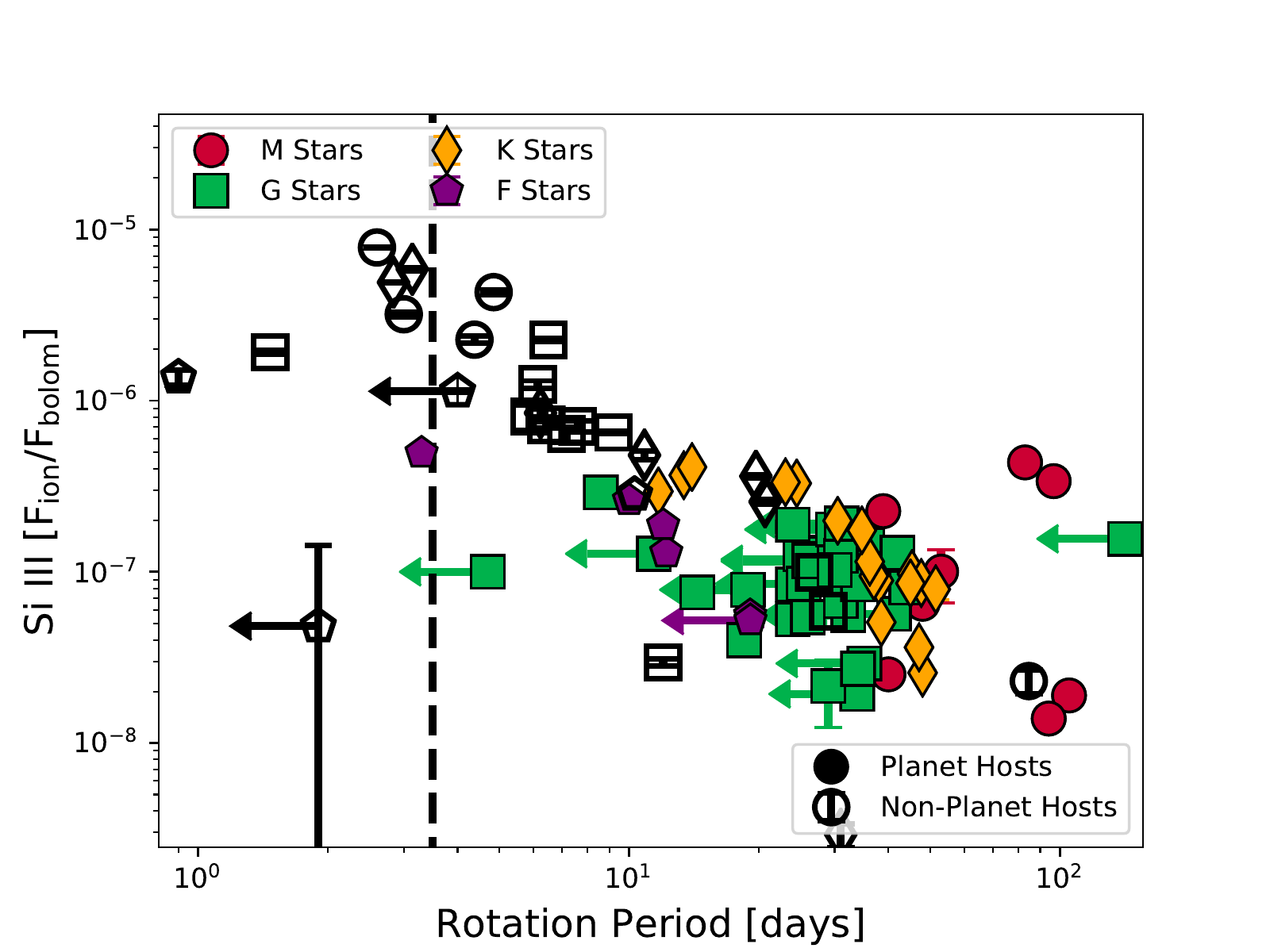} & 
\includegraphics[width=0.45\textwidth,angle=0,trim={.0in 0.0in 0.0in 0.0in},clip]{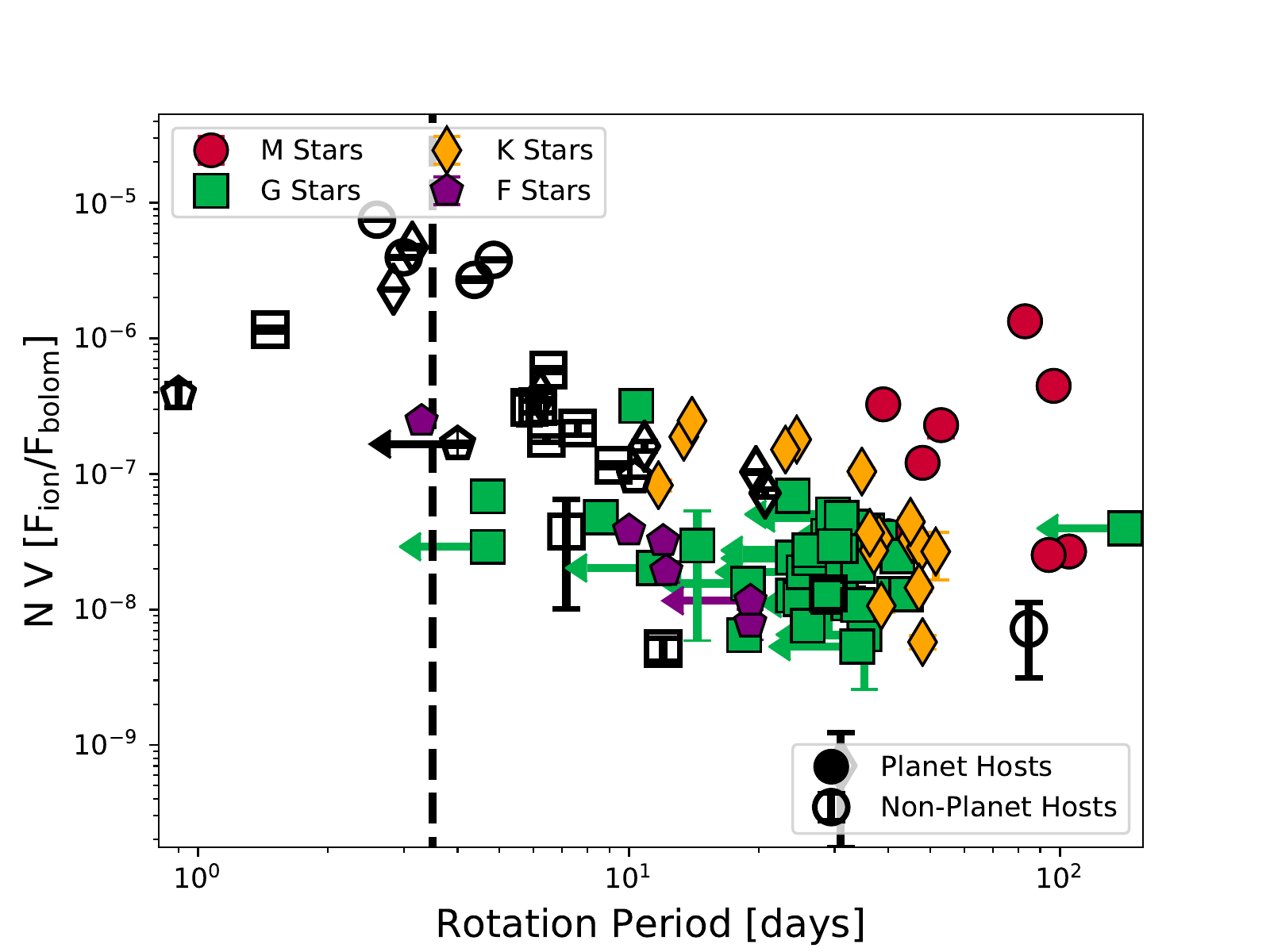} \\
\includegraphics[width=0.45\textwidth,angle=0,trim={.0in 0.0in 0.0in 0.0in},clip]{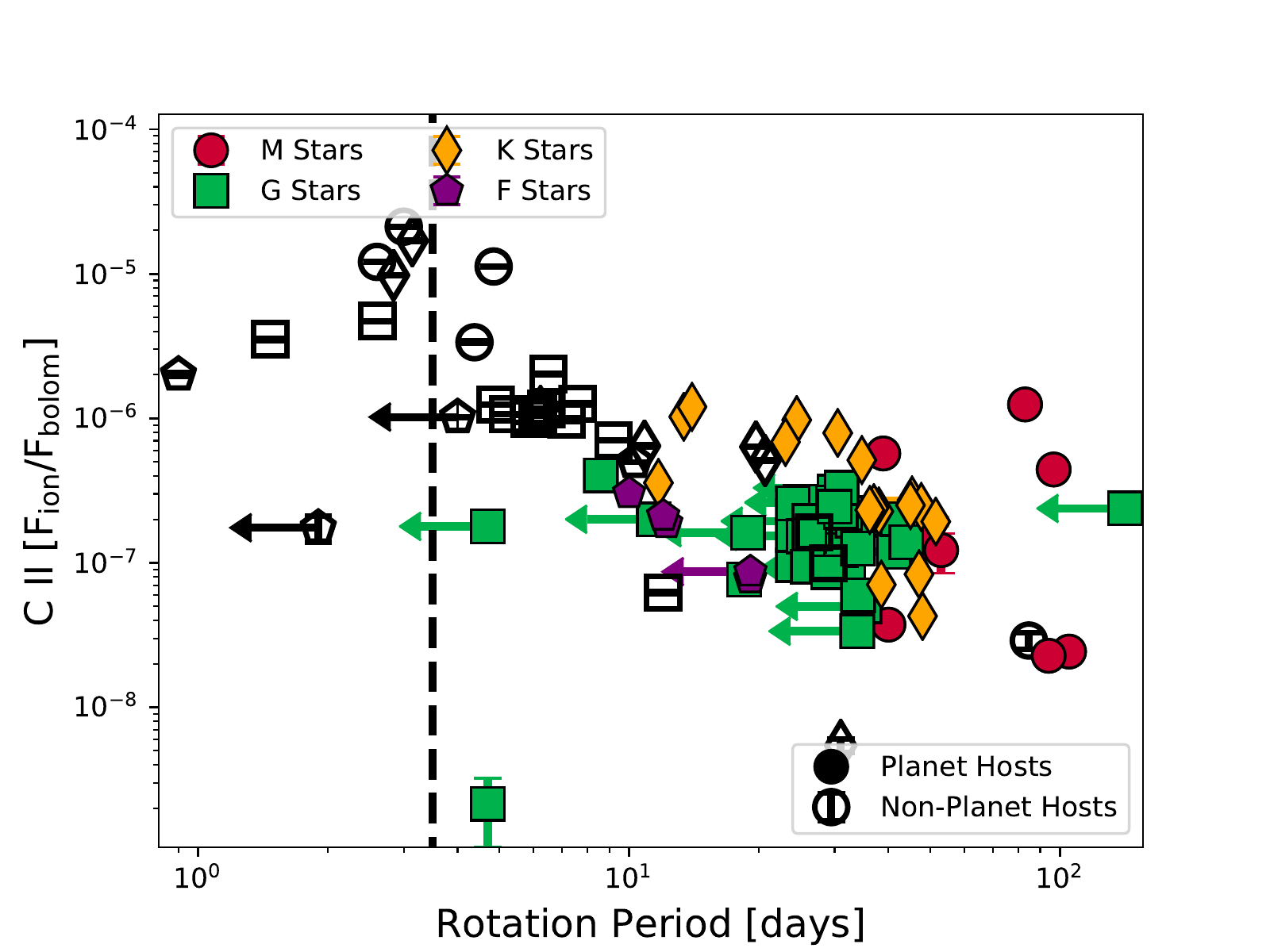} & 
\includegraphics[width=0.45\textwidth,angle=0,trim={.0in 0.0in 0.0in 0.0in},clip]{SiIV_fraclumvsProt.pdf}
\end{tabular}
\end{center}
\vspace{-0.2in}
\caption{UV activity levels as a function of stellar rotation period ($P_{rot}$) for (top left to lower right) \ion{Si}{3}~$\lambda$~1206~\AA, \ion{N}{5}~$\lambda$~1240~\AA, \ion{C}{2}~$\lambda$~1335~\AA, and \ion{Si}{4}~$\lambda$~1400~\AA.  Spectral types are given by different symbols (circles: M dwarfs, diamonds: K dwarfs, squares: G dwarfs, pentagons: F dwarfs) as shown in the legend.   }
\label{fig-proveit} 
\vspace{-0.15in}
\end{figure} 

\clearpage

\section{Principal Component Analysis of Multi-variate SPI Signals}

In this Appendix, we describe the methodology and calculations for the PCA analysis of the star-planet-interaction signal in our UV activity survey  (see the overview in Section  4.2).   
 First, the predictor variables 
 must be centered and scaled as 
\begin{equation}
\tilde{X}_{i, scaled} = \frac{x_i - \bar{x}}{L_x}, \\
L_x = \sqrt{ \sum \limits_{i=1}^{n} \left(x_i - \bar{x} \right)^2 } 
\end{equation}
where each $x_i$ is the predictor 
variable corresponding to an individual stellar system and $\bar{x}$ is the average of the entire sample. The scaling simplifies the problem by allowing us to calculate the correlations between scaled parameters as 
\begin{equation}
Cor \left(X_j, X_k \right) = \sum \limits_{i=1}^{n} \tilde{X}_{ij, scaled} \tilde{X}_{ik, scaled}
\end{equation}
where the indices $j$ and $k$ represent predictor variables (e.g. $j$ for $B-V$ and $k$ for $T_{eff}$). We construct a matrix by calculating correlation coefficients between each set of scaled predictor variables:

\[
\begin{blockarray}{ccccccc}
	& \mathbf{P_{rot}} & \mathbf{V} & \mathbf{B-V} & \mathbf{d} & \mathbf{T_{eff}} & \mathbf{\textbf{log} \left(SPI \right)} \\
	\begin{block}{c[cccccc]}
	\mathbf{P_{rot}} & 1 & 0.39 & 0.56 & -0.34 & -0.58 & -0.26 \\
	\mathbf{V} & 0.39 & 1 & 0.67 & 0.17 & -0.68 & 0.08 \\
	\mathbf{B-V} & 0.56 & 0.67 & 1 & -0.48 & -0.97 & -0.20 \\
	\mathbf{d} & -0.34 & 0.17 & -0.48 & 1 & 0.52 & 0.35 \\
	\mathbf{T_{eff}} & -0.58 & -0.68 & -0.97 & 0.52 & 1 & 0.20 \\
	\mathbf{\textbf{log} \left(SPI \right)} & -0.26 & 0.08 & -0.20 & 0.35 & 0.20 & 1 \\
	\end{block}
\end{blockarray}
\]

As expected, each parameter is perfectly correlated with itself, as evidenced by the coefficients of 1 along the diagonal of the correlation matrix. Next, we calculate the eigenvalues of the correlation matrix and place them in descending order $\left(\lambda = 3.22, 1.41, 0.72, 0.54, 0.085, 0.026 \right)$, where the largest eigenvalue is associated with the principal component that contributes most to the spread of fractional luminosities. The principal components can be constructed from the eigenvectors of the correlation matrix, in order from most to least significant:   
\begin{equation}
\begin{split}
PC_1 &= -0.41 \times P_{rot, scaled} - 0.37 \times V_{scaled} - 0.53 \times \left(B-V \right)_{scaled} \\
&+ 0.30 \times d_{scaled} + 0.54 \times T_{eff, scaled} + 0.18 \times \log \left(SPI \right)_{scaled} \\
PC_2 &= -0.059 \times P_{rot, scaled} + 0.58 \times V_{scaled} + 0.089 \times \left(B-V \right)_{scaled} \\
&+ 0.58 \times d_{scaled} - 0.077 \times T_{eff, scaled} + 0.57 \times \log \left(SPI \right)_{scaled} \\
PC_3 &= -0.26 \times P_{rot, scaled} - 0.24 \times V_{scaled} + 0.14 \times \left(B-V \right)_{scaled} \\
 &- 0.55 \times d_{scaled} - 0.16 \times T_{eff, scaled} + 0.73 \times \log \left(SPI \right)_{scaled} \\
PC_4 &= 0.87 \times P_{rot, scaled} - 0.20 \times V_{scaled} - 0.22 \times \left(B-V \right)_{scaled} \\
 &+ 0.0083 \times d_{scaled} + 0.19 \times T_{eff, scaled} + 0.34 \times \log \left(SPI \right)_{scaled} \\
PC_5 &= 0.017 \times P_{rot, scaled} - 0.64 \times V_{scaled} + 0.56 \times \left(B-V \right)_{scaled} \\
 &+ 0.50 \times d_{scaled} - 0.16 \times T_{eff, scaled} + 0.030 \times \log \left(SPI \right)_{scaled} \\
PC_6 &= 0.012 \times P_{rot, scaled} + 0.18 \times V_{scaled} + 0.57 \times \left(B-V \right)_{scaled} \\
 &- 0.16 \times d_{scaled} + 0.79 \times T_{eff, scaled} + 0.00026 \times \log \left(SPI \right)_{scaled} \\
\end{split}
\end{equation}

Although it is difficult to quantify the contribution of the individual predictor variables to each principal component, we can get a rough sense of which are significant by calculating the Spearman rank coefficients (see Table \ref{PC_correlations}). We find that the most dominant principal component $\left(PC_1 \right)$ is significantly correlated with $P_{rot}$, $B-V$, $d$, and $T_{eff}$. The SPI parameter is only weakly correlated with $PC_1$, but it contributes more to $PC_2$ and $PC_3$, which show smaller correlations with the other stellar parameters. $PC_4$, $PC_5$, and $PC_6$ are not strongly correlated with any of the predictor variables, indicating that they contribute less to the multiple linear regression and can be dropped from the analysis \citep{PeresNeto2005}. The new linear model becomes
\begin{equation}
\log \left(L_{ion} / L_{bol} \right) = \beta_0 + \left[ PC_1 \times \beta_{PC_1} \right] + \left[ PC_2 \times \beta_{PC_2} \right] + \left[ PC_3 \times \beta_{PC_3} \right]
\end{equation}

Table \ref{PC_linreg} lists the coefficients $\left( \beta \right)$ of the multiple linear regression analysis described by Equation C.4.

\begin{figure}[h]
\figurenum{C.1}
\begin{center}
\vspace{0.0in}
\begin{tabular}{cc}
\includegraphics[width=0.45\textwidth,angle=0,trim={.0in 0.0in 0.0in 0.0in},clip]{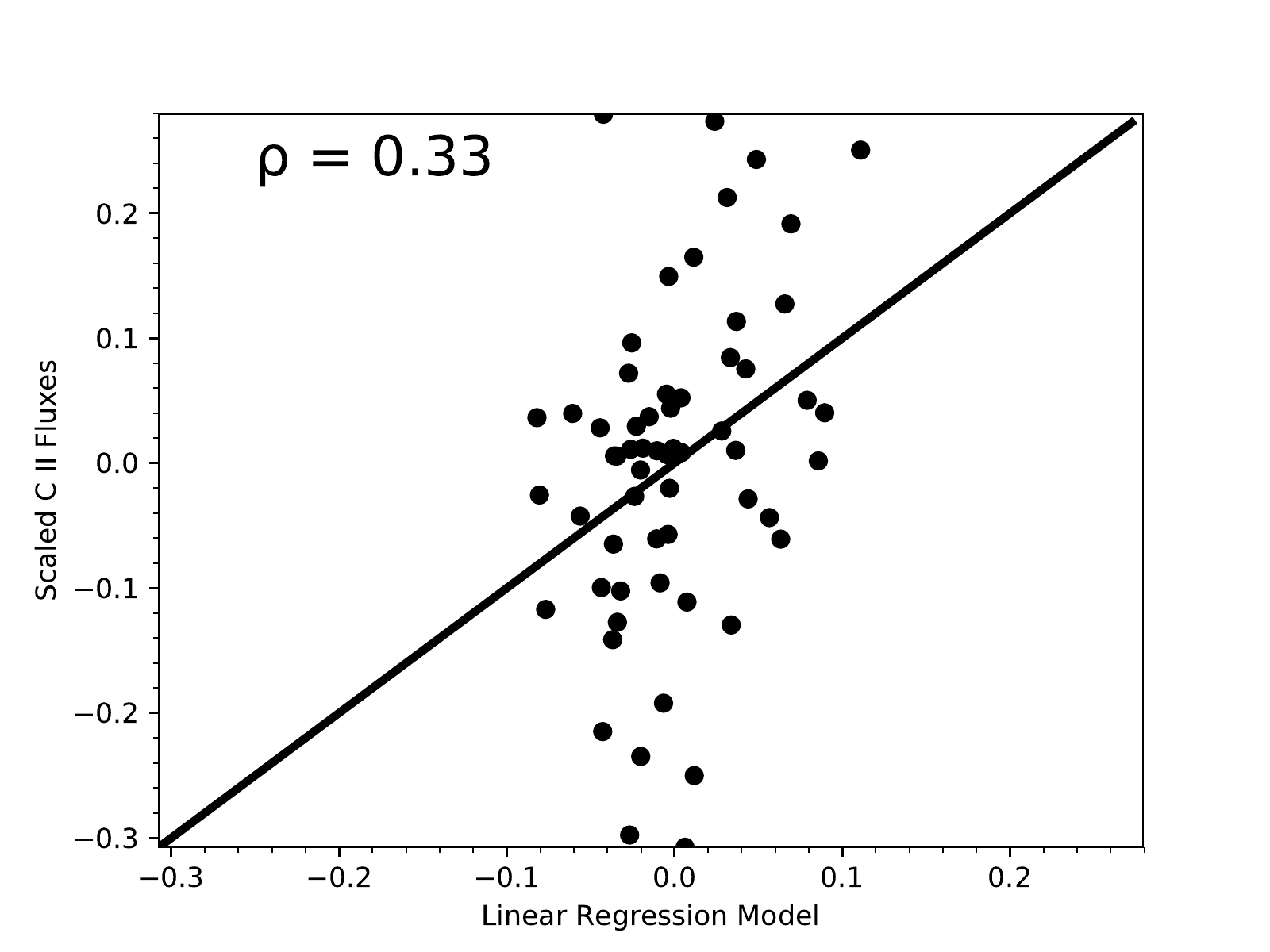} & 
\includegraphics[width=0.45\textwidth,angle=0,trim={.0in 0.0in 0.0in 0.0in},clip]{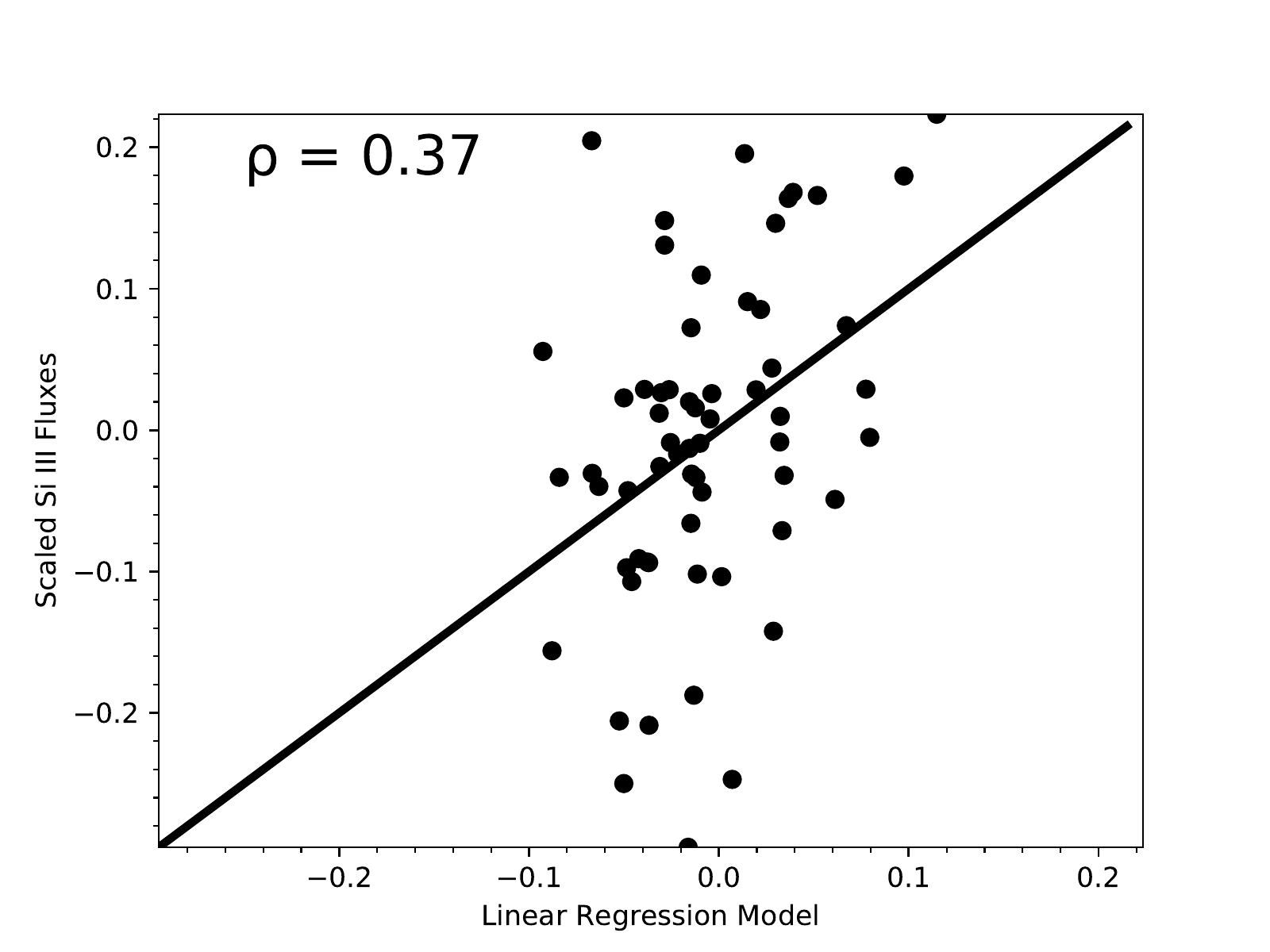} \\
\includegraphics[width=0.45\textwidth,angle=0,trim={.0in 0.0in 0.0in 0.0in},clip]{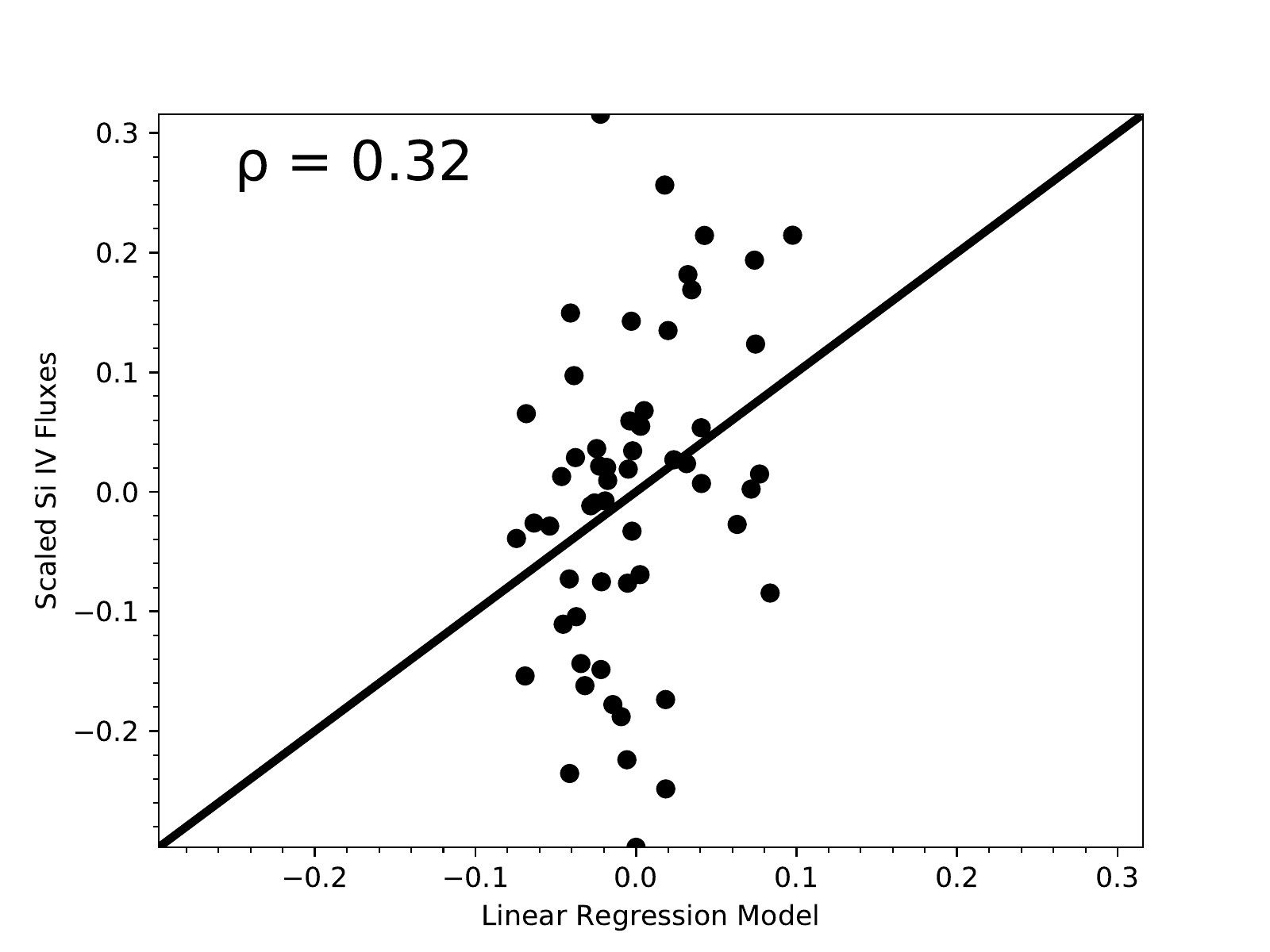} & 
\includegraphics[width=0.45\textwidth,angle=0,trim={.0in 0.0in 0.0in 0.0in},clip]{NV_PCAmodel.pdf}
\end{tabular}
\end{center}
\vspace{-0.2in}
\caption{Comparison of the linear regression model and relative ion fluxes for planet-hosting stars. The Spearman correlation coefficient ($\rho$) describes the agreement between the multivariate linear model and our observations, where we expect $|\rho| = 1$ for a perfect model. Linear regression plots for all four ions are displayed in Appendix C. }
\label{fig-proveit} 
\vspace{0.05in}
\end{figure} 

\subsection{Non-Planet Hosts}

The same analysis that we carried out for the planet-hosting stars was also applied to the sample of non-planet hosts, with the SPI parameter dropped as a predictor variable. The resulting correlation matrix, calculated from the scaled parameters, is

\[
\begin{blockarray}{cccccc}
	& \mathbf{P_{rot}} & \mathbf{V} & \mathbf{B-V} & \mathbf{d} & \mathbf{T_{eff}}  \\
	\begin{block}{c[ccccc]}
	\mathbf{P_{rot}} & 1 & -0.25 & -0.20 & -0.59 & -0.22 \\
	\mathbf{V} & -0.25 & 1 & 0.14 & 0.52 & -0.33 \\
	\mathbf{B-V} & -0.20 & 0.14 & 1 & -0.055 & -0.40 \\
	\mathbf{d} & -0.59 & 0.52 & -0.055 & 1 & 0.30 \\
	\mathbf{T_{eff}} & -0.22 & -0.33 & -0.40 & 0.30 & 1 \\
	\end{block}
\end{blockarray}
\]

corresponding to eigenvalues $\left[1.96, 1.58, 0.90, 0.36, 0.20 \right]$ and principal components 
\begin{equation}
\begin{split}
PC_1 &= 0.58 \times P_{rot, scaled} - 0.47 \times V_{scaled} - 0.09 \times \left(B-V \right)_{scaled} \\
&- 0.65 \times d_{scaled} - 0.14 \times T_{eff, scaled}  \\
PC_2 &= -0.06 \times P_{rot, scaled} - 0.38 \times V_{scaled} - 0.57 \times \left(B-V \right)_{scaled} \\
&+ 0.15 \times d_{scaled} + 0.71 \times T_{eff, scaled} \\
PC_3 &= -0.46 \times P_{rot, scaled} - 0.55 \times V_{scaled} + 0.65 \times \left(B-V \right)_{scaled} \\
&- 0.14 \times d_{scaled} + 0.22 \times T_{eff, scaled}  \\
PC_4 &= -0.65 \times P_{rot, scaled} - 0.14 \times V_{scaled} - 0.49 \times \left(B-V \right)_{scaled} \\
&- 0.31 \times d_{scaled} - 0.47 \times T_{eff, scaled} \\
PC_5 &= -0.17 \times P_{rot, scaled} + 0.56 \times V_{scaled} + 0.05 \times \left(B-V \right)_{scaled} \\
& - 0.66 \times d_{scaled} + 0.46 \times T_{eff, scaled} \\
\end{split}
\end{equation}

Table \ref{PC_correlations_noplanets} lists the correlation coefficients between the principal components and the stellar parameters. The first principal component is strongly correlated with $P_{rot}$ and $d$, while $B-V$ and $T_{eff}$ are more significant in $PC_2$. $P_{rot}$ alone is also correlated with $PC_3$ and $PC_4$. We drop $PC_5$ from the regression analysis, since this component is not strongly correlated with any of the stellar parameters, and find that the linear model appears to describe the sample of non-planet hosts better than the sample of planet hosts. We find larger Spearman rank coefficients between the models and observed fractional luminosities for all four elements (see Table \ref{PC_linreg_nonplanets}) in the non-planet hosts than in the planet hosting sample. Both $PC_1$ and $PC_2$ contribute significantly to the spread in the data for all ions except Si III, again in contrast with the planet-hosting sample.

\clearpage

\begin{deluxetable}{ccccccccccccc}
\tablecaption{Correlations between Stellar Parameters and Principal Components: Planet Hosts \label{PC_correlations}
}
\tablewidth{0 pt}
\tabletypesize{\scriptsize}
\tablehead{
\colhead{} & \rlap{\kern\tabcolsep$PC_1$} & \colhead{} & \rlap{\kern\tabcolsep$PC_2$} & \colhead{} & \rlap{\kern\tabcolsep$PC_3$} & \colhead{} & \rlap{\kern\tabcolsep$PC_4$} & \colhead{} & \rlap{\kern\tabcolsep$PC_5$} & \colhead{} & \rlap{\kern\tabcolsep$PC_6$} & \colhead{} \\
\colhead{} & \colhead{$\rho$} & \colhead{$p$-value} & \colhead{$\rho$} & \colhead{$p$-value} & \colhead{$\rho$} & \colhead{$p$-value} & \colhead{$\rho$} & \colhead{$p$-value} & \colhead{$\rho$} & \colhead{$p$-value} & \colhead{$\rho$} & \colhead{$p$-value} \\
}
\startdata
$P_{rot}$ & \bf{-0.798} & $2 \times 10^{-16}$ & -0.144 & 0.2 & -0.165 & 0.2 & 0.287 & 0.02 & 0.190 & 0.1 & -0.0128 & 0.9 \\
$V$ & -0.455 & $9 \times 10^{-5}$ & \bf{0.653} & $1 \times 10^{-9}$ & -0.256 & 0.03 & -0.173 & 0.2 & 0.293 & 0.01 & 0.0948 & 0.4 \\
$B-V$ & \bf{-0.929} & $2 \times 10^{-30}$ & -0.0144 & 0.9 & 0.160 & 0.2 & -0.168 & 0.2 & 0.354 & 0.003 & 0.0152 & 0.9 \\
$d$ & \bf{0.558} & $6 \times 10^{-7}$ & \bf{0.654} & $1 \times 10^{-9}$ & \bf{-0.522} & $4 \times 10^{-6}$ & 0.0368 & 0.8 & 0.120 & 0.3 & -0.0407 & 0.7 \\
$T_{eff}$ & \bf{0.927} & $3 \times 10^{-30}$ & 0.0807 & 0.5 & -0.199 & 0.1 & 0.189 & 0.1 & -0.285 & 0.02 & 0.136 & 0.3 \\
$\log \left( SPI \right)$ & 0.397 & $7 \times 10^{-4}$ & \bf{0.694} & $4 \times 10^{-11}$ & \bf{0.512} & $7 \times 10^{-6}$ & 0.410 & $5 \times 10^{-4}$ & 0.00216 & 0.986 & -0.0449 & 0.7 \\
\enddata
\tablecomments{$\rho$ is the Spearman rank coefficient, where $| \rho | = 1$ for a perfect correlation and $\rho = 0$ when the data are uncorrelated. The $p$-value indicates the likelihood of obtaining $|\rho|$ closer to 1, under the assumption that there is no correlation between the two parameters (i.e. the data are randomly distributed). Here we define ``significant" correlations here as having $| \rho | > 0.5$ (marked in bold).}
\end{deluxetable}

\begin{deluxetable}{cccc}
\tablecaption{Multiple Linear Regression Results: Principal Components as Predictor Variables for Planet Hosts \label{PC_linreg}
}
\tablewidth{0 pt}
\tabletypesize{\scriptsize}
\tablehead{
\colhead{Element} & \colhead{Predictor Variable} & \colhead{$\beta$} & \colhead{95\% Confidence Interval} \\
}
\startdata
N V & $PC_1$ & -0.2516 & (-0.356, -0.147) \\
 & $PC_2$ & 0.3193 & (0.123, 0.516) \\
 & $PC_3$ & 0.3928 & (0.166, 0.620) \\
 & y-intercept & -0.0092 & (-0.032, 0.014) \\
 & $\rho$ & 0.581 & $p = 1 \times 10^{-6}$ \\
\hline
Si III & $PC_1$ & 0.0562 & (-0.071, 0.184) \\
 & $PC_2$ & 0.2424 & (0.005, 0.480) \\
 & $PC_3$ & 0.3029 & (0.024, 0.582) \\
 & y-intercept & -0.0067 & (-0.035, 0.021) \\
 & $\rho$ & 0.370 & $p = 3 \times 10^{-3}$ \\
\hline
Si IV & $PC_1$ & -0.0012 & (-0.158, 0.156) \\
 & $PC_2$ & 0.288 & (-0.022, 0.598) \\
 & $PC_3$ & 0.252 & (-0.115, 0.619) \\
 & y-intercept & 0.0008 & (-0.034, 0.035) \\
 & $\rho$ & 0.325 & $p= 0.01$ \\
\hline
C II & $PC_1$ & 0.0307 & (-0.122, 0.183) \\
 & $PC_2$ & 0.268 & (-0.010, 0.546) \\
 & $PC_3$ & 0.309 & (-0.047, 0.665) \\
 & y-intercept & 0.0027 & (-0.030, 0.036) \\
 & $\rho$ & 0.327 & $p = 0.01$ \\
\enddata
\tablecomments{
$\rho$ is the Spearman rank coefficient, calculated between the multiple linear regression model and the observed fractional luminosities in a given element.}
\end{deluxetable}

\begin{deluxetable}{ccccccccccc}
\tablecaption{Correlations between Stellar Parameters and Principal Components: Non-Planet Hosts \label{PC_correlations_noplanets}
}
\tablewidth{0 pt}
\tabletypesize{\scriptsize}
\tablehead{
\colhead{} & \rlap{\kern\tabcolsep$PC_1$} & \colhead{} & \rlap{\kern\tabcolsep$PC_2$} & \colhead{} & \rlap{\kern\tabcolsep$PC_3$} & \colhead{} & \rlap{\kern\tabcolsep$PC_4$} & \colhead{} & \rlap{\kern\tabcolsep$PC_5$} & \colhead{} \\
\colhead{} & \colhead{$\rho$} & \colhead{$p$-value} & \colhead{$\rho$} & \colhead{$p$-value} & \colhead{$\rho$} & \colhead{$p$-value} & \colhead{$\rho$} & \colhead{$p$-value} & \colhead{$\rho$} & \colhead{$p$-value} \\
}
\startdata
$P_{rot}$ & \bf{0.796} & $1 \times 10^{-7}$ & -0.133 & $0.5$ & \textbf{-0.552} & 0.002 & \textbf{-0.566} & 0.001 & -0.0260 & 0.9 \\
$V$ & -0.365 & $0.05$ & -0.312 & 0.09 & -0.187 & 0.3 & 0.180 & 0.3 & -0.458 & 0.01 \\
$B-V$ & 0.230 & $0.2$ & \textbf{-0.868} & $5 \times 10^{-10}$ & 0.287 & 0.1 & -0.170 & 0.4 & 0.128 & 0.5 \\
$d$ & \textbf{-0.901} & $1 \times 10^{-11}$ & 0.381 & $0.04$ & -0.269 & 0.2 & -0.141 & 0.5 & -0.421 & 0.02 \\
$T_{eff}$ & -0.471 & $0.009$ & \textbf{0.905} & $7 \times 10^{-12}$ & -0.0376 & 0.8 & -0.0109 & 0.9 & -0.0245 & 0.9 \\
\enddata
\tablecomments{$\rho$ is the Spearman rank coefficient, where $| \rho | = 1$ for a perfect correlation and $\rho = 0$ when the data are uncorrelated. The $p$-value indicates the likelihood of obtaining $|\rho|$ closer to 1, under the assumption that there is no correlation between the two parameters (i.e. the data are randomly distributed). Here we define ``significant" correlations here as having $| \rho | > 0.5$ (marked in bold).}
\end{deluxetable}

\begin{deluxetable}{cccc}
\tablecaption{Multiple Linear Regression Results: Principal Components as Predictor Variables for Non-Planet Hosts \label{PC_linreg_nonplanets}
}
\tablewidth{0 pt}
\tabletypesize{\scriptsize}
\tablehead{
\colhead{Element} & \colhead{Predictor Variable} & \colhead{$\beta$} & \colhead{95\% Confidence Interval} \\
}
\startdata
N V & $PC_1$ & -0.5461 & (-0.810, -0.282) \\
 & $PC_2$ & -0.3460 & (-0.578, -0.114) \\
 & $PC_3$ & 0.8376 & (0.047, 1.628) \\
 & $PC_4$ & 0.4471 & (-0.327, 1.221) \\
 & y-intercept & -0.0161 & (-0.068, 0.036) \\
 & $\rho$ & 0.897 & $p = 3 \times 10^{-9}$ \\
\hline
Si III & $PC_1$ & -0.4578 & (-0.782, -0.134) \\
 & $PC_2$ & -0.2470 & (-0.527, 0.033) \\
 & $PC_3$ & 0.7144 & (-0.253, 1.682) \\
 & $PC_4$ & 0.2960 & (-0.639, 1.231) \\
 & y-intercept & -0.0029 & (-0.065, 0.059) \\
 & $\rho$ & 0.831 & $p = 1 \times 10^{-7}$ \\
\hline
Si IV & $PC_1$ & -0.4907 & (-0.739, -0.242) \\
 & $PC_2$ & -0.2423 & (-0.452, -0.033) \\
 & $PC_3$ & 0.6423 & (-0.093, 1.377) \\
 & $PC_4$ & 0.3728 & (-0.326, 1.071) \\
 & y-intercept & -0.0019 & (-0.047, 0.043) \\
 & $\rho$ & 0.833 & $p= 4 \times 10^{-8}$ \\
\hline
C II & $PC_1$ & -0.4987 & (-0.761, -0.237) \\
 & $PC_2$ & -0.2831 & (-0.508, -0.058) \\
 & $PC_3$ & 0.8234 & (0.042, 1.605) \\
 & $PC_4$ & 0.2921 & (-0.457, 1.041) \\
 & y-intercept & -0.0165 & (-0.064, 0.031) \\
 & $\rho$ & 0.846 & $p = 7 \times 10^{-9}$ \\
\enddata
\tablecomments{
$\rho$ is the Spearman rank coefficient, calculated between the multiple linear regression model and the observed fractional luminosities in a given element.}
\end{deluxetable}


  \bibliography{ms_spi24}

\end{document}